\documentclass[a4paper,11pt]{article}

\usepackage[utf8]{inputenc}

\usepackage{amsmath, amssymb, amsfonts, amsthm} 
\usepackage{graphicx}
\usepackage[usenames,dvipsnames]{color}
\usepackage{booktabs}
\usepackage{verbatim}
\usepackage{natbib}
\usepackage[titletoc,title]{appendix}
\usepackage[lined, boxed]{algorithm2e}
\usepackage[bf, font += footnotesize]{caption}
\usepackage[font += footnotesize]{subcaption}
\usepackage{pifont}
\usepackage{dashrule}
\usepackage{authblk} 
\usepackage[colorlinks = true, citecolor = blue, linkcolor = blue]{hyperref}

\usepackage{times}

\title{Estimation of a non-stationary model for annual precipitation
      in southern Norway using replicates of the spatial field} 
\date{}

\author[1]{Rikke Ingebrigtsen\thanks{Corresponding author}}
\author[2]{Finn Lindgren}
\author[1]{Ingelin Steinsland}
\author[3]{Sara Martino}
\affil[1]{Department of Mathematical Sciences, Norwegian University of Science and Technology (NTNU), Trondheim, N-7491, Norway}
\affil[2]{Department of Mathematical Sciences, University of Bath, Claverton Down, Bath, BA2~7AY, United Kingdom}
\affil[3]{SINTEF Energy Research, Trondheim, N-7491, Norway}

\begin{document}

\renewcommand\Authands{ and }
\renewcommand\Affilfont{\itshape\small}
\setlength{\affilsep}{0.7cm}

\maketitle

\begin{abstract}
Estimation of stationary dependence structure parameters using only
a single realisation of the spatial process, typically leads to inaccurate
estimates and poorly identified parameters. A common way to handle this is
to fix some of the parameters, or within the Bayesian framework, impose
prior knowledge. In many applied settings, stationary models are not
flexible enough to model the process of interest, thus non-stationary
spatial models are used. However, more flexible models usually means more
parameters, and the identifiability problem becomes even more challenging.
We investigate aspects of estimation of a Bayesian non-stationary spatial
model for annual precipitation using observations from multiple years.
The model contains replicates of the spatial field, which increases precision
of the estimates and makes them less prior sensitive. Using \texttt{R-INLA},
we analyse precipitation data from southern Norway, and investigate statistical
properties of the replicate model in a simulation study. The non-stationary spatial
model we explore belongs to a recently introduced class of stochastic partial
differential equation (SPDE) based spatial models. This model class allows for
non-stationary models with explanatory variables in the dependence structure.
We derive conditions to facilitate prior specification for these types of
non-stationary spatial models.

\end{abstract}



\section{Introduction}

At the core of any statistical analysis is the wish to learn about 
a process or phenomena based on available data.
The purpose of the analysis can be to gain insight about the process 
of interest and/or make predictions related to this process. In this 
paper, we study annual precipitation in southern Norway, and we aim 
both to learn about this process from data, and make predictions at 
spatial locations without observations. 

To be able to learn about a process from data we need a model that 
is 1) realistic, 2) interpretable, and 3) possible to draw inference 
from with the available data. For most realistic processes, compromises 
between these three aims are necessary. We can simplify the model, 
impose more knowledge or restrictions, or get more data. In this paper, 
we use a model that is based on physical understanding, but is very 
simplified. A Bayesian approach is taken, and knowledge about the 
system is expressed through carefully chosen prior distributions. 
Further, we are able to utilise more data to learn about the 
precipitation process by extending the model such that several years 
of annual precipitation observations can be used. 

The precipitation process is driven by humidity, changes in 
circulations, weather patterns and temperature, 
as well as interactions with the topography. Southern 
Norway is separated by the mountain range Langfjella: to the west there 
is a mountainous coastline dominated by large fjords, while eastern 
Norway has a more gentle landscape consisting of valleys and lowlands. 
This topographical difference is reflected in the climate. Humid oceanic 
winds hit the west coast and are forced to ascend due to the mountains. 
The result is that the western part of Norway receives high amounts of 
precipitation, while the eastern part is relatively dry being located 
in the ``rain shadow''. This phenomenon is known as orographic 
precipitation. Statistical modelling and spatial prediction of 
precipitation in Norway are challenging tasks 
\citep{Orskaug2011, Ingebrigtsen2014, Dyrrdal2014}. In some areas there 
are large variations in the amount of precipitation within relatively 
short distances, while other areas are more homogeneous. These features 
can be explained by the physics of the precipitation process in a complex 
and diverse terrain. 

The main purpose of precipitation interpolation in Norway is as input to 
hydrological forecasting models to predict run-off either for flood warnings 
or to better schedule hydro-power production. Because of the topography, most 
of the catchment (where prediction is of interest) is often in 
mountainous areas, while the precipitation gauges (observations) are 
located in lower areas due to easier and cheaper maintenance. It is not 
uncommon that all the closest precipitation gauges are at lower elevation 
than the lowest point of the catchment of interest. Since changes in 
elevation is one of the driving forces of precipitation, this causes a 
problem of non-preferential sampling, and to do spatial interpolation we 
will need to do extrapolation with respect to elevation. This calls for 
a model with a physical basis, and a way to tackle this challenge is to 
include elevation in the model. 

A statistical model for annual precipitation over southern Norway should 
have two important properties. We have already argued that elevation should 
be included in the model. In addition, as locations close in space are more 
alike than locations further apart, the model should also include a spatial 
process. If the dependence structure of a spatial process changes within the 
domain it is defined, the process is non-stationary. Because of the complex 
Norwegian topography, it is reasonable to use a non-stationary spatial process, 
and that the non-stationarity depends on the topography.

A flexible and popular framework for statistical modelling is hierarchical 
models. The general scheme consists of two levels. The first level is the 
likelihood part, which specifies the model for the observations given a 
latent field. The second level specifies the model for the latent field, 
which can consist of several terms, e.g.\ a term for elevation and a spatial 
process. Both levels of the hierarchy contain parameters we would like to 
infer something about based on the observations. With a likelihood approach 
to inference, identifiability (or estimability) might become an issue. It is 
not always possible to estimate all parameters, but rather only some of them, 
or a function of them. Furthermore, one often does not know whether there are 
identifiablity issues \citep{Lele2010}. 

A Bayesian approach is often taken within the hierarchical modelling framework. 
In the Bayesian paradigm the identifiability challenge is solved by 
introducing a third level in the hierarchy; prior models for the parameters. 
Prior specifications for hierarchical model parameters are challenging for 
applied Bayesians, and common priors might have undesired properties 
\citep{Gelman2006,Sorbye2014,Simpson2014}. Especially if one is not aware of 
the identification properties, priors can give unintended inference consequences.  

Gaussian random fields (GRFs) play an important role in spatial statistics; they 
are commonly used to model the spatial process itself, or as building blocks in 
hierarchical models \citep[see e.g.][]{Banerjee2004, Cressie2011}. 
The GRF is characterised by a mean function and a covariance function. In this 
paper, we use the stochastic partial differential equation (SPDE) approach 
introduced in \citet{Lindgren2011}, where it was shown that Gaussian Markov 
random field (GMRF) approximations to GRFs with the Mat\'ern covariance function 
can be derived from an SPDE formulation. The Markovian approximation enables fast 
simulations and inference, and integrated nested Laplace approximations 
\citep[INLA,][]{Rue2009} can be applied.

A constant mean GRF is stationary if the covariance function is invariant to spatial shifts.
For many spatial processes a stationarity assumption is unrealistic, and there 
has been great interest in the literature to develop non-stationary spatial 
models, included contributions towards having explanatory variables in the 
dependence structure. Deformation methods obtain a non-stationary model by 
deformation of a latent space where the process is stationary 
\citep{Sampson1992}, and the axes of this latent space can be defined by 
explanatory variables \citep{Schmidt2011}. Another approach is spatial 
convolution models, where smoothing kernels are convolved with a white 
noise process \citep{Higdon1999, Paciorek2006}, or with a spatial dependent 
process \citep{Fuentes2002}, to obtain non-stationarity. \citet{Reich2011} 
extends the convolution model from \citet{Fuentes2002} and use spatial 
covariates as kernels, while \citet{Neto2014} extends the convolution 
approaches in \citet{Higdon1999} and \citet{Paciorek2006} to incorporate 
covariates in the covariance structure. The SPDE formulation of 
\citet{Lindgren2011} also provides ways to easily introduce non-stationarity 
by letting the parameters or operators vary in space 
\citep{Lindgren2011, Bolin2011, Fuglstad2014, Fuglstad2014-precip, 
Ingebrigtsen2014}. We will use the approach in \citet{Ingebrigtsen2014}, 
where the SPDE formulation is used, and non-stationarity is obtained by 
covariance parameters that depends on spatial explanatory variables. 

A spatial model with an unknown spatially varying mean and a general non-stationary 
covariance is completely unidentifiable from one realisation of the process. 
Consider two models, where 1) the observations are exact observations of the spatially 
varying mean, and 2) the observations come from a model with a constant mean 
and a spatial field with a complex non-stationary dependence structure. 
The difference between these two models first appears when there are observed 
several replicates of the process. 
For model 1) the observations will for a given location be identical for all replicates. 
For model 2) the observations will vary between replicates according to the covariance 
model. Hence, the estimated model from one realisation is completely determined by 
the precise model choice and priors. Further, for spatial Gaussian models with known mean and 
a stationary Mat{\'e}rn covariance function, it has been shown that for 
a fixed domain and in-fill asymptotics there are no consistent estimators for 
all covariance parameters \citep{Zhang2004, Kaufman2013}, i.e.\ the variance of the 
estimators do not converge to zero as the number of observations increases. 
This means that even for a stationary model with known mean, there is limited 
information about some of the dependence parameters over a fixed domain from 
one realisation of the process. Therefore, spatial models, including specification 
of priors, have to be carefully chosen. Further, whenever possible, replicates of 
the process should be used to make inference. From a practical perspective, 
it is also easier to observe replicates of the spatial process, than it is to 
establish new sampling locations.
  
This paper is motivated by \citet{Ingebrigtsen2014}, and the challenges reported 
there. \citet{Ingebrigtsen2014} suggest to use a Bayesian non-stationary SPDE based 
spatial model for annual total precipitation in southern Norway. The model has 
interpretable parameters, and the non-stationary model is shown to be superior 
to the corresponding stationary model with respect to the deviance information 
criterion \citep[DIC,][]{DIC} and predictive performance. 
However, numerical problems and uncomfortable large prior sensitivity were reported. 
In a simulation study, some of the posterior mean estimates and credible intervals 
showed poor statistical properties, i.e.\ bias and low coverage. The model was 
interpretable and realistic, but it was not possible to make satisfactory inference 
with the available data. This inspired us to understand the model class better, to 
enable us to specify better prior distributions, and to explore the opportunity to 
use more data to learn about the underlying non-stationary process. 
In this paper, we propose a framework for prior specification for the non-stationary 
SPDE based models with explanatory variables in the dependence structure. Further, 
we propose and test a model for annual precipitation that allows for replicates, 
such that multiple years of annual precipitation observations can be used to make inference.
The replicate model is set up such that the model for one replicate 
corresponds to the model in \citet{Ingebrigtsen2014}. 
Properties of the model and study system are explored in a simulation 
study.

This paper is organised as follows. In Section~\ref{sec:SPDE}, we present 
the SPDE approach to spatial statistics and how this approach is used to 
define the non-stationary GRF model with spatial explanatory variables in 
the dependence structure. Section~\ref{sec:Model} presents annual 
precipitation data from southern Norway, the suggested replicate model for 
annual precipitation, prior specification for the SPDE parameters, and a 
brief outline of the inference procedure and model evaluation. The annual 
precipitation dataset is analysed in Section~\ref{sec:Data_analysis}, and 
Section~\ref{sec:Simulation_study} presents the simulation study. We end 
the paper with a discussion in Section~\ref{sec:Discussion}.

\section{The SPDE approach to spatial statistics}\label{sec:SPDE}

Classical geostatistical models for point referenced data can easily
become computer intensive as the computation time increases cubically
with the number of observation locations. The computational bottleneck
comes from the need to perform matrix operations with large and dense
covariance matrices. For gridded data and for graph models, Gaussian
Markov random fields (GMRF) have been used as computationally
efficient alternatives, since they allow the computations to be
carried out using a sparse precision matrix, i.e.\ the inverse of the
covariance matrix \citep{GMRF}.  \citet{Lindgren2011} suggest an
approach where the two approaches are unified, by specifying a spatial
Gaussian random field (GRF) model via a stochastic partial
differential equation (SPDE) formulation instead of explicitly
defining a covariance function.  Many SPDE models are also Markov
models, or can be approximated with Markov models, which means that
the spatial model itself can be defined in continuous space, while the
computations can be carried out efficiently due to properties of
discrete GMRFs \citep[see e.g.][]{GMRF}. However, it turns out that
the benefits of the SPDE approach are not only computational; the SPDE
formulation also provides a framework for introducing non-stationarity
to the dependence structure \citep{Lindgren2011, Bolin2011,
Ingebrigtsen2014, Fuglstad2014}.

The basis of the SPDE approach is the equation  
\begin{equation}\label{eq:spde-original}
 (\kappa^2 - \Delta)^{\alpha/2}(\tau x(\boldsymbol{s})) = \mathcal{W}(\boldsymbol{s}),
\quad \boldsymbol{s}\in\mathbb{R}^d,
\end{equation}
whose stationary solutions $x(\boldsymbol{s})$ are GRFs with Mat\'ern covariance functions
for every $\alpha > d/2$, \citep{Whittle1954, Whittle1963} and GMRFs
when $\alpha$ is an integer.  In the formulation above, $\mathcal{W}$
is spatial Gaussian white noise, $\Delta$ is the Laplacian, $\kappa >
0$ controls the spatial range, $\tau > 0$ controls the variance, and
$\alpha$ controls the smoothness.
The Mat\'ern 
covariance function between locations $\boldsymbol{s}_1$ and $\boldsymbol{s}_2$ in 
$\mathbb{R}^d$ is given by
\begin{equation}\label{eq:matern}
C(\boldsymbol{s}_1,\boldsymbol{s}_2)=\frac{\sigma^2}{2^{\nu-1}\Gamma(\nu)}
(\kappa\|\boldsymbol{s}_2-\boldsymbol{s}_1\|)^\nu K_\nu(\kappa\|\boldsymbol{s}_2-\boldsymbol{s}_1\|), 
\end{equation}
where $K_\nu$ is the modified Bessel function of the second kind and order $\nu>0$, 
$\kappa$ is a positive scaling parameter, and $\sigma^2$ is the marginal variance. 
The parameters in the SPDE and the Mat\'ern covariance function of its solution are coupled; the 
smoothness parameter of the Mat\'ern function is $\nu = \alpha - d/2$, and the 
marginal variance is 
\begin{equation}\label{eq:margvar}
 \sigma^2 = \frac{\Gamma(\nu)}{\Gamma(\alpha)(4\pi)^{d/2}\kappa^{2\nu}\tau^2}.
\end{equation}

The SPDE above have stationary solutions with Mat{\'e}rn covariance. However, since 
properties of the random field can be characterised by an SPDE rather than a 
covariance function, we can modify the SPDE to obtain GRFs with more flexible 
dependence structures. One way to introduce more flexibility is allowing the 
parameters to change with location. Such local specification of parameters is 
possible due to the local nature of the differential operators.  

The non-stationary SPDE based GRF model in \citet{Ingebrigtsen2014} is defined via
\begin{equation}\label{eq:spde}
 (\kappa(\boldsymbol{s})^2-\Delta)(\tau(\boldsymbol{s}) x(\boldsymbol{s})) =
\mathcal{W}(\boldsymbol{s}), \quad \boldsymbol{s} \in \mathbb{R}^2,
\end{equation}
which is a generalisation of the stationary model with $d=2$ and $\alpha=2$.  The smoothness of the solutions depend on the smoothness of the non-stationary parameter functions.  For piecewise continuous $\kappa(\boldsymbol{s})$ and $\tau(\boldsymbol{s})$, both bounded away from zero, the product $\tau(\boldsymbol{s})x(\boldsymbol{s})$ will have at least piecewise H{\"o}lder continuity $1-\epsilon$ for any $\epsilon>0$, corresponding to $\nu=1$ in the stationary case.  The continuity of $x(\boldsymbol{s})$ is then determined via the continuity of $\tau(\boldsymbol{s})$, since it acts as a direct and local scaling factor.
Further, for deterministic basis
functions $b_{\tau, j}(\cdot)$ and $b_{\kappa, j}(\cdot)$, and weight parameters
$\boldsymbol{\theta}$, log-linear models for $\tau(\boldsymbol{s})$ and
$\kappa(\boldsymbol{s})$ are given by
\begin{equation}\label{eq:kappa-tau}
 \log \tau(\boldsymbol{s}) = \theta_{\tau, 1} + \sum_{j} b_{\tau, j}(\boldsymbol{s})\theta_{\tau, j}, \quad
 \log \kappa(\boldsymbol{s}) = \theta_{\kappa, 1} + \sum_{j} b_{\kappa, j}(\boldsymbol{s})\theta_{\kappa, j}.
\end{equation}
The basis functions can be spherical harmonics or B-splines as in \citet{Lindgren2011},
or spatial explanatory variables as in \citet{Ingebrigtsen2014}.
Spatially rough explanatory variables may benefit from pre-smoothing, which in practice has the effect of moving some of the local structure into the observation noise.

The non-stationary SPDE based model has the computational advantages mentioned 
in the beginning of this section. It is possible to derive a GMRF approximation 
to the non-stationary GRF, which speeds up the computations considerably. The 
approximation is found using a finite element method (FEM) \citep[see e.g.][]{FEM}. 
The spatial domain is discretised into non-intersecting triangles with $m$ nodes 
and the approximation to the infinite dimensional GRF is built on the finite 
basis function representation
\begin{equation}\label{eq:basis_repr}
 x(\boldsymbol{s})=\sum_{i=1}^m \psi_i(\boldsymbol{s})w_i, 
\end{equation}
with deterministic basis functions $\{\psi_i\}_{i=1,\dots,m}$ and weights 
$\boldsymbol{w}=(w_1, \dots, w_m)$. 
The approximation properties depend on the quality of the
triangulation mesh \citep{Lindgren2011}, with generally
increasing accuracy for decreasing triangle edge lengths.
The Gaussian weights are chosen so that 
the representation in \eqref{eq:basis_repr} approximates the distribution of 
the solution to the SPDE in \eqref{eq:spde-original} (in a stochastically 
weak sense). \citet{Lindgren2011} use piecewise polynomial basis functions 
with compact support to obtain a Markov approximation, and the sparse structure 
of the precision matrix $\boldsymbol{Q}$ of $\boldsymbol{w}$ is determined by the triangulation. 
The elements of $\boldsymbol{Q}$ are functions of the 
$\theta$'s in \eqref{eq:kappa-tau} that define the spatially varying covariance 
parameters $\tau$ and $\kappa$, i.e.\
\begin{equation}\label{eq:SPDE_Q}
 \boldsymbol{Q} = \boldsymbol{T}(\boldsymbol{K}^2\boldsymbol{C}\boldsymbol{K}^2 + \boldsymbol{K}^2\boldsymbol{G} + 
\boldsymbol{G}\boldsymbol{K}^2 +\boldsymbol{G}\boldsymbol{C}^{-1}\boldsymbol{G})\boldsymbol{T},
\end{equation}
where $\boldsymbol{T}$ and $\boldsymbol{K}$ are diagonal matrices with 
$T_{ii}=\tau(\boldsymbol{s}_i)$ and $K_{ii}=\kappa(\boldsymbol{s}_i)$, 
and $i$ is an index over the $m$ nodes in the triangulation. The matrices 
$\boldsymbol{C}$ and $\boldsymbol{G}$ are finite element structure matrices, 
where $\boldsymbol{C}$ is diagonal, with 
$C_{ii}=\int \psi_i(\boldsymbol{s})\,d\boldsymbol{s}$, and $\boldsymbol{G}$ 
is sparse positive semi-definite, with 
$G_{ij}=\int \nabla\psi_i(\boldsymbol{s}) \cdot \nabla\psi_j(\boldsymbol{s}) \,d\boldsymbol{s}$.

\section{Annual precipitation: data, model, and inference}\label{sec:Model}

This section begins with a presentation of the annual precipitation data 
from southern Norway. Then, the stationary and non-stationary annual 
precipitation models that were compared in \citet{Ingebrigtsen2014} 
are extended to include data from multiple years. New conditions for 
prior specification of the dependence structure parameters are derived. 
Further, a brief outline of the inference procedure using integrated 
nested Laplace approximation \citep[INLA,][]{Rue2009} is given. The 
section ends with a presentation of how the models are evaluated.  

\subsection{Data}\label{sec:data}

\begin{table}
 \footnotesize
 \centering
 \caption{Five years of annual total precipitation data from southern Norway. 
          The data are from the Norwegian Meteorological Institute's database eKlima. 
          The unit of the annual total precipitation is metres.}
 \label{tab:data}
 \begin{tabular}{cccccc}
 \toprule
 year      & observations & mean   & median &  minimum  & maximum  \\
 \midrule
 2008-2009 &     233      & 1.2778 & 1.0495 &  0.3560   & 3.6886 \\ 
 2009-2010 &     222      & 1.0091 & 0.8930 &  0.3055   & 2.8376 \\ 
 2010-2011 &     194      & 1.1825 & 0.9879 &  0.4008   & 3.5118 \\ 
 2011-2012 &     141      & 1.6450 & 1.2938 &  0.4051   & 4.5472 \\ 
 2012-2013 &     199      & 1.1393 & 0.9514 &  0.3120   & 3.4627 \\ 
 \bottomrule
 \end{tabular}
\end{table}

\begin{figure}
 \centering
  \begin{subfigure}{0.48\textwidth}
   \includegraphics[width=\textwidth]{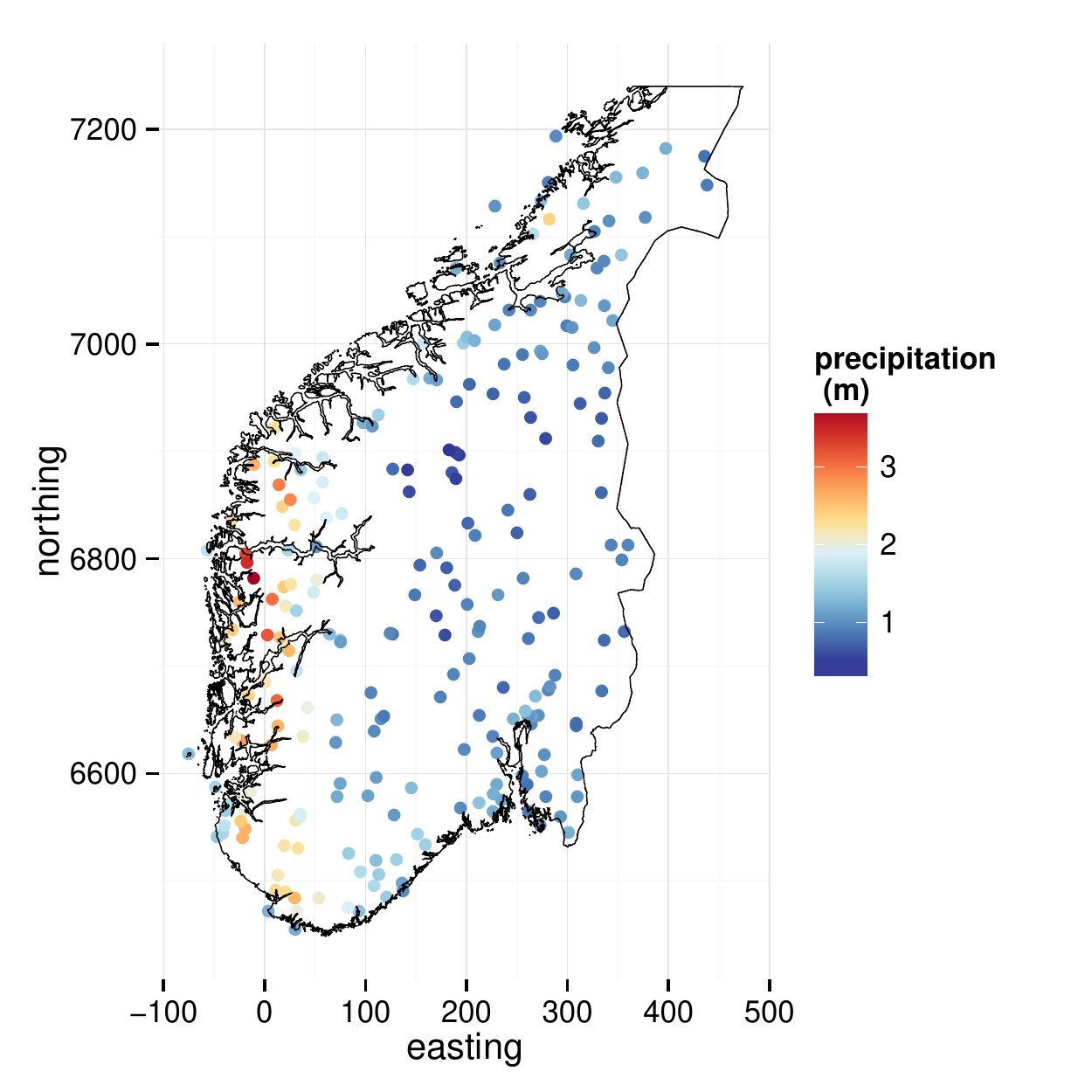}
   \caption{Annual total precipitation in 2008-2009}
   \label{fig:observations}
  \end{subfigure}
  ~
  \begin{subfigure}{0.48\textwidth}
   \includegraphics[width=\textwidth]{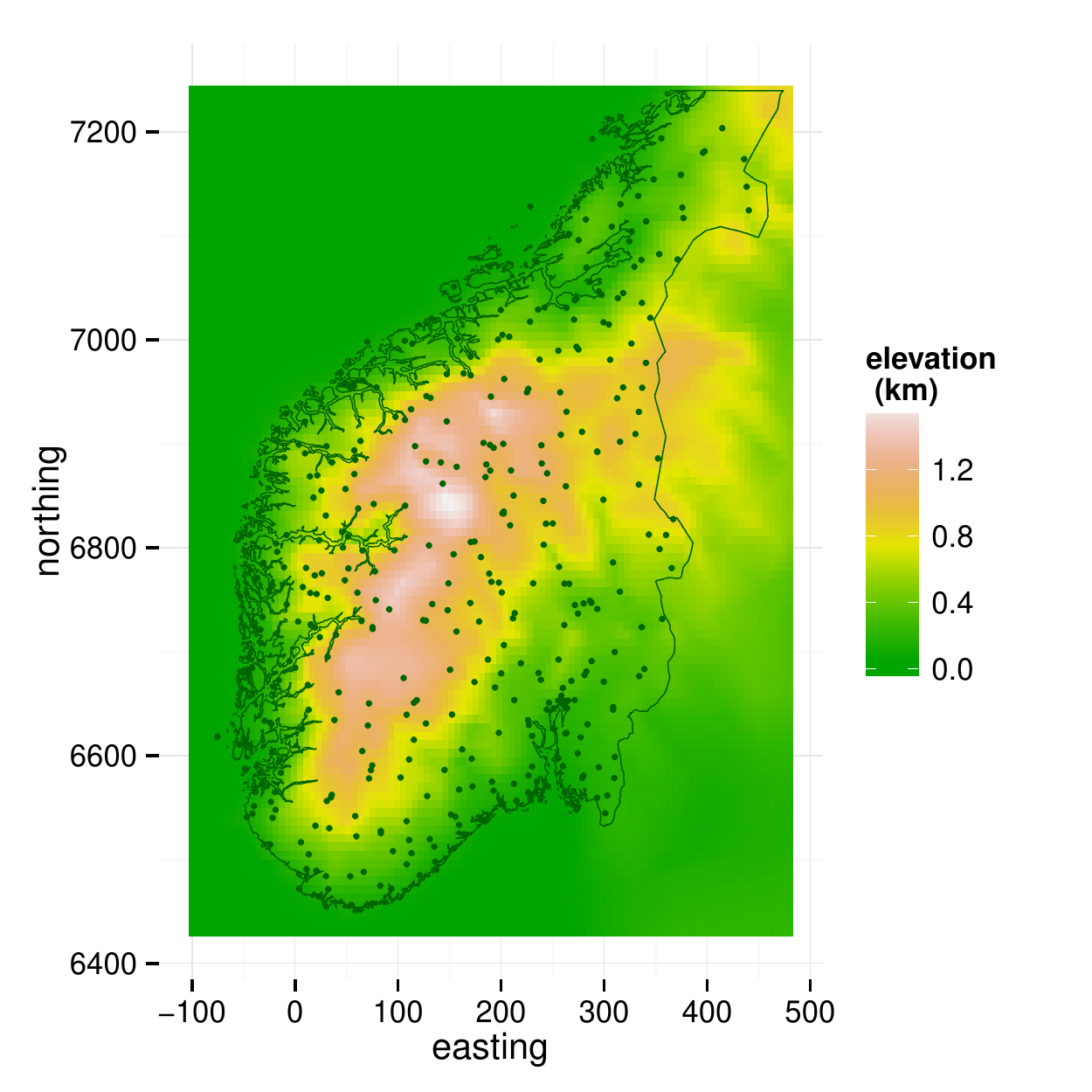}
  \caption{Elevation}
  \label{fig:elevation}
  \end{subfigure} 
 \caption{\ref{fig:observations}) 
          Observations of annual total precipitation in 2008-2009 (from the Norwegian Meteorological Institute). 
          \ref{fig:elevation})
          Smoothed elevation map of southern Norway \citep[based on the digital elevation model GLOBE,][]{GLOBE}
          with the 371 measurement locations indicated with dots.  
          The coordinate reference system is UTM33 and distances are in km.}
 \label{fig:data}
\end{figure}

The data analysed in this paper is annual total precipitation in 
southern Norway. Daily precipitation observations from stations/gauges 
in the 16 counties south of and included Nord-Tr{\o}ndelag, over the 
five year period 2008-09-01 -- 2013-08-31, were obtained from eKlima, 
a database provided by the Norwegian Meteorological Institute 
(www.eKlima.no). Five annual datasets were created by aggregating the daily 
observations and removing stations with incomplete records. The 
five year dataset contains observations from 371 different 
stations. However, the number of stations, and which stations  
there are measurements from, varies from year to year. 
Table~\ref{tab:data} contains a summary of the five year dataset. 
An elevation map with the locations of the 371 measurement 
stations, and a spatial plot of the annual total from 2008-2009, 
are found in Figure~\ref{fig:data}. The triangulation of the domain 
and the smoothed elevation model are the same as in \citet{Ingebrigtsen2014}.
The observed spatial pattern, with highest amounts of precipitation 
in western Norway, is characteristic for the Norwegian climate. 
Thus, the pattern seen in Figure~\ref{fig:observations} is similar 
for all of the years in the study. Consecutive years have 
fewer observations than 2008-2009. This is due to a combination of 
missing data and that several stations have been taken out of operation.

\subsection{Precipitation model with replicates}\label{sec:Model-precipitation}       

Consider a spatial domain $\mathcal{D}$ and $n$ observation 
locations $\{\boldsymbol{s}_1, \dots, \boldsymbol{s}_n\}$ in 
$\mathcal{D}$, in our case, southern Norway and the locations 
of the Norwegian Meteorological Institute's weather stations 
in southern Norway. The observed annual total precipitation 
$y_{ij}$ at station $i$ and year $j$ is modelled as 
\begin{equation}\label{eq:model_repl}
y_{ij} = \beta_j + \beta_h h(\boldsymbol{s}_i) + x_j(\boldsymbol{s}_i) + \epsilon_{ij}, 
\quad i = 1, \dots, n, \quad j = 1, \dots, r,
\end{equation}
where $\beta_j$ is a year specific intercept, $h(\boldsymbol{s}_i)$ 
is the elevation at location $\boldsymbol{s}_i$, and $\beta_h$ is a 
linear effect of elevation, assumed to be common for all years. The 
spatial structure is incorporated through a zero mean GRF $x(\boldsymbol{s})$, 
which is modelled using the SPDE based model defined in Section~\ref{sec:SPDE} 
and has parameters $\boldsymbol{\theta}$. The spatial fields 
$x_j(\boldsymbol{s})$, $j = 1, \dots, r$, are assumed to be independent 
realisations, or replicates, of this GRF model. Additive measurement error 
is included as independent, identically distributed noise terms 
$\epsilon_{ij}$. It is assumed that the $\epsilon_{ij}$'s are independent 
of the other model components, and that they are normally distributed 
with zero mean and precision $\tau_\epsilon$. 

One of the main objectives of this study is to gain insight into 
how the use of replicates influences the inference. Hence, the replicate 
model is set up such that the parameters have the same interpretation 
regardless of whether the model is fitted to only one realisation or to
several replicates. The amount of precipitation differs between years 
(Table~\ref{tab:data}), therefore, the intercept is allowed to vary 
from year to year. If the model only had a common intercept, the between 
year variation would be captured by the random spatial term $x_j(\boldsymbol{s})$
and its parameters $\boldsymbol{\theta}$ would not have the same interpretation 
with one realisation and with replicates. The linear effect of elevation in the 
mean $\beta_h$ is modelled as common for all years because it is expected to be 
a coefficient related to the physical process. 

An alternative model for observations from several years, 
would be to have two spatial terms: 1) a temporal constant 
field $c(\boldsymbol{s})$ that could be interpreted as the 
climatology, and 2) annual fields $x_j(\boldsymbol{s})$ to 
model the between year variation. 
This model would not be identifiable with one realisation. 
Furthermore, the spatial random term $x_j(\boldsymbol{s})$  
have a different interpretation in this model than in the model in \eqref{eq:model_repl} 
since it describes spatial deviation from the climatology and 
not overall spatial variation. 
A model including the climatology term $c(\boldsymbol{s})$ 
can be useful if one wants to do forecasting or interpolate 
the climatology with uncertainty. In addition, the model 
in \eqref{eq:model_repl} has only one explanatory variable (elevation), 
both in the mean and in the dependence structure. A simple model like 
this makes the results more transparent. However, if the main purpose 
was to predict and/or model precipitation it would be appropriate to 
search for more explanatory variables, e.g.\ other topographical  
variables such as aspect or surface roughness.
Because our main focus is on estimation and the use of replicates, 
alternative models are not explored in this paper. 

The GRF $x$ in \eqref{eq:model_repl} is a solution 
to the SPDE \eqref{eq:spde}, with 
dependence structure parameters $\boldsymbol{\theta}$ as given 
in \eqref{eq:kappa-tau}. We introduce models with different dependence 
structures for $x$: 1) A stationary 
model where the GRF has Mat\'ern covariance and parameters
$\boldsymbol{\theta}_\text{S}=(\theta_\tau, \theta_\kappa)$,
\begin{equation}
\label{eq:kappa_tau_S}
 \log \tau = \theta_\tau \quad \text{and} \quad
 \log \kappa = \theta_\kappa.
\end{equation}
2) A non-stationary model where the GRF has parameters 
$\boldsymbol{\theta}_\text{NS}=(\theta_{\tau, 1},\! \theta_{\tau, h},\! \theta_{\kappa, 1},\! \theta_{\kappa, h})$, 
and elevation is included in the 
dependence structure as a log-linear effect on $\tau$ and $\kappa$  
\begin{equation}
\label{eq:kappa_tau_NS}
 \log \tau(\boldsymbol{s}) = \theta_{\tau, 1} +  h(\boldsymbol{s})\,\theta_{\tau, h}  \quad \text{and} \quad
 \log \kappa(\boldsymbol{s}) = \theta_{\kappa, 1} + h(\boldsymbol{s})\,\theta_{\kappa, h}.
\end{equation}
Note that the non-stationary model is stationary if $\theta_{\tau, h} = \theta_{\kappa, h} = 0$. 
We will use $x_{\text{S}}$ to denote the stationary GRF and 
$x_{\text{NS}}$ to denote the non-stationary GRF, and refer 
to the stationary and non-stationary precipitation models 
as the model in \eqref{eq:model_repl} with the GRF being 
stationary or non-stationary, respectively. However, the 
overall structure of the model does not change using the different 
GRF models, only the dependence structure of the SPDE based spatial 
model. Thus, if the subscript is not specified, we refer to both models.  

To make the model identifiable, i.e.\ to separate the linear 
effects $\beta_j$ and $\beta_h$ from the spatial effect $x_j$, 
the following linear orthogonality constraints are imposed 
on the spatial fields
\begin{align}
 \int_{\mathcal{D}} x_j(\boldsymbol{s}) \,d\boldsymbol{s} = 0, j = 1, \dots, r, \label{eq:zero_constr} \\
 \sum_{j=1}^r \int_{\mathcal{D}} h(\boldsymbol{s}) x_j(\boldsymbol{s})\,d\boldsymbol{s}  = 0. \label{eq:h_constr}
\end{align}
This is known as restricted spatial regression \citep{Hanks2015}, 
and the constraints are imposed for both the stationary and the non-stationary 
model. Restricting the random field to be orthogonal to spatial covariates 
(here elevation) changes the interpretation of the spatial field to be a 
Bayesian version of restricted maximum likelihood; 
It is the spatial effect after the linear effect of elevation in the mean is accounted for.
If precipitation data from only 
one year is used, i.e.\ $r=1$, the replicate model in 
\eqref{eq:model_repl} corresponds to the model in 
\citet{Ingebrigtsen2014}, and this model will be 
referred to as the individual model.  

Inference with the annual precipitation models is carried 
out under the Bayesian paradigm. Thus, prior distributions 
need to be specified for all model parameters, i.e.\ the 
regression coefficients $\beta_1, \dots, \beta_r$, $\beta_h$, 
the precision of the measurement error $\tau_\epsilon$, and 
the dependence structure parameters $\boldsymbol{\theta}$.  
For the regression parameters we use vague Gaussian priors; 
all $\beta$'s are $\mathcal{N}(0, 100^2)$. The 
measurement error precision $\tau_\epsilon$ is Gamma 
distributed with shape parameter 2 and rate parameter 0.02.
Prior specification for $\boldsymbol{\theta}$ follows in the next section. 

\subsection{Prior specification for SPDE parameters}\label{sec:prior}

Assigning priors to the SPDE parameters $\boldsymbol{\theta}_{\text{S}}$ 
and $\boldsymbol{\theta}_{\text{NS}}$ is a bit intricate;  
they control the spatial dependence structure, but do not 
have a direct physical interpretation. Further, the simulation 
study in \citet{Ingebrigtsen2014} demonstrated the difficulty 
of estimating covariance parameters based on one realisation 
of the spatial field. The spatial correlation range parameters, 
$\theta_{\kappa,1}$ and $\theta_{\kappa, h}$, used to sample 
datasets were not well recovered, and the posterior credible 
interval coverages were as low as 5\% and 26\% 
for $\theta_{\kappa, 1}$ and $\theta_{\kappa, h}$, respectively.
The coverage was higher for $\theta_{\tau, 1}$ and $\theta_{\tau, h}$; 
84\% for both parameters. In addition to low coverage, there 
was bias towards the prior means, which indicates prior sensitivity 
and show that priors for $\boldsymbol{\theta}_{\text{NS}}$ 
should be chosen with care, and not be too informative. 
Also, since the stationary and non-stationary models are compared, 
priors in the two models should be based on similar assumptions 
about the dependence structure of the underlying spatial process. 

For the stationary GRF with Mat{\'e}rn covariance and smoothness
parameter $\nu=1$, the marginal standard deviation is given by
\begin{equation}\label{eq:sigma_S}
 \sigma_{\text{S}} = 1/(\sqrt{4\pi}\tau\kappa), 
\end{equation}
and we define the spatial correlation range to be
\begin{equation}\label{eq:rho_S}
 \rho_{\text{S}} = \sqrt{8}/\kappa. 
\end{equation}
With $\rho_{\text{S}}$ defined as above, this is the distance 
where the correlation has dropped to 0.13. 

Recall from \eqref{eq:kappa_tau_S} that $\theta_\tau = \log\tau$ 
and $\theta_\kappa=\log\kappa$. We assume a priori that 
$\theta_\tau\sim \mathcal{N}(\mu_\tau, \sigma_\tau^2)$ and 
$\theta_\kappa \sim \mathcal{N}(\mu_\kappa, \sigma_\kappa^2)$, 
and that $\theta_\tau$ and $\theta_\kappa$ are independent. 
The parameters $(\mu_\tau,\sigma_\tau^2)$ and $(\mu_\kappa,\sigma_\kappa^2)$ 
need to be specified. However, since we have knowledge about the 
magnitude of the annual total precipitation and distances in 
southern Norway, it would be easier to specify prior parameters 
for the distributions of $\sigma_{\text{S}}$ and $\rho_{\text{S}}$ 
rather than for $\theta_\tau$ and $\theta_\kappa$. From properties 
of the log-normal distribution it follows that
\begin{equation*}
  \rho_{\text{S}}\sim\log\mathcal{N}(\log\sqrt{8}-\mu_\kappa,\sigma_\kappa^2),
\end{equation*}
and
\begin{equation*}
  \sigma_{\text{S}}\sim\log\mathcal{N}(-\log\sqrt{4\pi}-\mu_\tau-\mu_\kappa,\sigma_\tau^2+\sigma_\kappa^2). 
\end{equation*}
The $p$-quantiles of the log-normal distributions for the correlation range and 
marginal standard deviation are
\begin{equation*}
 \rho_{\text{S}}(p) = \sqrt{8}\exp{(-\mu_\kappa+\sigma_\kappa\Phi^{-1}(p))},
\end{equation*}
and
\begin{equation*}
 \sigma_{\text{S}}(p) = \frac{1}{\sqrt{4\pi}}\exp{(-\mu_\tau-\mu_\kappa+\sqrt{\sigma_\tau^2+\sigma_\kappa^2}\Phi^{-1}(p))},
\end{equation*}
where $0 \leq p \leq 1$, and $\Phi(\cdot)$ is the cumulative distribution 
function for the standard normal distribution. To choose priors we can now 
specify two quantiles of $\rho_\text{S}$ and $\sigma_\text{S}$, e.g.\ 
the median and 0.9-quantile, and then solve the corresponding four 
equations for $(\mu_\tau,\sigma_\tau^2)$ and $(\mu_\kappa,\sigma_\kappa^2)$.  

The relationships between the marginal standard deviation 
and correlation range, and the SPDE parameters $\tau$ and 
$\kappa$ in \eqref{eq:sigma_S} and \eqref{eq:rho_S} are 
only valid in the stationary case. In the non-stationary 
case, we can obtain nominal approximations for 
$\sigma_{\text{NS}}(h)$ and $\rho_{\text{NS}}(h)$ as 
functions of the elevation $h = h(\boldsymbol{s})$. 
Similar to the stationary case, the Gaussian priors are assigned 
to $\boldsymbol{\theta}_\text{NS}$ such that
\begin{align*}
 \theta_{\tau, 1} &\sim \mathcal{N}(\mu_{\tau, 1}, \sigma_{\tau, 1}^2), \quad    
 \theta_{\tau, h} \sim \mathcal{N}(\mu_{\tau, h}, \sigma_{\tau, h}^2),   \\ 
 \theta_{\kappa, 1} &\sim \mathcal{N}(\mu_{\kappa, 1}, \sigma_{\kappa, 1}^2),  \quad
 \theta_{\kappa, h} \sim \mathcal{N}(\mu_{\kappa, h}, \sigma_{\kappa, h}^2).
\end{align*}
The $\theta$'s are assumed to be independent, which yields 
the following marginal prior distributions of the nominal 
range and nominal standard deviation at elevation $h$  
\begin{equation*}
\rho_\text{NS}(h)\sim\log\mathcal{N}(\log\sqrt{8}-\mu_{\kappa,1}-h\mu_{\kappa,h}, 
\sigma_{\kappa,1}^2+h^2\sigma_{\kappa,h}^2),  
\end{equation*} 
and
\begin{equation*}
\sigma_\text{NS}(h)\sim\log\mathcal{N}(-\log\sqrt{4\pi}-\mu_{\tau,1}-\mu_{\kappa,1}-h(\mu_{\tau,h}+\mu_{\kappa,h}),
\sigma_{\tau, 1}^2 + \sigma_{\kappa, 1}^2 + h^2(\sigma_{\tau, h}^2 + \sigma_{\kappa, h}^2)), 
\end{equation*}
where $h = h(\boldsymbol{s})$. 

The non-stationary nominal prior distributions change with 
elevation. To make them similar to the stationary prior 
distributions we set up the following coherence conditions;
\begin{enumerate}
 \item $\sigma_\text{NS}(0) \overset{d}{=} \sigma_\text{S}$ and 
       $\rho_\text{NS}(0) \overset{d}{=} \rho_\text{S}$,
 \item $\mu_{\tau, h} = \mu_{\kappa, h} = 0$,  
 \item given a reference elevation $h_0$, $c_\rho$ is the 
       coefficient of variation for the ratio $\rho_\text{NS}(h_0)/\rho_\text{NS}(0)$, 
       and $c_\sigma$ is the coefficient of variation for the ratio 
       $\sigma_\text{NS}(h_0)/\sigma_\text{NS}(0)$.
\end{enumerate}
Here, $h_0$, $c_\rho$, $c_\sigma$, $\sigma_\text{S}$, and $\rho_\text{S}$ are used 
to define the prior for $\boldsymbol{\theta}_\text{NS}$. 
The first condition states that the stationary prior distributions and the 
non-stationary nominal prior distributions are equal at sea level. The 
second condition has two interpretations: i) the medians of the 
non-stationary prior distributions for the nominal range and 
standard deviation are invariant to elevation, ii) the priors are 
centred around no influence, i.e.\ a priori we do not make any 
assumption of which effect elevation has on the spatial correlation 
range and marginal standard deviation. The third condition is introduced 
to control how much the priors are allowed to change with elevation. We 
look at the relative change in $\rho_\text{NS}(h)$ and $\sigma_\text{NS}(h)$ 
from sea level to a chosen reference elevation $h_0$, and control the change 
by specifying the coefficients of variation.

From the three coherence conditions and the stationary prior parameters 
$(\mu_\tau,\sigma_\tau^2)$ and $(\mu_\kappa,\sigma_\kappa^2)$, 
it follows that
\begin{align*}
 \mu_{\tau, 1}    & = \mu_\tau,    \quad \sigma_{\tau, 1}^2    =  \sigma_\tau^2,\\
 \mu_{\tau, h}    & = 0, \         \quad \sigma_{\tau, h}^2    = \frac{1}{h_0^2}\log\left(\frac{c_\sigma^2 + 1}{c_\rho^2 + 1}\right),\\
 \mu_{\kappa, 1}  & = \mu_\kappa,  \quad \sigma_{\kappa, 1}^2  =  \sigma_\kappa^2, \\
 \mu_{\kappa, h}  & = 0, \         \quad \sigma_{\kappa, h}^2  =  \frac{1}{h_0^2}\log(c_\rho^2 + 1),
\end{align*}
for a given reference elevation $h_0$ and coefficients of variation $c_\sigma$ 
and $c_\rho$. To ensure positive variance, $c_\sigma > c_\rho$, i.e.\ the 
relative change in $\sigma_\text{NS}(h)$ needs to be allowed to vary more than 
the relative change in $\rho_\text{NS}(h)$. 

\begin{figure}[!h]
  \centering
   \begin{subfigure}{0.49\textwidth}
    \includegraphics[width=\textwidth]{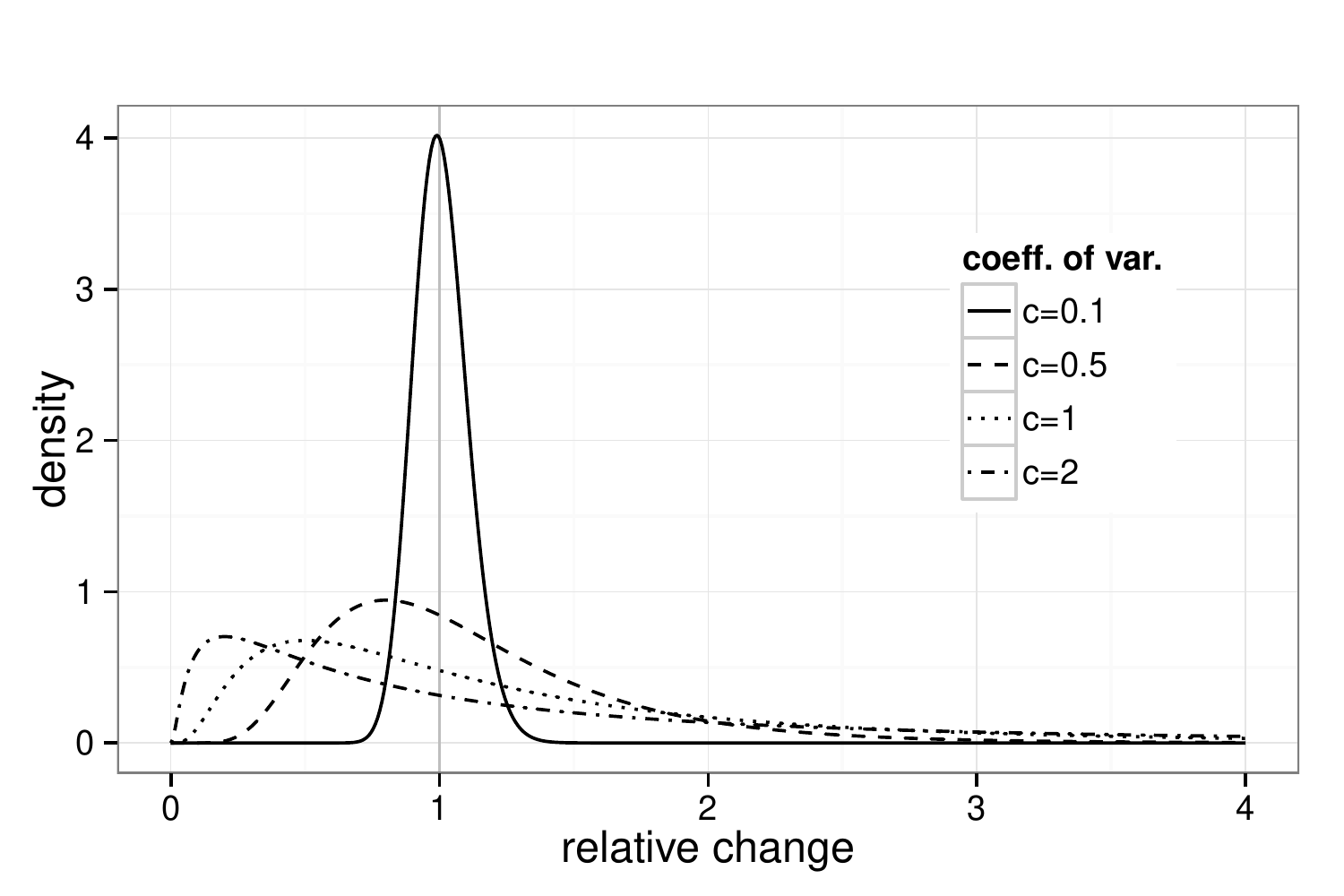}
    \caption{Density of the relative change in the non-stationary nominal 
             priors from sea level to $h_0$ for different values of 
             coefficient of variation.}
    \label{fig:rel_change_density}
   \end{subfigure}

   \begin{subfigure}{0.49\textwidth}
    \includegraphics[width=\textwidth]{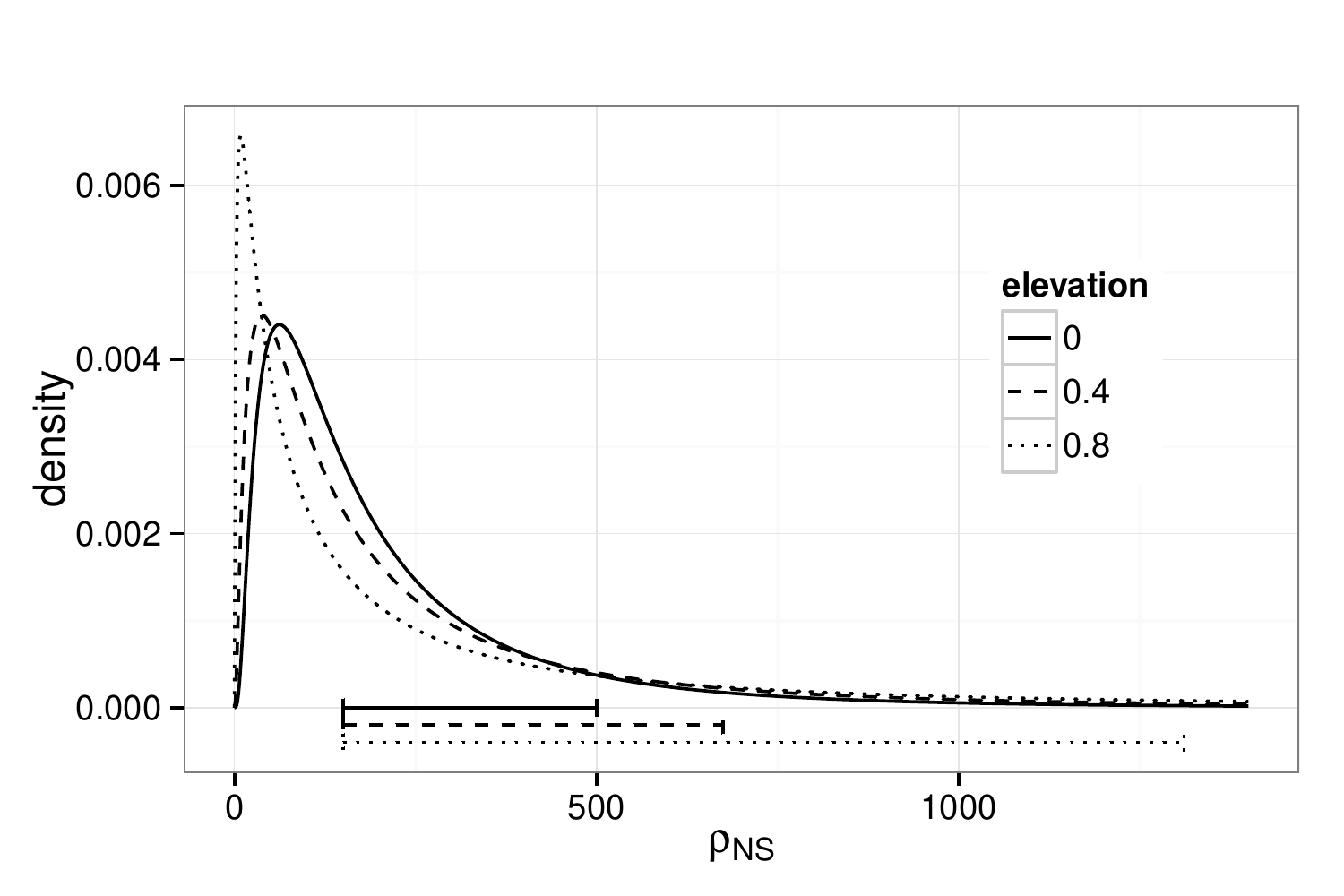}
    \caption{Prior density of $\rho_\text{NS}(h)$ when $h = 0, h_0, 2h_0$.}
    \label{fig:prior_rho}
   \end{subfigure}
   \begin{subfigure}{0.49\textwidth}
    \includegraphics[width=\textwidth]{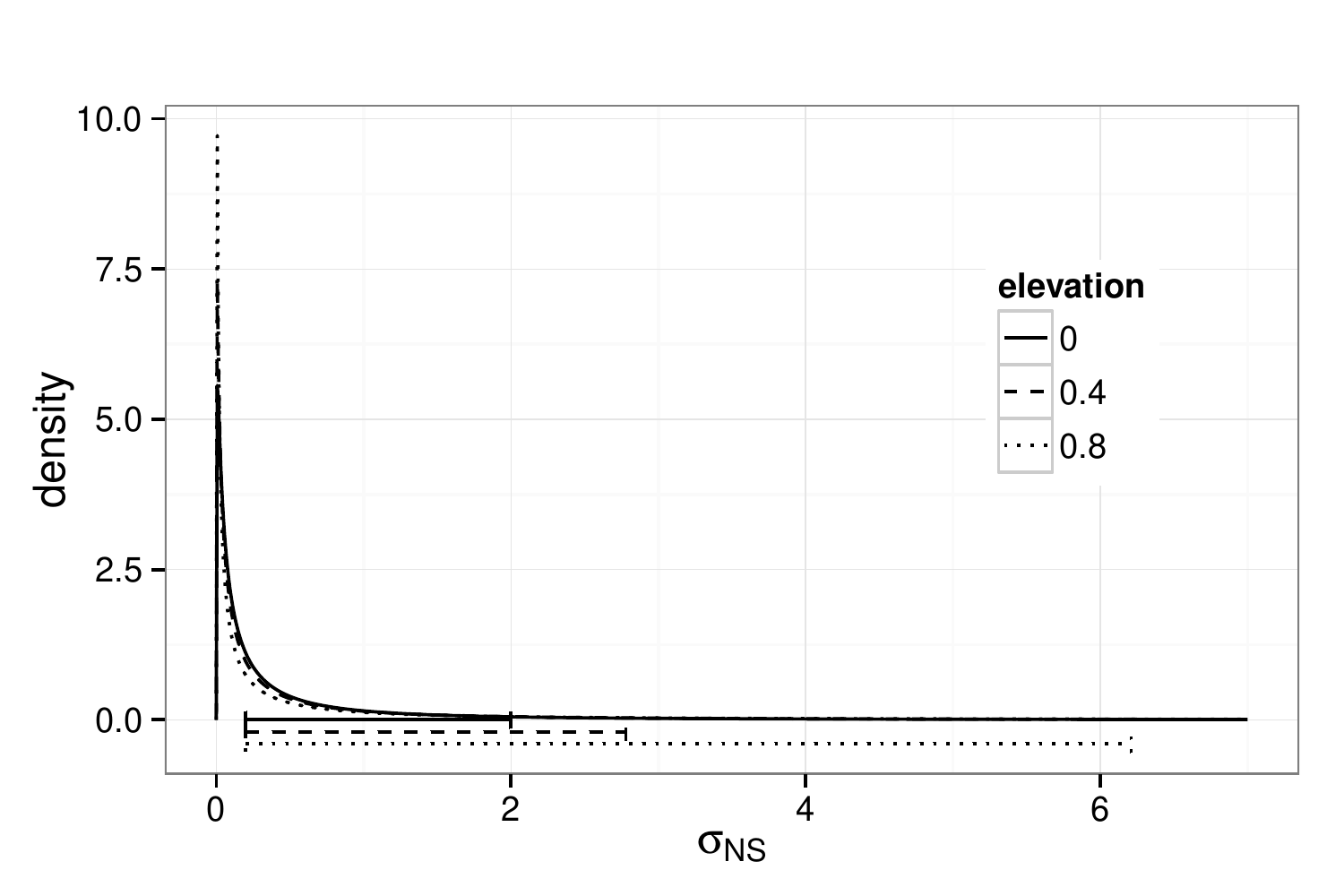}
    \caption{Prior density of $\sigma_\text{NS}(h)$ when $h = 0, h_0, 2h_0$.}
    \label{fig:prior_sigma}
   \end{subfigure}
   \caption{\ref{fig:rel_change_density}) is the density of the relative changes
            $\rho_\text{NS}(h_0)/\rho_\text{NS}(0)$ and 
            $\sigma_\text{NS}(h_0)/\sigma_\text{NS}(0)$ 
            for different values of the coefficient of variation. 
            \ref{fig:prior_rho}) and \ref{fig:prior_sigma}) are the prior densities for 
            $\rho_\text{NS}(h)$ and $\sigma_\text{NS}(h)$ at different elevations $h$. 
            The reference elevation $h_0$ is 0.4 km, and the coefficients of variation 
            are $c_\rho = 0.8$ and $c_\sigma = 1.3$. The 0.5- and 0.9-quantiles at sea 
            level are set to 150 km and 500 km for $\rho_\text{NS}$, and 0.2 m and 2 m 
            for $\sigma_\text{NS}$. The intervals given by the 0.5- and 0.9-quantiles 
            are indicated with horizontal lines at the bottom of the plots. 
            }
  \label{fig:priors}
\end{figure}

The coefficients of variation are used to set the prior variance 
for the non-stationarity parameters $\theta_{\tau, h}$ and $\theta_{\kappa, h}$. 
In the limiting case, when $c_\sigma$ and $c_\rho$ approach zero, the 
prior variances also approach zero, and the model is essentially assumed 
to be stationary. Increasing the coefficients of variance, means increasing 
the prior variance. 
The relative changes in nominal range and standard deviation are log-normally 
distributed, i.e.\ 
$\rho_\text{NS}(h_0)/\rho_\text{NS}(0)\sim\log\mathcal{N}(0, \log(c_\rho^2 + 1))$ and 
$\sigma_\text{NS}(h_0)/\sigma_\text{NS}(0)\sim\log\mathcal{N}(0, \log(c_\sigma^2 + 1))$. 
The medians of both ratios are one, thus, the distributions are centred around no change. 
Figure~\ref{fig:rel_change_density} show the density for $\rho_\text{NS}(h_0)/\rho_\text{NS}(0)$
and $\sigma_\text{NS}(h_0)/\sigma_\text{NS}(0)$ for different values of the 
coefficient of variation ($c$). When $c=0.1$, the density is narrow and close 
to symmetric around one. Increasing $c$, increases the prior probability that the 
range/standard deviation is e.g.\ doubled/halved from sea level to $h_0$. 
Note that similar approaches for prior specification can be applied 
to any type of explanatory variable. 

The analysis domain, southern Norway, is approximately 500 km $\times$ 700 km. 
We set the prior median range ($\rho_\text{S}$) to 150 km and 0.9-quantile 
to 500 km. This yields $\mu_\kappa = -3.97$ and $\sigma_\kappa^2 = 0.88$. 
Further, after considering the variation in the precipitation data, 
the prior median standard deviation ($\sigma_\text{S}$) is set to 
0.2 m and 0.9-quantile to 2 m, giving $\mu_\tau = 4.31$ and $\sigma_\tau^2 = 2.35$.    
The average value of the smoothed elevation model is close to 0.4 km, 
and this value is used as the reference elevation $h_0$. We set $c_\rho = 0.8$ 
and $c_\sigma = 1.3$, and get $\sigma_{\tau, h}^2 = 3.09$ and 
$\sigma_{\kappa, h}^2 = 3.09$. The prior densities for $\rho_\text{NS}(h)$ 
and $\sigma_\text{NS}(h)$, for this choice of parameters, and at different 
elevations $h$, can be seen in Figure~\ref{fig:priors}. The prior values 
are chosen as a compromise between having wide enough priors, but narrow 
enough to restrict the prior range not to be too long. 
Having a $\mathcal{N}(0, 3.09)$-prior for $\theta_{\kappa, h}$ might seem strict. 
However, in Figure~\ref{fig:prior_rho} it can be seen that the 0.9-quantile of 
$\rho_\text{NS}$ at $h=0.8$ is around 1300 km, i.e.\ almost two times the 
length of the domain. 

\subsection{Bayesian inference with \texttt{R-INLA}}

The annual precipitation models are fitted under the Bayesian 
paradigm and target for the inference are posterior predictive 
distributions at spatial locations of interest and posterior 
distributions for $\beta_1, \dots, \beta_r$, $\beta_h$, 
$\tau_\epsilon$, and $\boldsymbol{\theta}$. The prior 
distributions assigned to these model parameters were specified 
in Sections~\ref{sec:Model-precipitation} and \ref{sec:prior}. 
In addition, the weight vector 
$\boldsymbol{w}$ from the basis representation of the GRF in 
\eqref{eq:basis_repr} is assigned a GMRF prior with precision 
matrix as in \eqref{eq:SPDE_Q}. The stationary and non-stationary 
precipitation models fit into the latent Gaussian model framework 
from \citet{Rue2009} and approximations to the posterior marginal 
densities, posterior predictive densities as well as measures of 
model fit can be obtained with integrated nested Laplace approximation (INLA). 
In fact, because the likelihood is Gaussian, the approximations are 
exact up to numerical integration error.

We refer to \citet[][Section~4.5]{Ingebrigtsen2014} for details on 
inference with our precipitation models using INLA, and to 
\citet{Lindgren2014} for a thorough introduction to Bayesian spatial 
modelling with the SPDE approach and INLA. Both the INLA methodology 
and SPDE based spatial models are available in R \citep{Rlanguage}, 
in the R package \texttt{R-INLA} (see www.r-inla.org). All inference was carried 
out using this software.

\subsection{Model evaluation}\label{sec:evaluation}

To assess how well each model fits the observations, we use the deviance 
information criterion \citep[DIC,][]{DIC}. DIC is a model selection 
tool for Bayesian hierarchical models. It is a measure that combines 
goodness-of-fit in terms of deviance and complexity in terms of 
effective number of parameters. 
The deviance is defined as $D = - 2\log L$, 
where $L$ is the likelihood.
DIC is given by the expression
\begin{equation*}
 \text{DIC} = \bar{D} + p_D.
\end{equation*}
$\bar{D}$ is the expectation of the deviance over the posterior distribution.
The effective number of parameters $p_D$ is the posterior mean deviance minus 
the deviance of the posterior mean of the parameters. 
Models with smaller DIC have better support in the data. 
A commonly used rule of thumb is that a difference in DIC 
greater than ten is considered a significant difference.
DIC has been criticised, but is also frequently used 
(see \citet{DIC-2014} for a summary). DIC performed very well for 
selecting the correct model in the simulation study in 
\citet{Ingebrigtsen2014}, so we choose to evaluate model 
fit of the stationary and non-stationary precipitation 
models with DIC. Refer to \citet{Holand2013} for a description of how DIC is 
computed in \texttt{R-INLA}. 

To evaluate predictive performance we run leave-one-out cross-validation. 
All annual observations at one station are held out of the analysis and 
the remaining stations are used for both model fitting and interpolating/predicting the amount 
of precipitation, for each year, at the location of the removed station. 
This is repeated for all 371 stations. We compute the the root mean square 
error (RMSE) for each year 
\begin{equation*}
 \text{RMSE}_j = \sqrt{\frac{1}{n_j}\sum_{i=1}^{n_j} (\hat{y}_{ij} - y_{ij})^2},
\end{equation*}
where $n_j$ is the number of observations in year $j$, 
$y_{ij}$ is the observation at station $i$ and year $j$, 
and $\hat{y}_{ij}$ is the mean of the posterior predictive 
distribution when the model was fitted to data from all 
stations except station $i$. As a summary score we use 
the average RMSE
\begin{equation*}
 \overline{\text{RMSE}} = \frac{1}{r}\sum_{j=1}^{r} \text{RMSE}_j. 
\end{equation*}  

In addition to RMSE, we compute the continuous ranked probability 
score (CRPS), which unlike RMSE assesses the whole posterior 
predictive distribution, and not only the predictive mean. 
CRPS is defined as
\begin{equation*}
  \text{CRPS}(F, y) = \int_{-\infty}^\infty (F(u) - 1\{y \leq u\})^2 du,
\end{equation*}
where $F$ is the predictive cumulative distribution function and $y$ is the
observed value \citep{crps_gneiting}.
Both RMSE and CRPS are on the same scale as the observations, i.e.\
metres in our case.  Similarly to for RMSE, we define the average
CRPS as
\begin{equation*}
  \overline{\text{CRPS}} =
    \frac{1}{r}\sum_{j=1}^{r} \frac{1}{n_j} \sum_{i=1}^{n_j}
    \text{CRPS}(F_{ij},y_{ij}),
\end{equation*}
where $F_{ij}$ is the leave-one-out posterior predictive cumulative
distribution function for observation $y_{ij}$.

The true leave-one-out posterior predictive distribution for each
$y_{ij}$ is a continuous Gaussian mixture, but to simplify the
numerical calculations, the predictive distributions are approximated
with a Gaussian distribution, which allows CRPS to be computed
analytically with the function available in the R-package
\texttt{verification} \citep{verification}.  The approximation uses
the leave-one-out posterior mean $\hat{y}_{ij}$ of the linear
predictor $\eta_{ij} = \beta_j + \beta_h h(\boldsymbol{s}_i) +
x_j(\boldsymbol{s}_i)$, and the predictive variance is taken to be the
sum of the leave-one-out posterior variance of $\eta_{ij}$ and the
posterior expectation of the observation noise variance
$1/\tau_\epsilon$.  Hence, the uncertainty of all parameters except
$\tau_\epsilon$ is integrated out.

\section{Data analysis}\label{sec:Data_analysis}

\begin{table}[!h]
 \footnotesize
 \centering
 \caption{Difference in deviance information criterion (DIC) when the 
          annual precipitation models were fitted with stationary and 
          non-stationary dependence structures. Individual models were 
          fitted to the five years separately, and replicate models 
          to all five years. $\Delta$DIC is defined as DIC for the 
          stationary model minus DIC for the non-stationary model.}
 \label{tab:dic}
 \begin{tabular}{cccccccc}
  \toprule
   model       & individual & individual & individual & individual & individual && replicate  \\
   year(s)     &  2008-2009 &  2009-2010 &  2010-2011 &  2011-2012 &  2012-2013 &&  2008-2013 \\
                         \cmidrule(r){2-6}                \cmidrule(r){7-8}                 
   $\Delta$DIC &  55  & 51  & 90  &  10  &  7 &&  159 \\
  \bottomrule
 \end{tabular}
\end{table}

The stationary and non-stationary replicate models from 
Section~\ref{sec:Model-precipitation}, with the priors 
specified in Section~\ref{sec:prior}, were fitted to the 
southern Norway annual precipitation data from 2008-2013. 
Five years of observations are available, so the number of 
replicates $r$ is five. For comparison, individual models 
($r=1$) were fitted to the five years separately. 
Table~\ref{tab:dic} contains the difference in DIC between 
the stationary and non-stationary models. DIC favours a 
non-stationary dependence structure for the replicate model 
and all individual models, although the difference is small 
for 2011-2012 and 2012-2013. 

\begin{table}
 \footnotesize
 \centering
 \caption{Model predictions evaluated with leave-one-out 
          cross-validation. Reported is CRPS averaged over 
          stations and years, and RMSE averaged over years.}
 \label{tab:predictive_performance}
 \begin{tabular}{ccccccc}
  \toprule
        && \multicolumn{2}{c}{$\overline{\text{CRPS}}$} && \multicolumn{2}{c}{$\overline{\text{RMSE}}$} \\
        \cmidrule(r){3-4}                \cmidrule(r){6-7}  
           && stationary & non-stationary &&  stationary & non-stationary \\
   \cmidrule(r){3-4}                \cmidrule(r){6-7}  
 individual models && 0.14360 & \emph{0.13816}   && 0.26555 & \emph{0.25315} \\
 replicate model   && 0.14377 & 0.14358          && 0.26456 & 0.25672 \\  
\bottomrule
 \end{tabular}
\end{table}
   
The main motivation for introducing non-stationarity 
with elevation as an explanatory variable in the 
dependence structure, was to improve spatial prediction 
of precipitation in difficult and mountainous terrain. 
Predictive performance is evaluated using the leave-one-out 
cross-validation described in Section~\ref{sec:evaluation}. 
Table~\ref{tab:predictive_performance} contains CRPS 
averaged over stations and the five years, and RMSE 
averaged over the five years. The best prediction 
results are obtained when individual models with 
non-stationary dependence structure are fitted to 
each annual dataset separately. The non-stationary 
replicate model is ranked second best according to 
both RMSE and CRPS, but the difference in average CRPS 
between the non-stationary replicate model and 
stationary individual models is minimal.  

\begin{figure}
 \centering
  \begin{subfigure}{0.48\textwidth}
   \includegraphics[width=\textwidth]{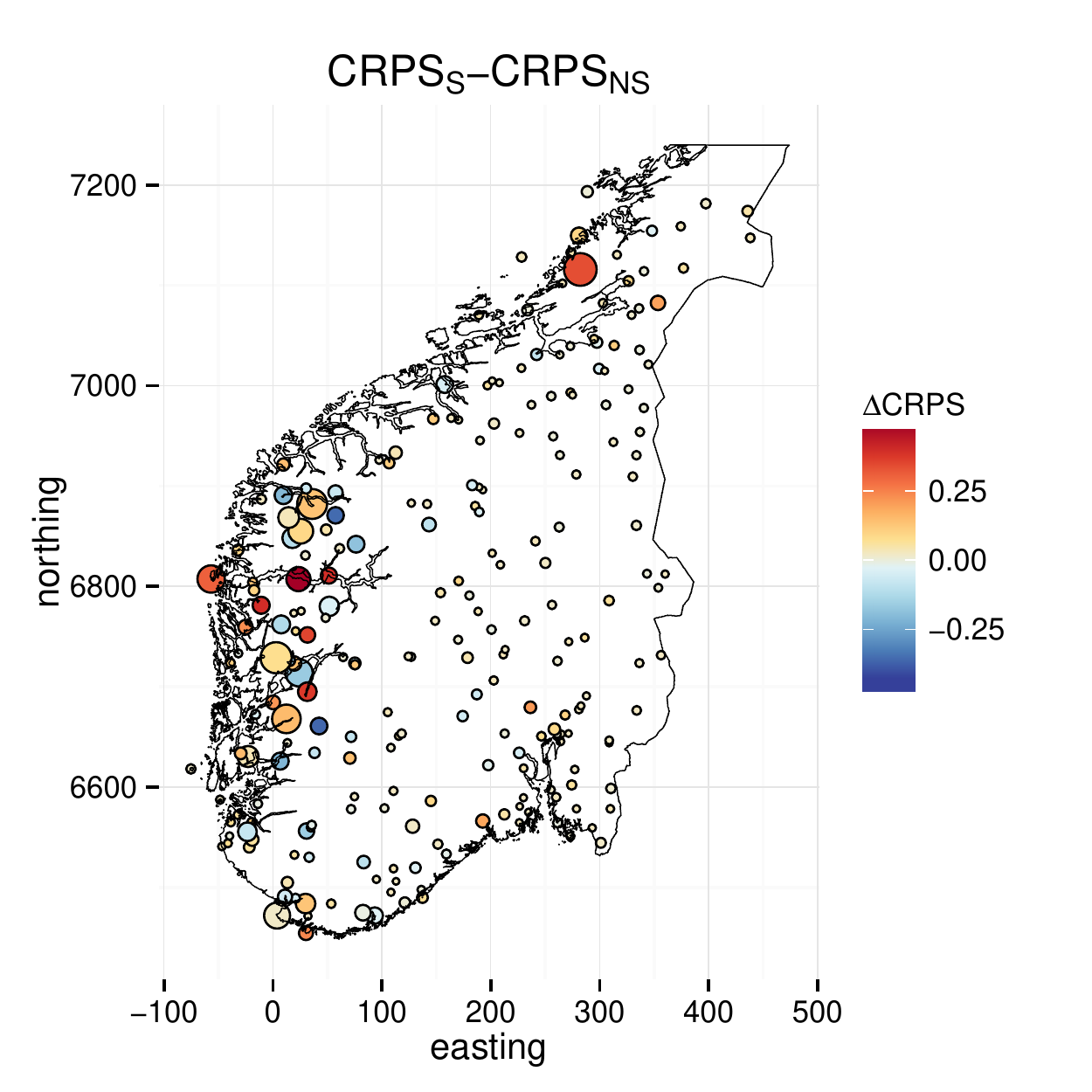}
   \caption{}
   \label{fig:crps_diff_model}
  \end{subfigure}
   ~
  \begin{subfigure}{0.48\textwidth}
   \includegraphics[width=\textwidth]{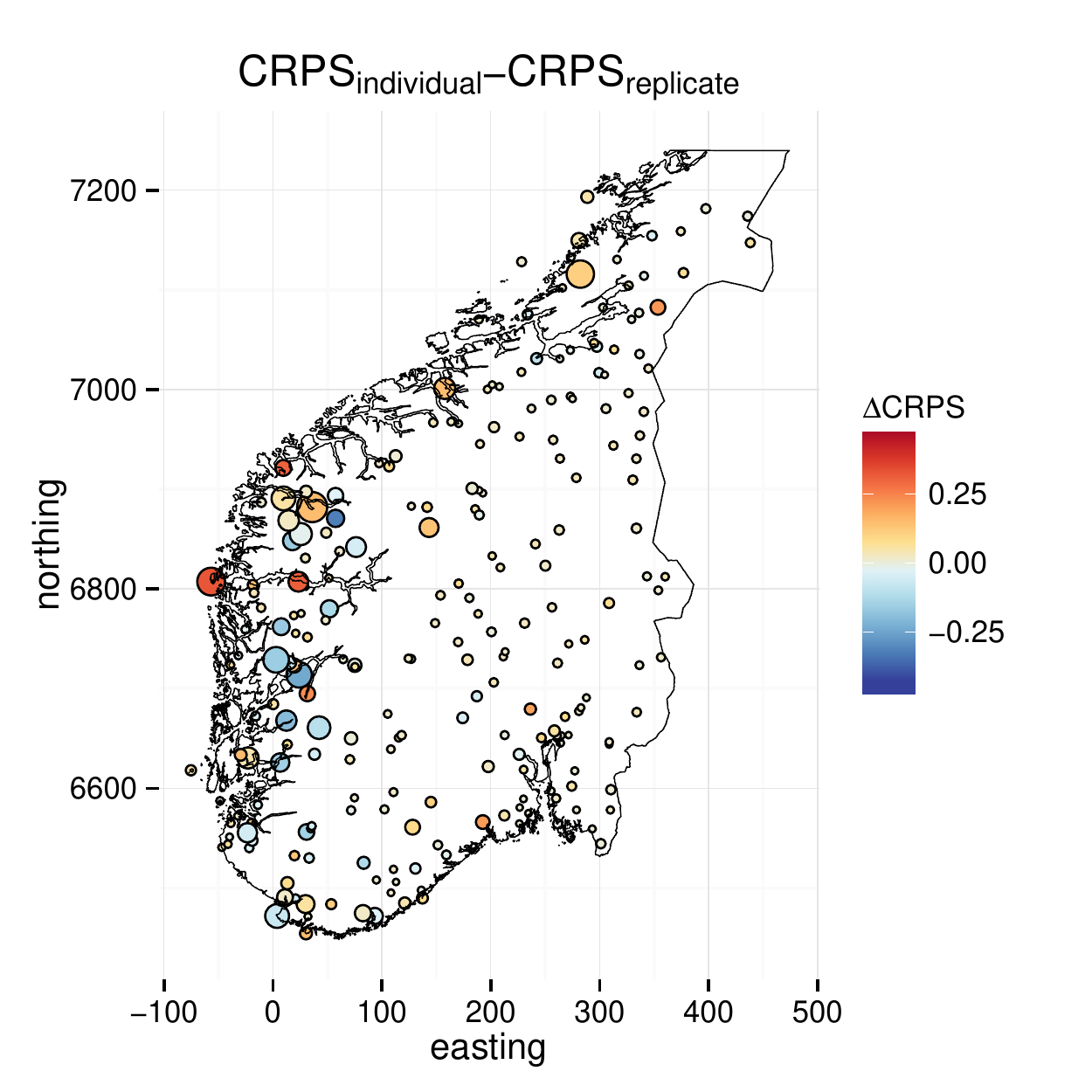}
    \caption{}
   \label{fig:crps_diff_replicates}
  \end{subfigure}%
 \caption{Maps of CRPS in 2008-2009 at each of the 233 weather stations 
          computed with leave-one-out cross-validation. The colour scale 
          indicates the difference in CRPS between stationary and 
          non-stationary dependence structures in the replicate model 
          (\ref{fig:crps_diff_model}), and between using the non-stationary 
          individual model and non-stationary replicate model 
          (\ref{fig:crps_diff_replicates}). The size of the points 
          indicates the magnitude of the CRPS (average of the two models 
          compared). Smaller size means better predictions.  
          }
 \label{fig:crps_map}
\end{figure}
  
The differences in average CRPS are small. 
However, looking at CRPS values at individual 
stations shows that the picture is more complex. The differences 
are much higher than the average suggests because one model is not 
uniformly better at all stations. Figure~\ref{fig:crps_map} compares 
CRPS when different dependence structures were used in the replicate 
model (\ref{fig:crps_diff_model}), and when the replicate or individual 
model were used and the dependence structure was non-stationary 
(\ref{fig:crps_diff_replicates}). The spatial plots are for the year 
2008-2009, and the difference in CPRS at each station is indicated 
with colour, while the size of the circles indicates the magnitude 
of the CRPS.

The largest circles, hence largest CRPS values, are located in 
western Norway, i.e.\ this is where the predictions miss the most. 
This is also where the amount of precipitation is high, and the 
topography is complex. In 2008-2009, the non-stationary replicate 
model has lower CRPS than the stationary replicate model at 
48\% of the stations, this is where the circles are filled with 
red. For the individual model this number is 57\%. Many of the 
circles filled with red are located at difficult locations in 
western Norway, but the picture is not clear and there are 
locations at the west coast where the stationary model gives 
better predictions than the non-stationary model. The difference 
in CRPS between the individual and replicate models is smaller 
than the difference between dependence structures. The individual 
model is better at 62\% of the stations in 2008-2009. This is 
where the circles are filled with blue in 
Figure~\ref{fig:crps_diff_replicates}.

DIC, $\overline{\text{CRPS}}$, and $\overline{\text{RMSE}}$ all 
favour a non-stationary dependence structure. The non-stationary 
model will therefore be in focus for the rest of this section.
Using annual precipitation data from more years did not improve 
predictions. However, it is not surprising that the individual 
model fitted to one annual dataset is better at predicting 
the amount of precipitation that particular year. Recall that 
the motivation for introducing non-stationarity was a more 
physics based model and possible improved predictions, but the 
motivation for introducing replicates was preciser estimates 
with better statistical properties (less biased and prior sensitive) 
for the dependence structure parameters. 

\begin{figure}[!h]
 \centering
  \begin{subfigure}{0.4\textwidth}
   \includegraphics[width=\textwidth]{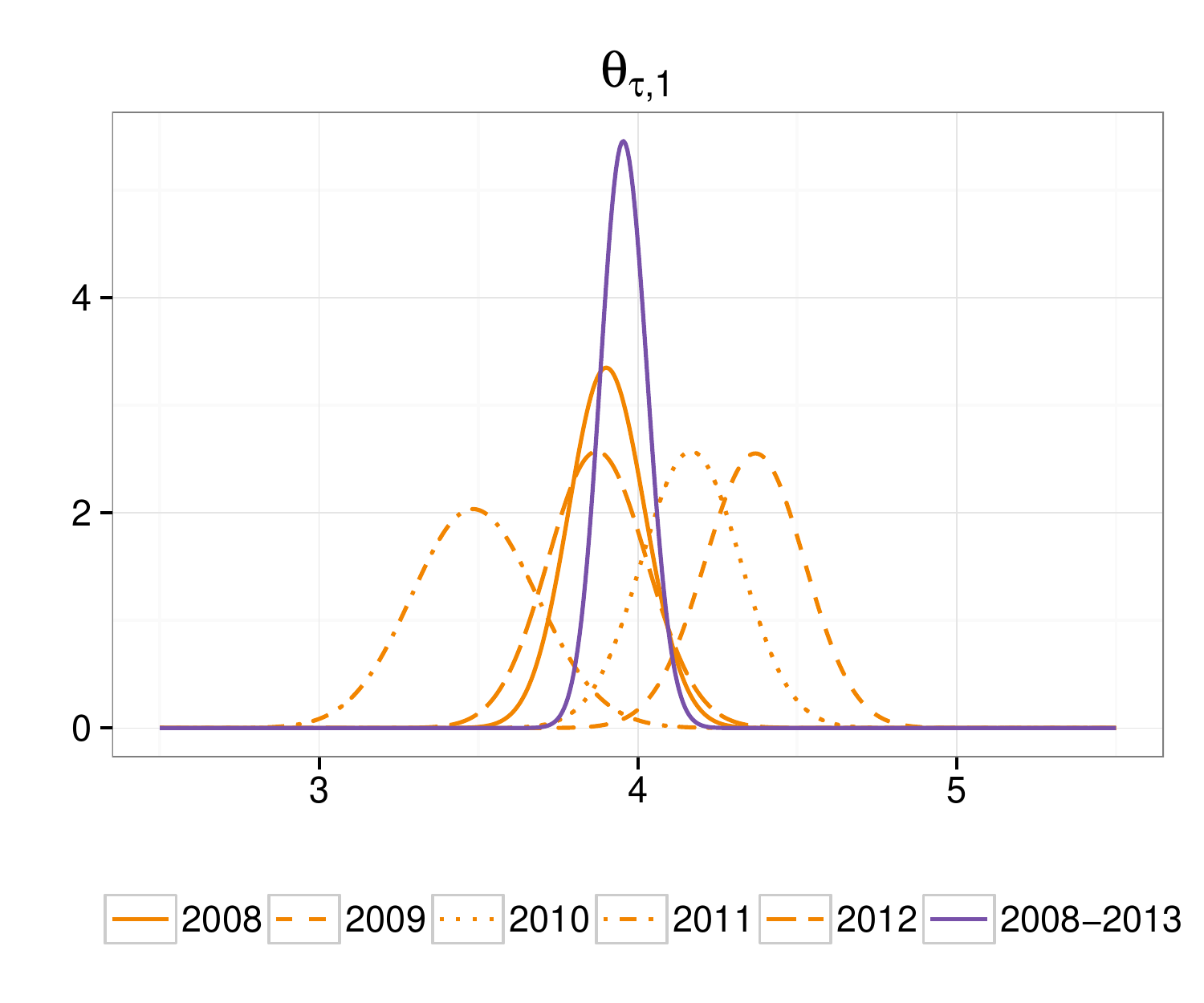}
   \caption{} 
   \label{fig:posterior_theta1_nonstat}
  \end{subfigure}
  \begin{subfigure}{0.4\textwidth}
   \includegraphics[width=\textwidth]{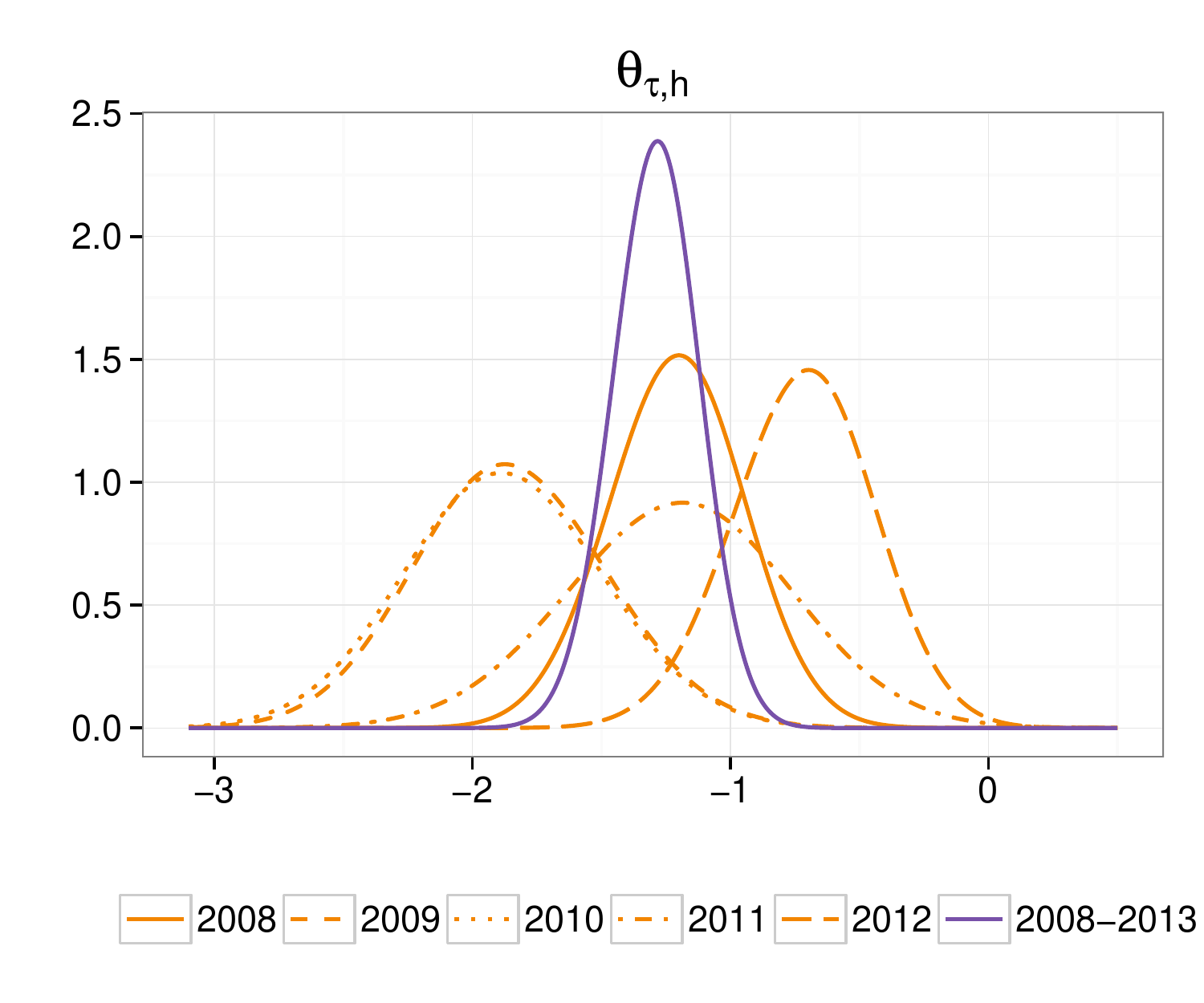}
   \caption{} 
   \label{fig:posterior_theta2_nonstat}
  \end{subfigure}
 
  \begin{subfigure}{0.4\textwidth}
   \includegraphics[width=\textwidth]{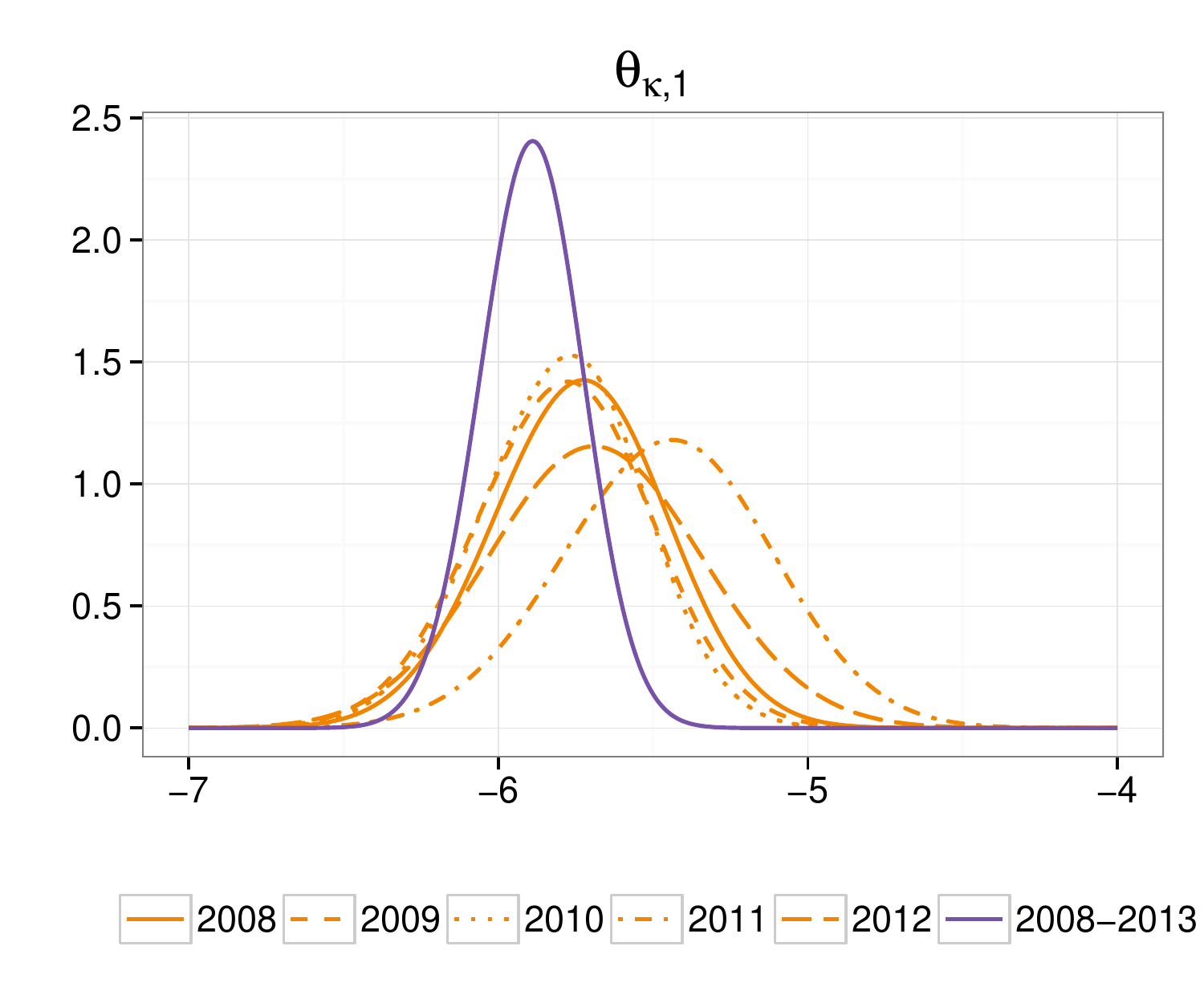}
   \caption{} 
   \label{fig:posterior_theta3_nonstat}
  \end{subfigure}
  \begin{subfigure}{0.4\textwidth}
   \includegraphics[width=\textwidth]{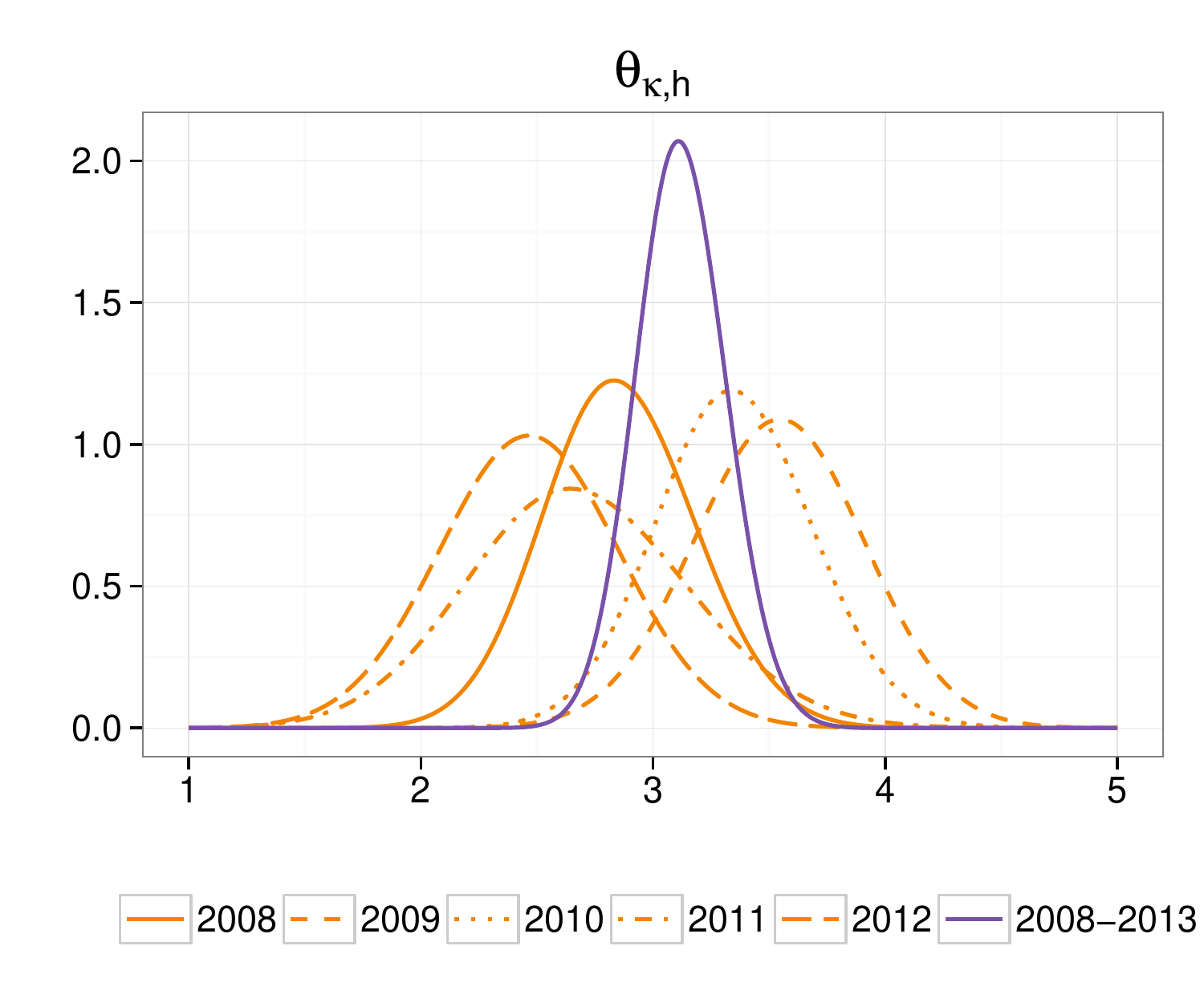}
   \caption{} 
   \label{fig:posterior_theta4_nonstat}
  \end{subfigure}
  \caption{Posterior marginal densities for the dependence structure parameters 
           $\boldsymbol{\theta}_\text{NS}$ in the non-stationary precipitation model. 
           The individual model was fitted to each annual dataset and the replicate 
           model was fitted to the five years together.  
          }
 \label{fig:marginals_theta_nonstationary}
\end{figure}

Figure~\ref{fig:marginals_theta_nonstationary} show the posterior
marginal densities for $\boldsymbol{\theta}_\text{NS}$ when individual 
models were fitted to the annual data, and when the replicate model 
was fitted to all five years. The posterior marginal densities 
with the replicate model are narrower than the individual posteriors, 
i.e.\ the estimates have higher precision. Also, the replicate posteriors 
for $\theta_{\tau, 1}$, $\theta_{\tau, h}$, and $\theta_{\kappa, h}$ 
are centred among the individual priors; the 
replicate model produces a kind of average estimate. The situation 
is different for $\theta_{\kappa, 1}$. For this parameter, the replicate posterior 
is shifted away from the prior mean (-3.97) and towards 
smaller values and longer spatial correlation range at sea level. 

\begin{figure}[!h]
 \centering
\begin{subfigure}{0.45\textwidth}
  \includegraphics[width=\textwidth]{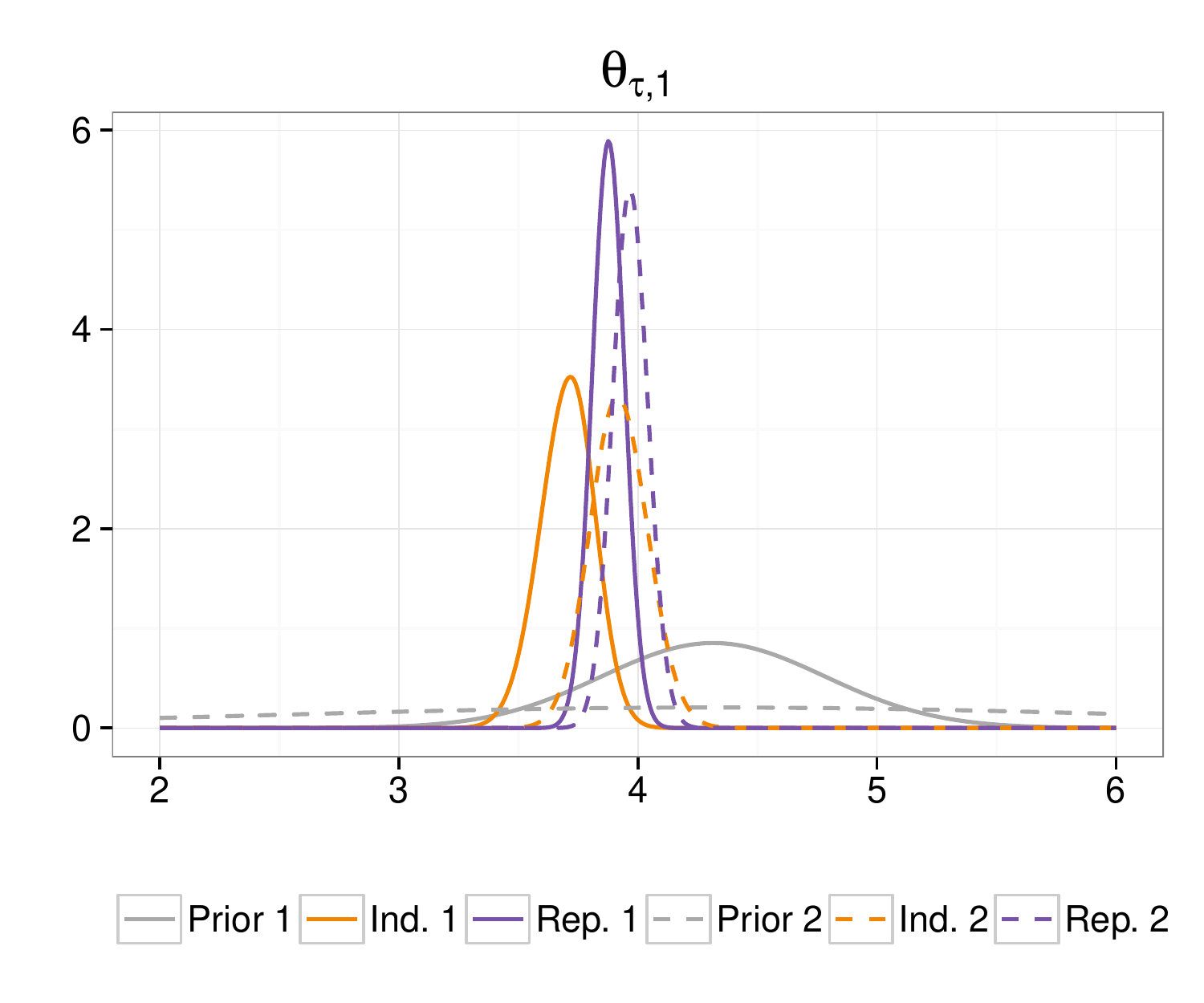}
  \caption{}
  \label{fig:posterior_prior_theta1_nonstat}
 \end{subfigure}%
 \begin{subfigure}{0.45\textwidth}
  \includegraphics[width=\textwidth]{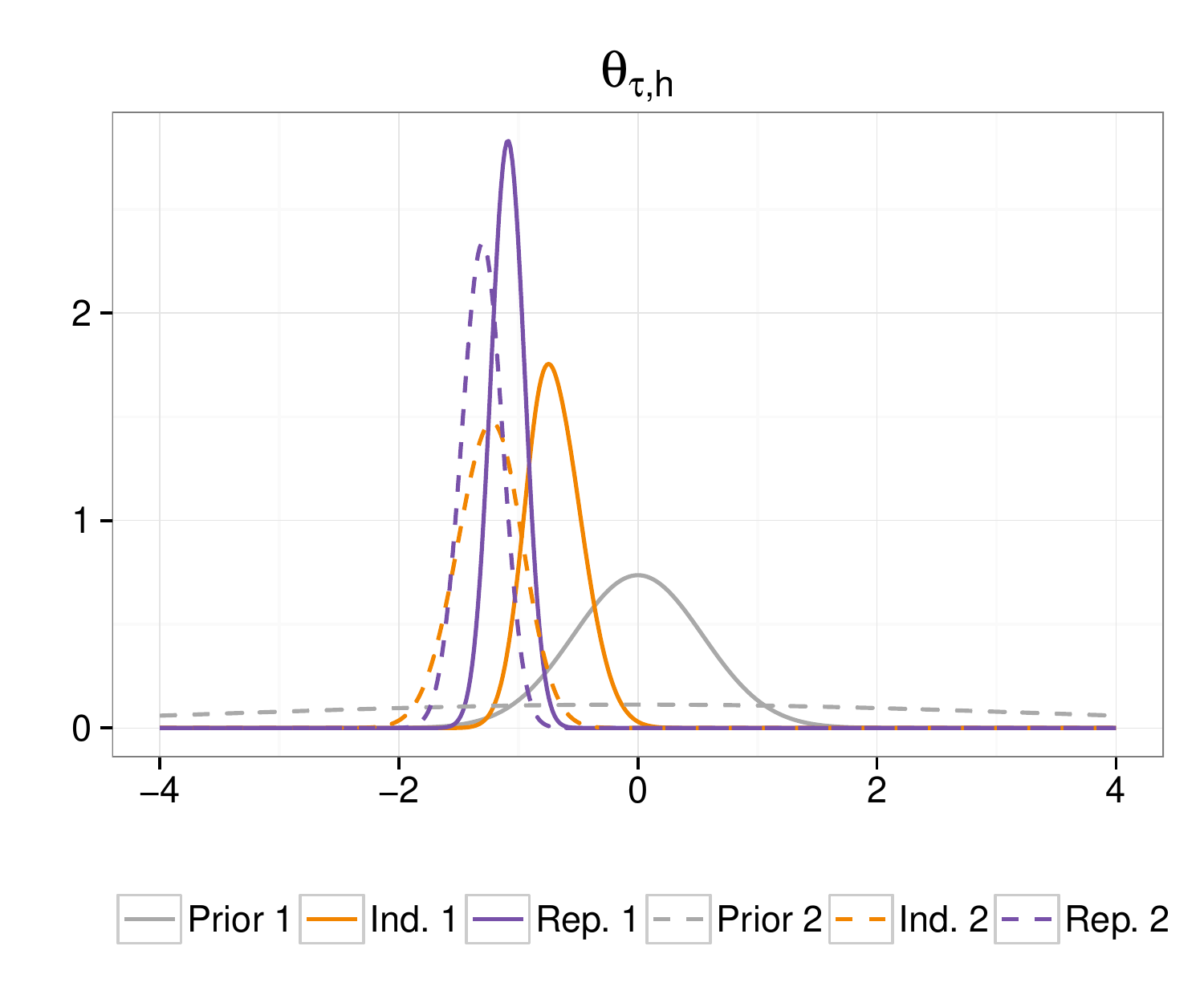}
  \caption{}
  \label{fig:posterior_prior_theta2_nonstat}
 \end{subfigure}%

 \begin{subfigure}{0.45\textwidth}
  \includegraphics[width=\textwidth]{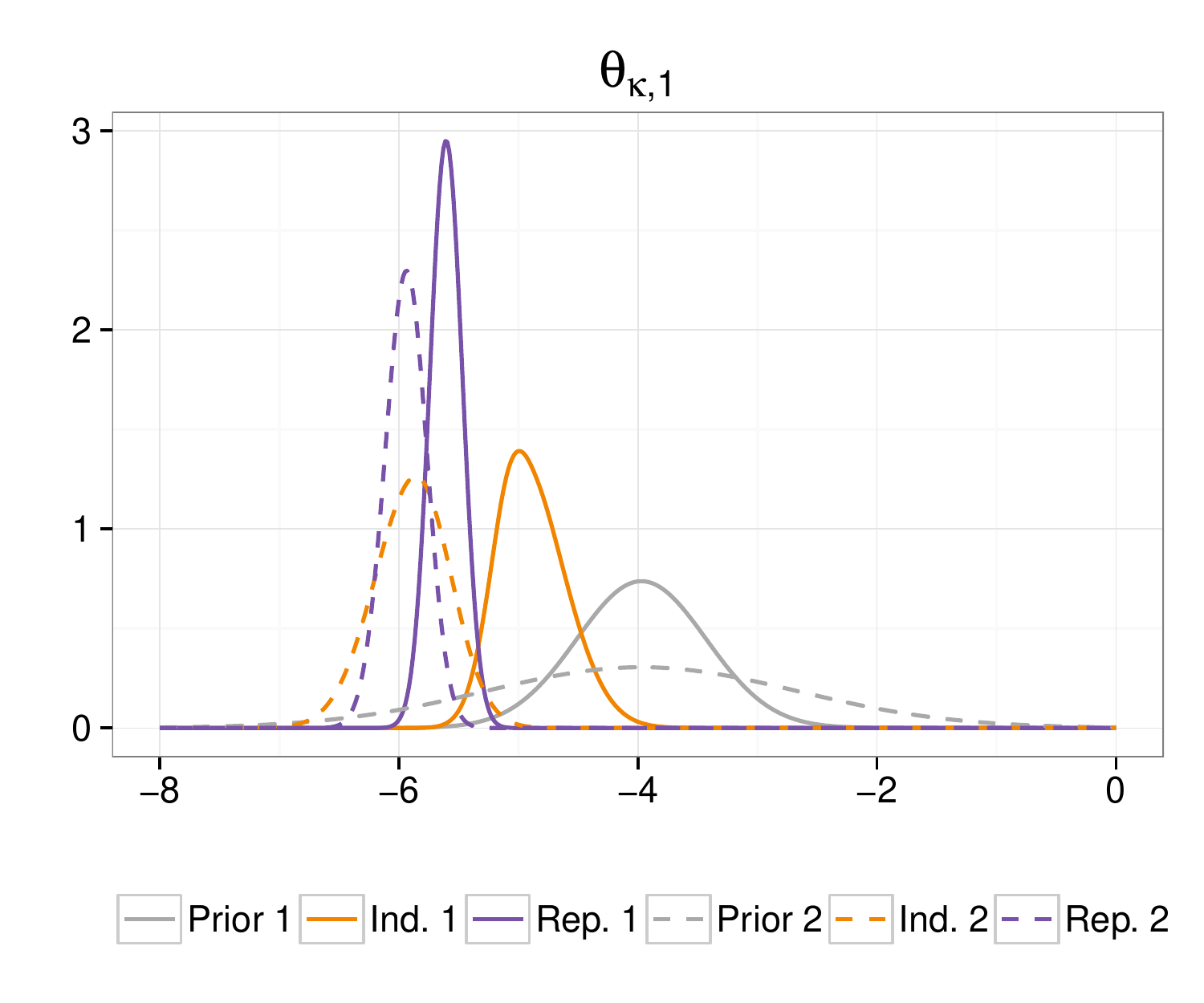}
  \caption{}
  \label{fig:posterior_prior_theta3_nonstat}
 \end{subfigure}%
 \begin{subfigure}{0.45\textwidth}
  \includegraphics[width=\textwidth]{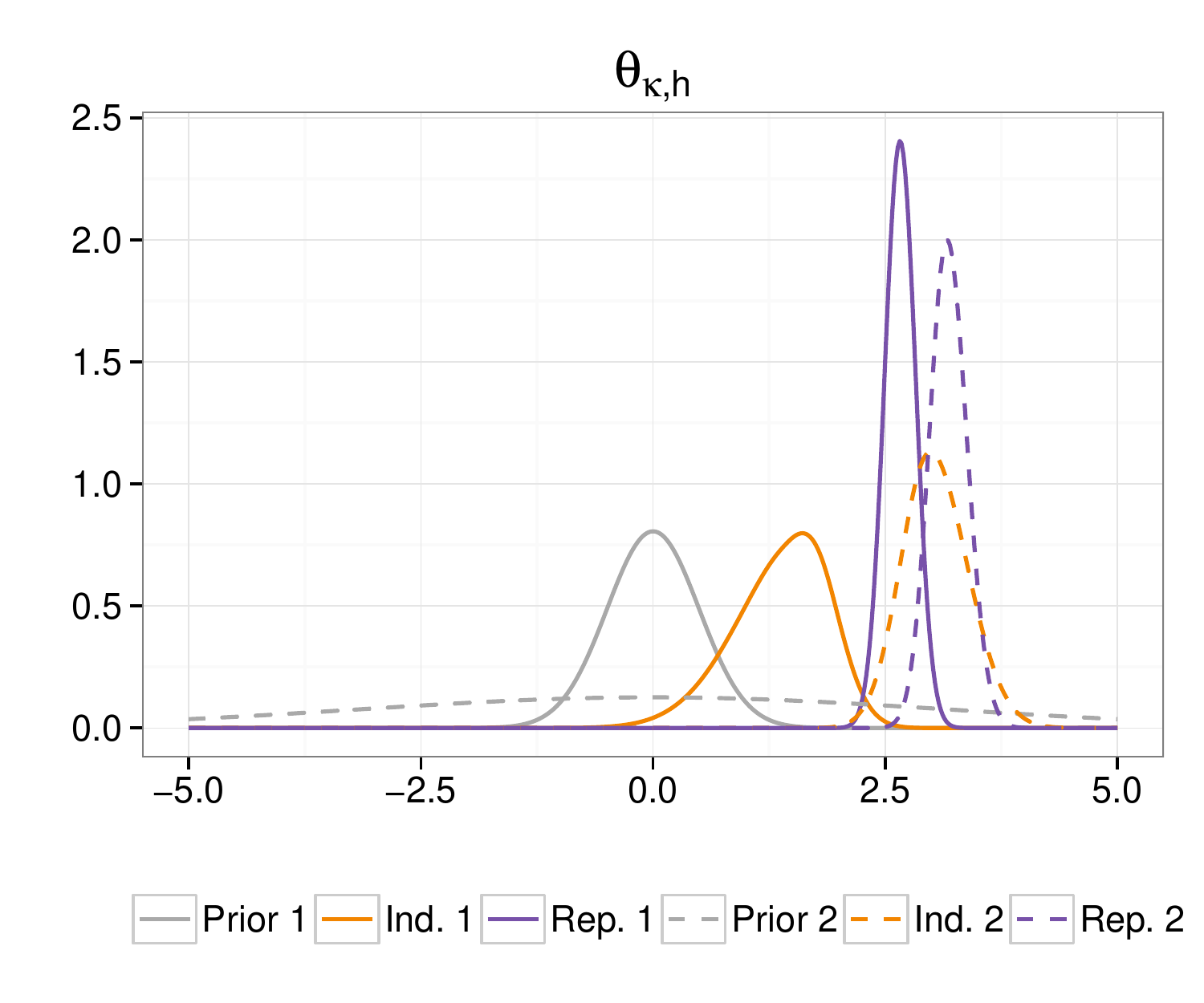}
  \caption{}
  \label{fig:posterior_prior_theta4_nonstat}
 \end{subfigure}%
 \caption{Posterior marginal densities for $\boldsymbol{\theta}_\text{NS}$ when two different 
          priors were used: 1) informative (solid lines) and 2) vague (dashed lines). Posterior marginals 
          with the individual model (data: 2008-2009) are in orange and replicate model (data: 2008-2013) in purple. 
          The priors are in grey.}
 \label{fig:theta_prior_posterior}
\end{figure}

From Figure~\ref{fig:posterior_theta3_nonstat} it seems like the 
posterior marginals from the individual models are shifted towards 
the prior mean. We investigate this further by fitting the non-stationary 
individual models and replicate model with an informative prior and 
a vague prior. The prior mean values are kept fixed, but the variances 
are decreased/increased by changing the 0.9-quantiles of $\rho_\text{S}$ and 
$\sigma_\text{S}$, and the coefficients of variation $c_\rho$ and $c_\sigma$. 
For the informative prior, the 0.9-quantile of $\rho_\text{S}$ is set to 300 km, 
and the 0.9-quantile of $\sigma_\text{S}$ is set to 0.5 m. Further, $c_\rho = 0.2$ 
and $c_\sigma=0.3$. For the vague prior, the 0.9-quantiles are 800 km and 
4 m, and $c_\rho = 2$ and $c_\sigma=6$. In Figure~\ref{fig:theta_prior_posterior}, 
posteriors from models with the different priors are compared. We also compare  
posteriors from the individual model fitted to the 2008-2009 observations, 
and posteriors from the replicate model fitted to the 2008-2013 observations.

\begin{figure}[!h]
 \centering
  \begin{subfigure}{0.32\textwidth}
   \includegraphics[width=\textwidth]{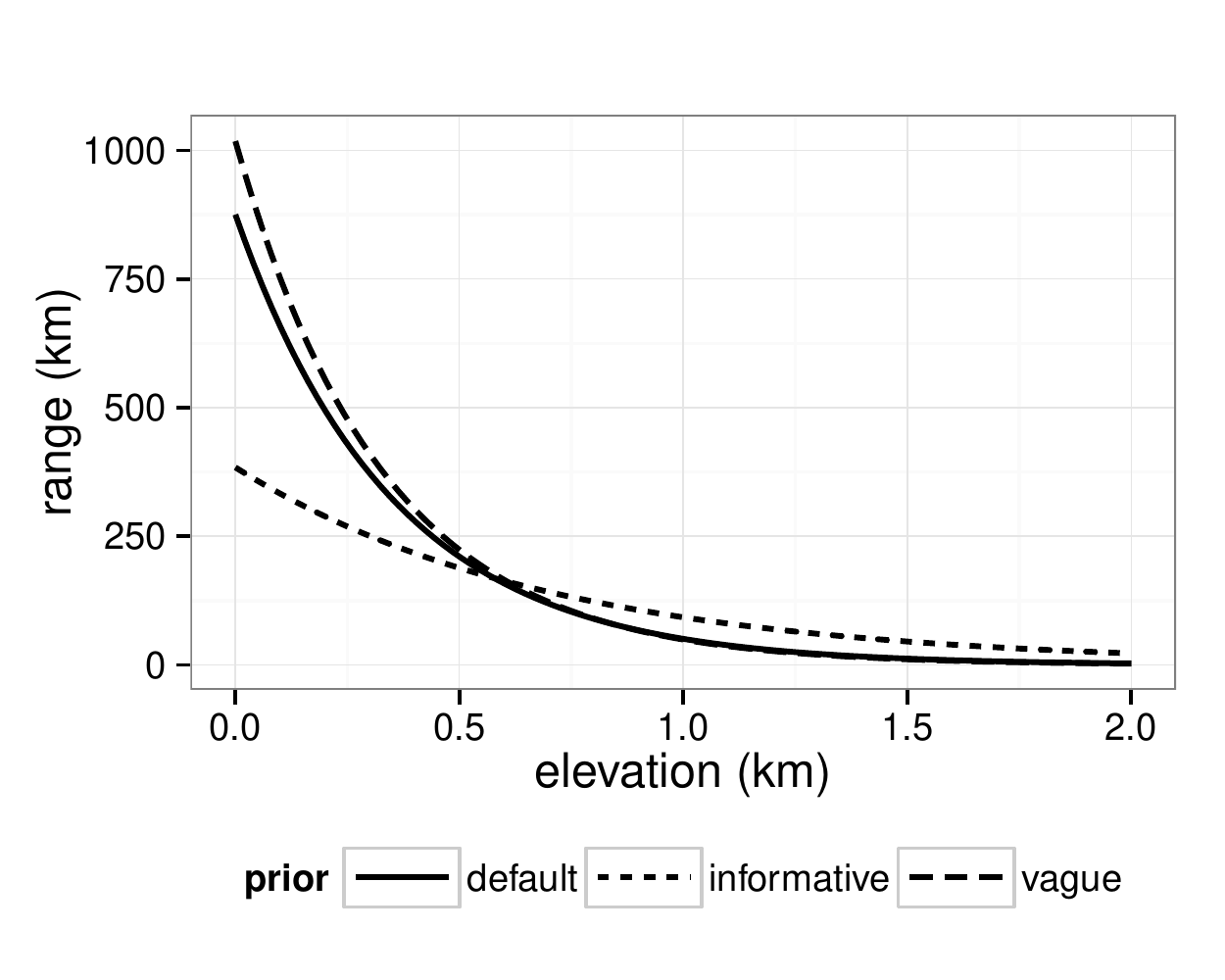}
   \caption{Individual model (2008-2009)}
   \label{fig:range_vs_elevation_year_1}
  \end{subfigure}
  \begin{subfigure}{0.32\textwidth}
   \includegraphics[width=\textwidth]{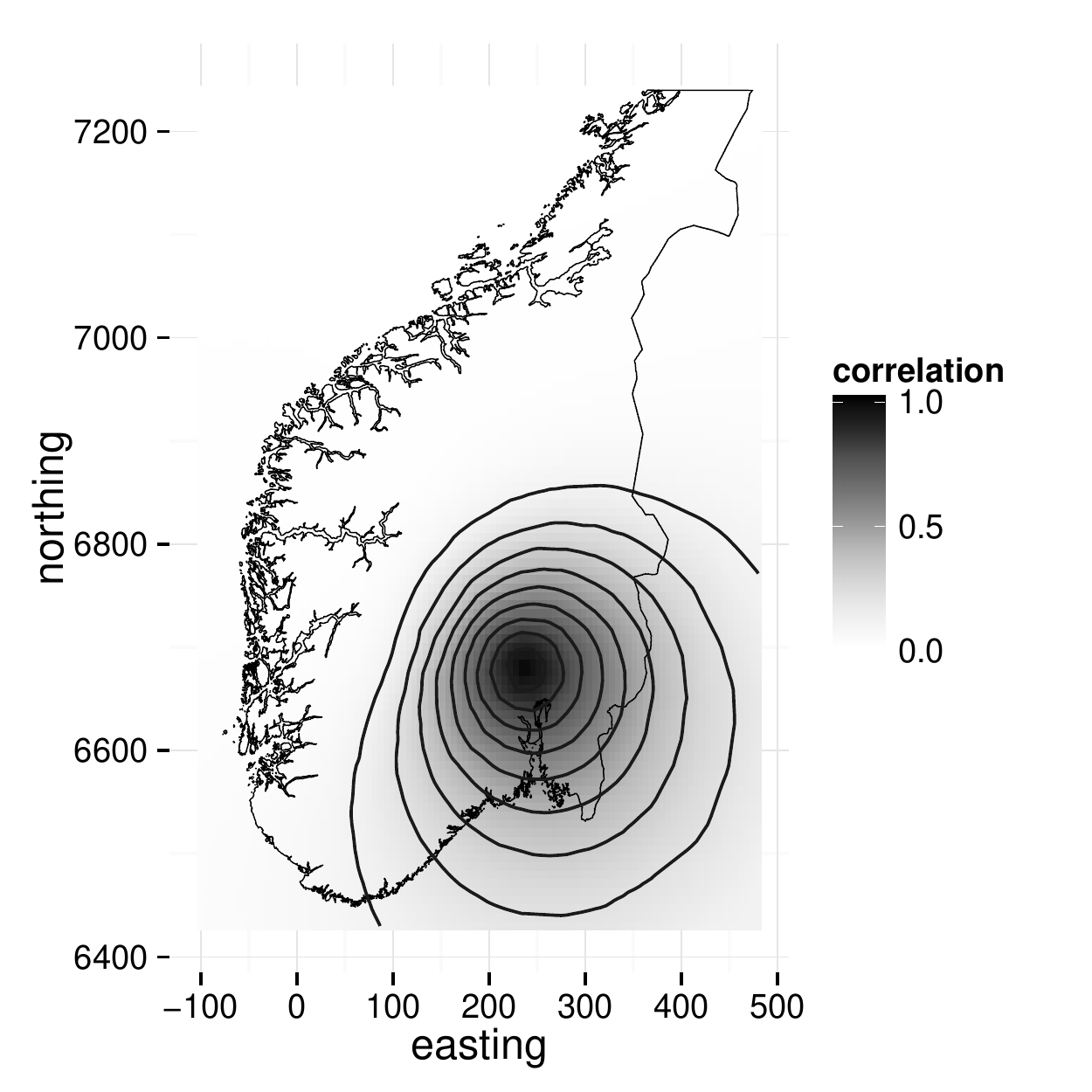}
   \caption{Informative (2008-2009)}
   \label{fig:corr_indiv_informative}
  \end{subfigure}
  \begin{subfigure}{0.32\textwidth}
   \includegraphics[width=\textwidth]{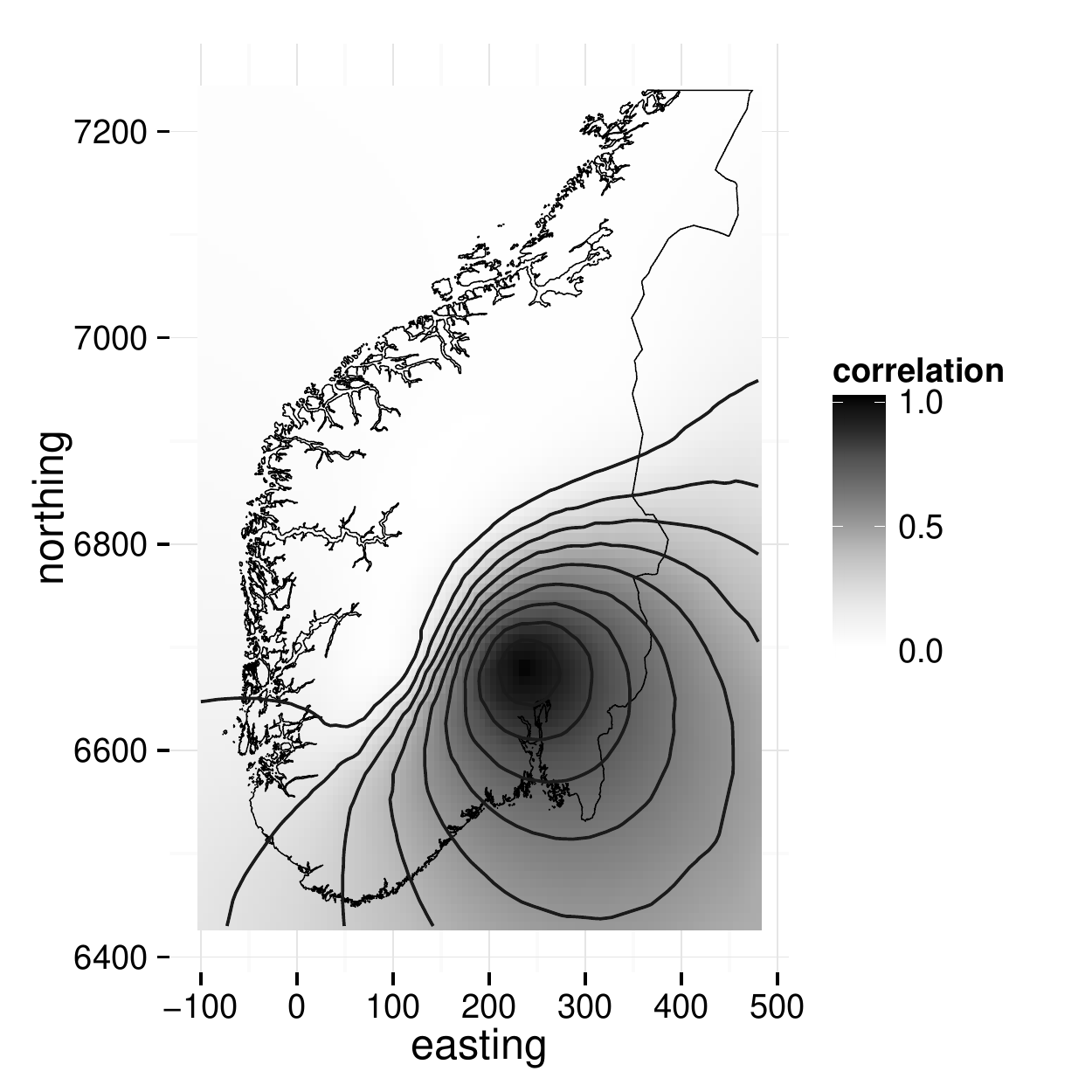}
   \caption{Vague (2008-2009)}
   \label{fig:corr_indiv_vague}
  \end{subfigure}
  
  \begin{subfigure}{0.32\textwidth}
   \includegraphics[width=\textwidth]{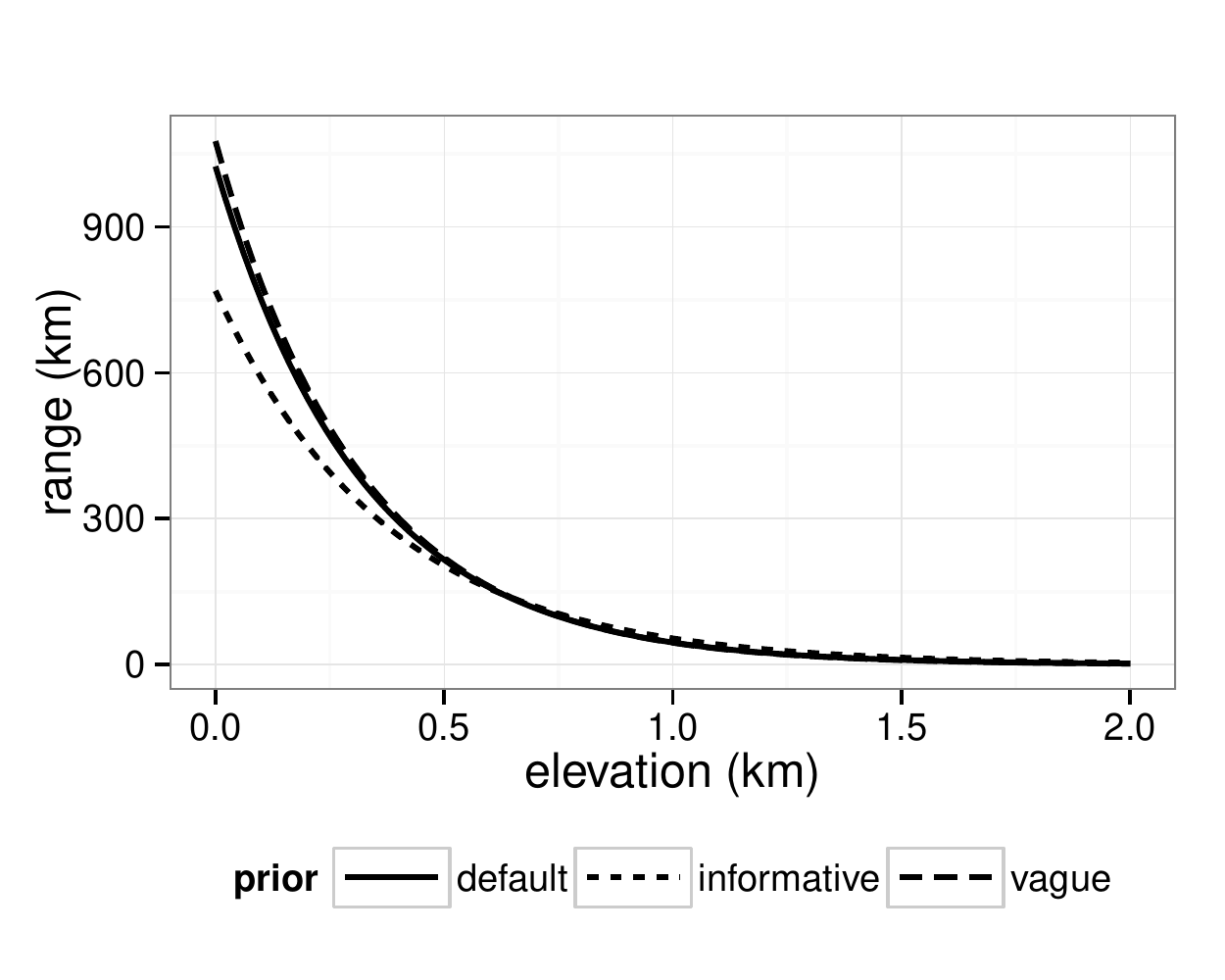}
   \caption{Replicate model (2008-2013)}
   \label{fig:range_vs_elevation_repl}
  \end{subfigure}
  \begin{subfigure}{0.32\textwidth}
   \includegraphics[width=\textwidth]{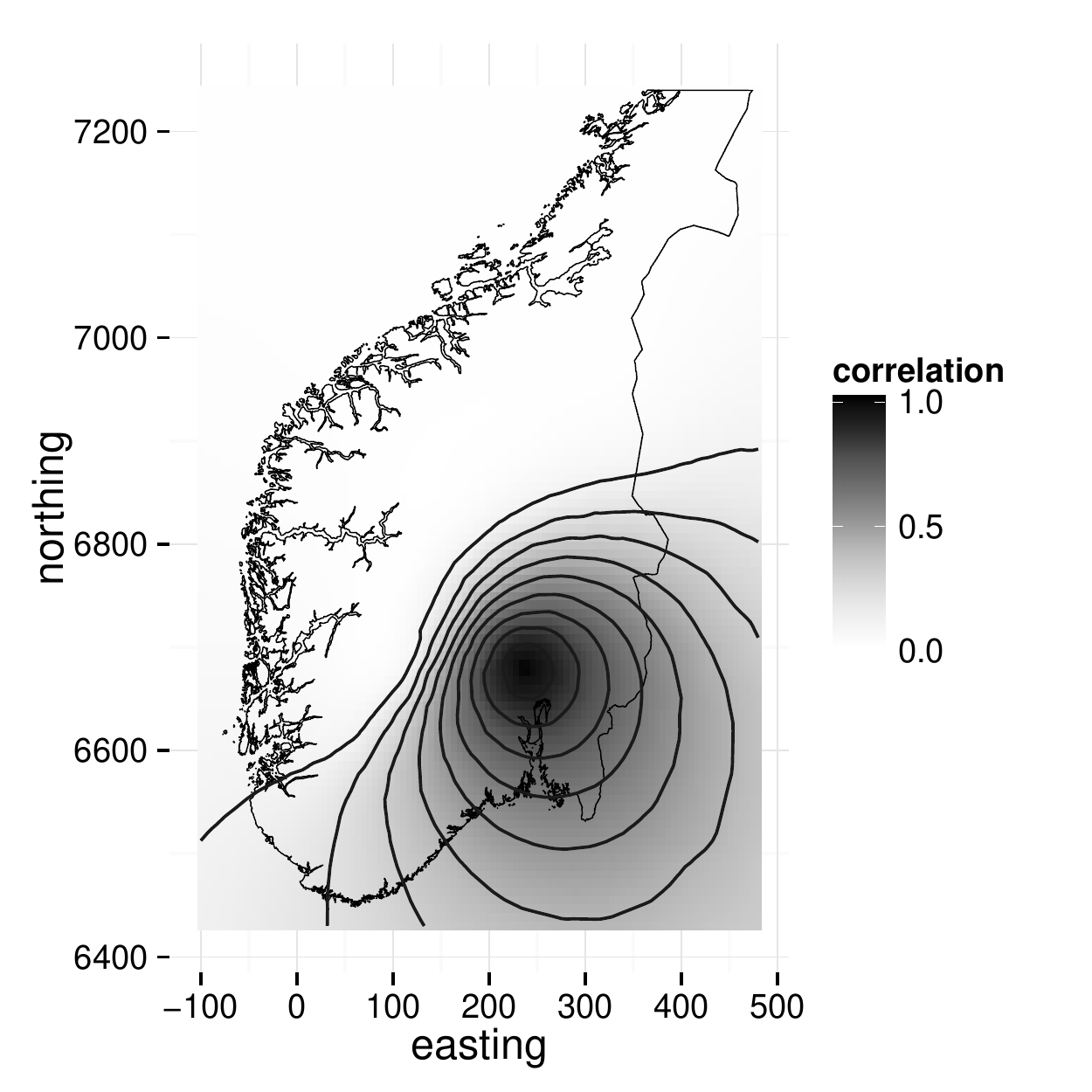}
   \caption{Informative (2008-2013)}
   \label{fig:corr_repl_informative}
  \end{subfigure}
  \begin{subfigure}{0.32\textwidth}
   \includegraphics[width=\textwidth]{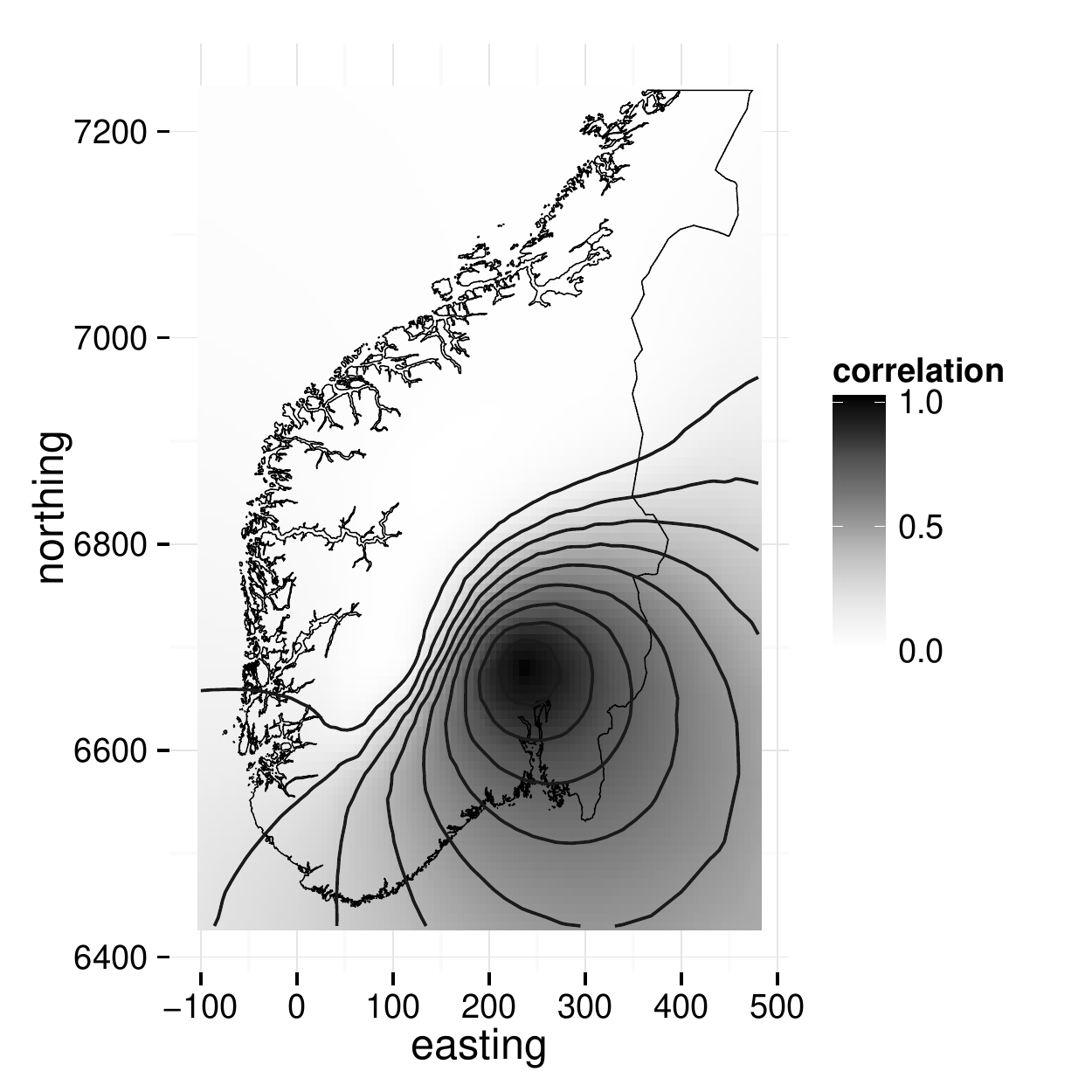}
   \caption{Vague (2008-2013)}
   \label{fig:corr_repl_vague}
  \end{subfigure}
 \caption{Spatial correlation range as a function of elevation, 
          and between a reference location south-east in Norway and all 
          other locations in the domain. Individual model in panel 
          \ref{fig:range_vs_elevation_year_1}-\ref{fig:corr_indiv_vague} 
          and replicate model in panel 
          \ref{fig:range_vs_elevation_repl}-\ref{fig:corr_repl_vague}. 
          Comparison between the informative and vague priors. The default 
          prior referred to in panel \ref{fig:range_vs_elevation_year_1} 
          and \ref{fig:range_vs_elevation_repl} is the one defined in Section~\ref{sec:prior}. }
 \label{fig:correlation_range}
\end{figure}

The dependence structure parameters are more sensitive to the 
choice of priors when there are no replicates of the spatial 
field. Also, the range parameters $\theta_{\kappa, 1}$ and 
$\theta_{\kappa, h}$ are less robust than the variance 
parameters $\theta_{\tau, 1}$ and $\theta_{\tau, h}$. 
In Figure~\ref{fig:correlation_range} we show how the 
estimated spatial correlation range is influenced by different priors.
With the individual model (panels \ref{fig:range_vs_elevation_year_1}-\ref{fig:corr_indiv_vague}) the prior is very influential 
on the spatial correlation. At sea level, the difference in 
correlation range is longer than the domain. With the replicate 
model (panels \ref{fig:range_vs_elevation_repl}-\ref{fig:corr_repl_vague}), the difference is much smaller. 

A small study of the sensitivity with respect to $c_\rho$ 
and $c_\sigma$ can be found in Appendix~\ref{app:Prior_sensitivity}. 
The conclusion is that both estimates and cross-validated 
predictive scores are robust to the choices of $c_\rho$ and $c_\sigma$.

\section{Simulation study}\label{sec:Simulation_study}

The purpose of this simulation study is to investigate statistical 
properties of the spatial replicate model for annual precipitation. 
Parameter estimates and spatial predictions are assessed when 
increasing number of replicates of the spatial field are used for 
inference. Datasets are sampled from both the stationary and 
non-stationary models, and models with both dependence structures 
are fitted to all datasets. The sampled fields respect the 
identifiability constraints in \eqref{eq:zero_constr} and \eqref{eq:h_constr}, 
by use of the general constrained 
sampling methods from \citet{GMRF}.
We investigate how well DIC performs as 
a model choice criterion, and evaluate how well the true parameters 
are recovered by looking at posterior mean values, credible interval 
coverage, and RMSE. Further, we compare predictive performance of the 
stationary and non-stationary models as a function of the number of 
replicates, as well as pointwise in space.

\subsection{Set-up}\label{sec:Simulation_study_set-up}

Annual precipitation datasets were sampled from the replicate 
model defined in Section~\ref{sec:Model-precipitation}, with
both stationary and non-stationary dependence structures. The 
spatial domain, triangulation, and smoothed elevation model were the 
same as in the data analysis section (Section~\ref{sec:Data_analysis}). 
We used the 233 stations from the 2008-2009 dataset as observation 
locations. The chosen parameter values are based on the analysis of 
the southern Norway data: The year specific intercepts 
$\beta_1,\beta_2,\dots,\beta_r$ were given a common value of 
$\beta_0 = 0.6$, the linear effect of elevation was set to 
$\beta_h = 0.4$, and the precision of the measurement error was set 
to $\tau_\epsilon = 40$. These parameters were equal for the 
stationary and non-stationary datasets. The dependence structure 
parameters were set to $\boldsymbol{\theta}_\text{S} = (3.5, -4.5)$ 
to obtain datasets based on stationary spatial fields, and  
$\boldsymbol{\theta}_\text{NS} = (3.9, -1.3, -5.9, 3.1)$ to obtain 
datasets based on non-stationary spatial fields.
 
We assume we have data from $r$ years at the 233 stations, 
and let the number of replicates $r$ range from one 
to ten. The results we present in the next section are 
based on 250 sampled datasets. For $r=1$ this means 250 annual 
datasets. However, for $r>1$, one dataset means $r$ annual datasets, 
so in total we sample $250 \times r$ sets of observations. 

The stationary and non-stationary models with the priors defined 
in Section~\ref{sec:prior} were fitted to all sampled datasets, 
i.e.\ $r$ sets of annual total precipitation observations at the 
233 stations with the underlying fields being both stationary and 
non-stationary. The fitted 
models were used to predict annual total precipitation on the 
entire spatial domain (on the triangulation) and the predictions 
were compared to the sampled precipitation fields. 
Results follow in the next section, and in 
Appendix~\ref{app:Figures}.

\subsection{Results}\label{sec:Simulation_study_results}

\begin{figure}[!h]
 \centering
   \begin{subfigure}{0.49\textwidth}
    \includegraphics[width=\textwidth]{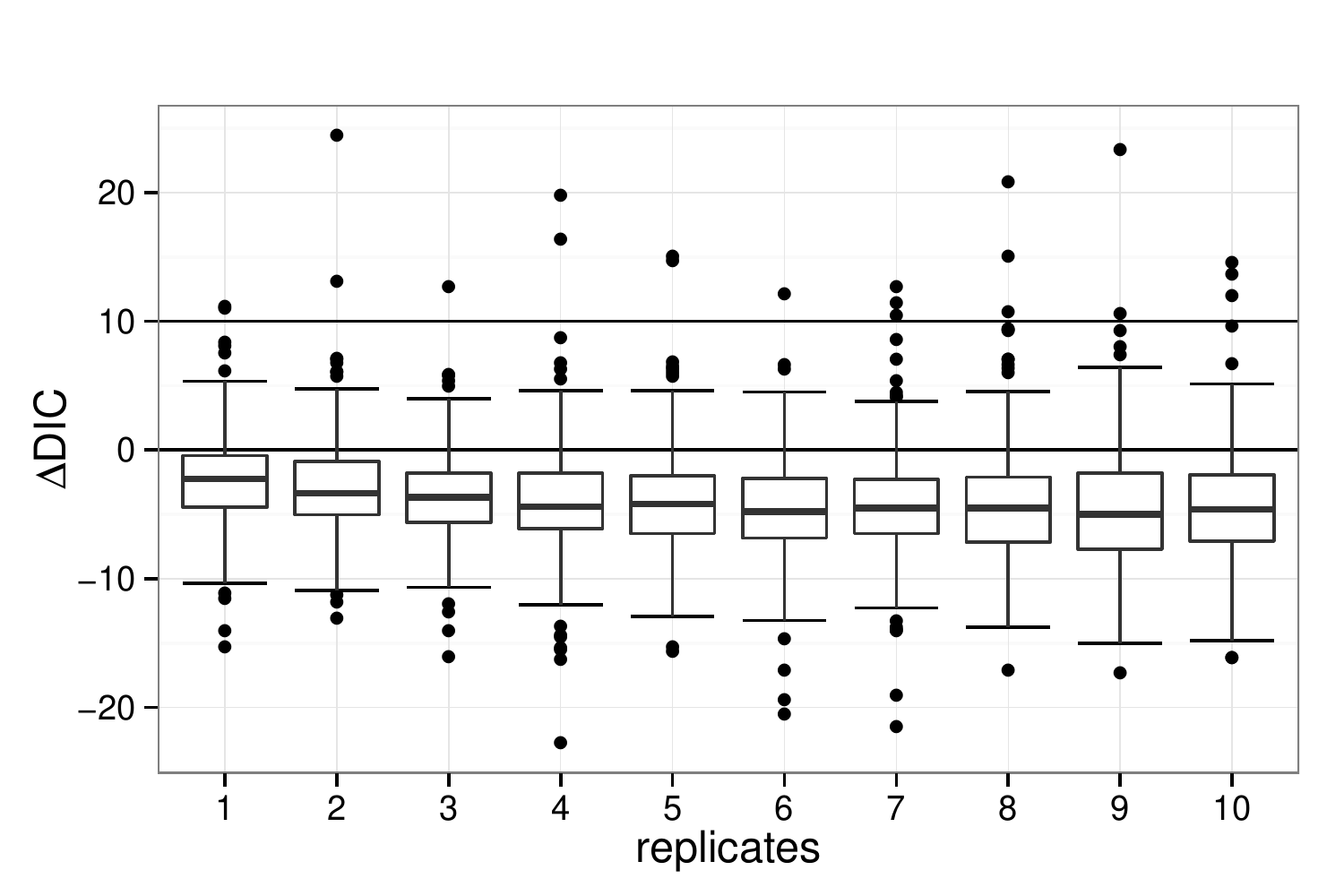}
    \caption{Stationary datasets}
    \label{fig:delta_dic_stat}
   \end{subfigure}%
    ~
  \begin{subfigure}{0.49\textwidth}
   \includegraphics[width=\textwidth]{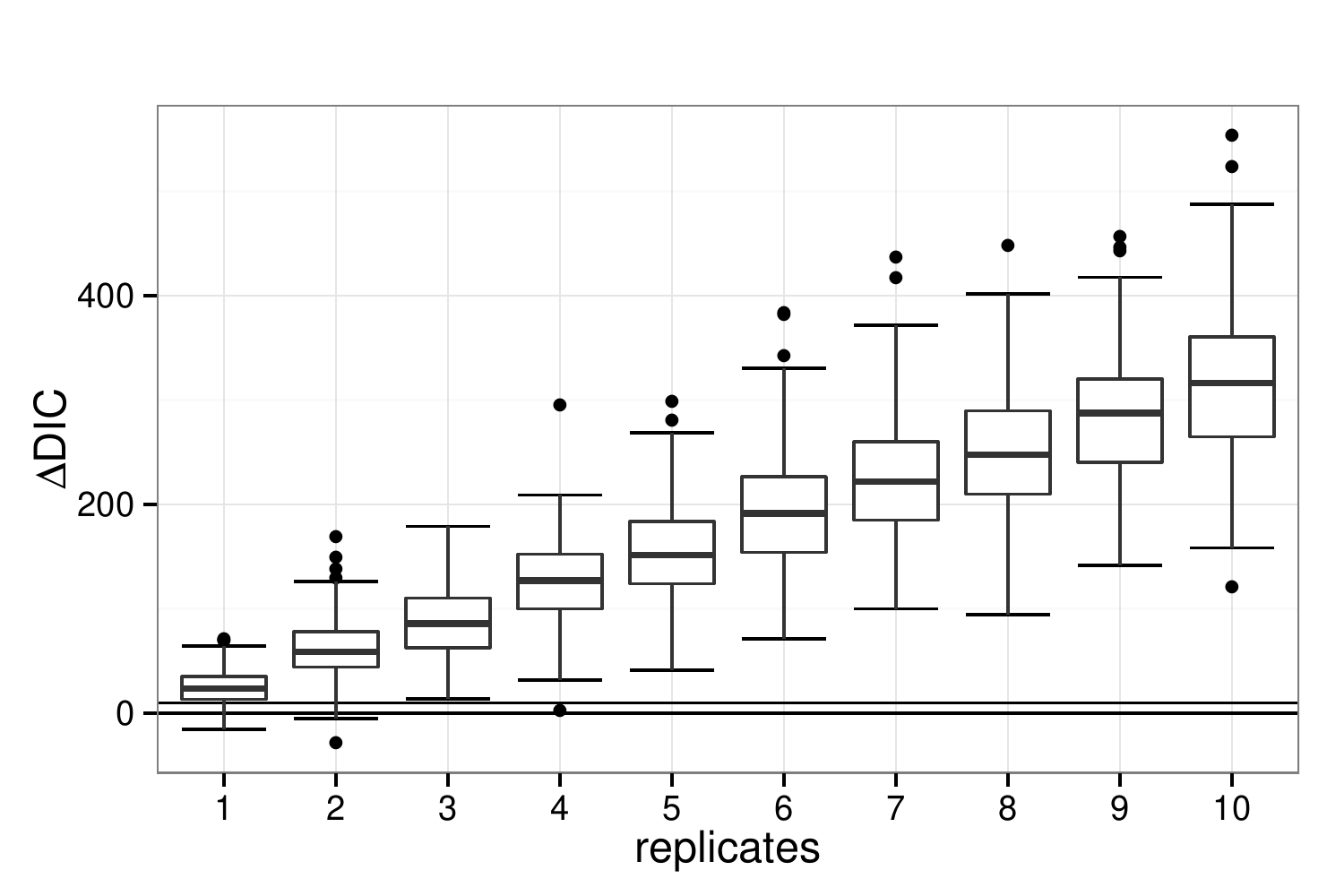}
   \caption{Non-stationary datasets}
   \label{fig:delta_dic_nonstat}
  \end{subfigure}
  \caption{Box plots of the difference in DIC as a function of the number of replicates. 
           Stationary and non-stationary models were fitted to datasets sampled from the 
           stationary model (\ref{fig:delta_dic_stat}) and non-stationary model 
           (\ref{fig:delta_dic_nonstat}). $\Delta$DIC is the DIC of the stationary model minus 
           DIC of the non-stationary model. $\Delta$DIC of zero and ten are indicated with 
           horizontal lines.}
 \label{fig:delta_dic}
\end{figure}

Figure~\ref{fig:delta_dic} contains boxplots of $\Delta$DIC, defined 
as DIC of the stationary model minus DIC of the non-stationary model. 
In the left panel (\ref{fig:delta_dic_stat}) the datasets are sampled 
from the stationary model and in the right panel (\ref{fig:delta_dic_nonstat}) 
from the non-stationary model. The model with the lowest DIC should be 
preferred and a rule of thumb is that a complex model should be preferred 
over a simpler model if the difference is greater than ten. We have indicated 
$\Delta\text{DIC}=10$ with a horizontal line and note that DIC almost always 
favour the correct model. Note also that when the datasets are non-stationary, 
the ability to select the correct model improves when the number of replicates 
increases. When the datasets are 
from the non-stationary model, the misclassification rate is 0.16 for one 
replicate, 0.05 for two replicates, one dataset is misclassified when $r=4$, 
and zero for all other values of $r$. 

The rest of this section will focus on results when the datasets 
are sampled from a non-stationary model, and on the dependence 
structure parameters $\boldsymbol{\theta}_\text{NS}$. Results 
for the other parameters; $\beta_0$, $\beta_h$, and $\tau_\epsilon$, 
and results when the true model is stationary can be found in 
Appendix~\ref{app:Figures}. 
In Figure~\ref{fig:mean_theta_nonstat_data} we have plotted 
the average of the posterior mean values for $\theta_{\tau,1}$, 
 $\theta_{\tau,h}$, $\theta_{\kappa,1}$, and $\theta_{\kappa,h}$ 
estimated from the 250 datasets. The shaded areas indicate the 
range of the posterior mean values over all simulations. 
The overall trend is that 
the average values approach the true values when the number of 
replicates increases and the posterior mean values become more 
concentrated around the truth. The stationary model with 
dependence structure parameters $\theta_\tau$ and 
$\theta_\kappa$ is, as expected, not able to recover the 
non-stationary parameters $\theta_{\tau,1}$ and $\theta_{\kappa,1}$ 
and using replicates does not reduce the bias. 
However, in Figure~\ref{fig:mean_theta_stat_data} in 
Appendix~\ref{app:Stationary_datasets} it can be seen 
that the non-stationary model is able to adapt to the 
stationary datasets.  

\begin{figure}[!h]
 \centering
  \begin{subfigure}{0.45\textwidth}
   \includegraphics[width=\textwidth]{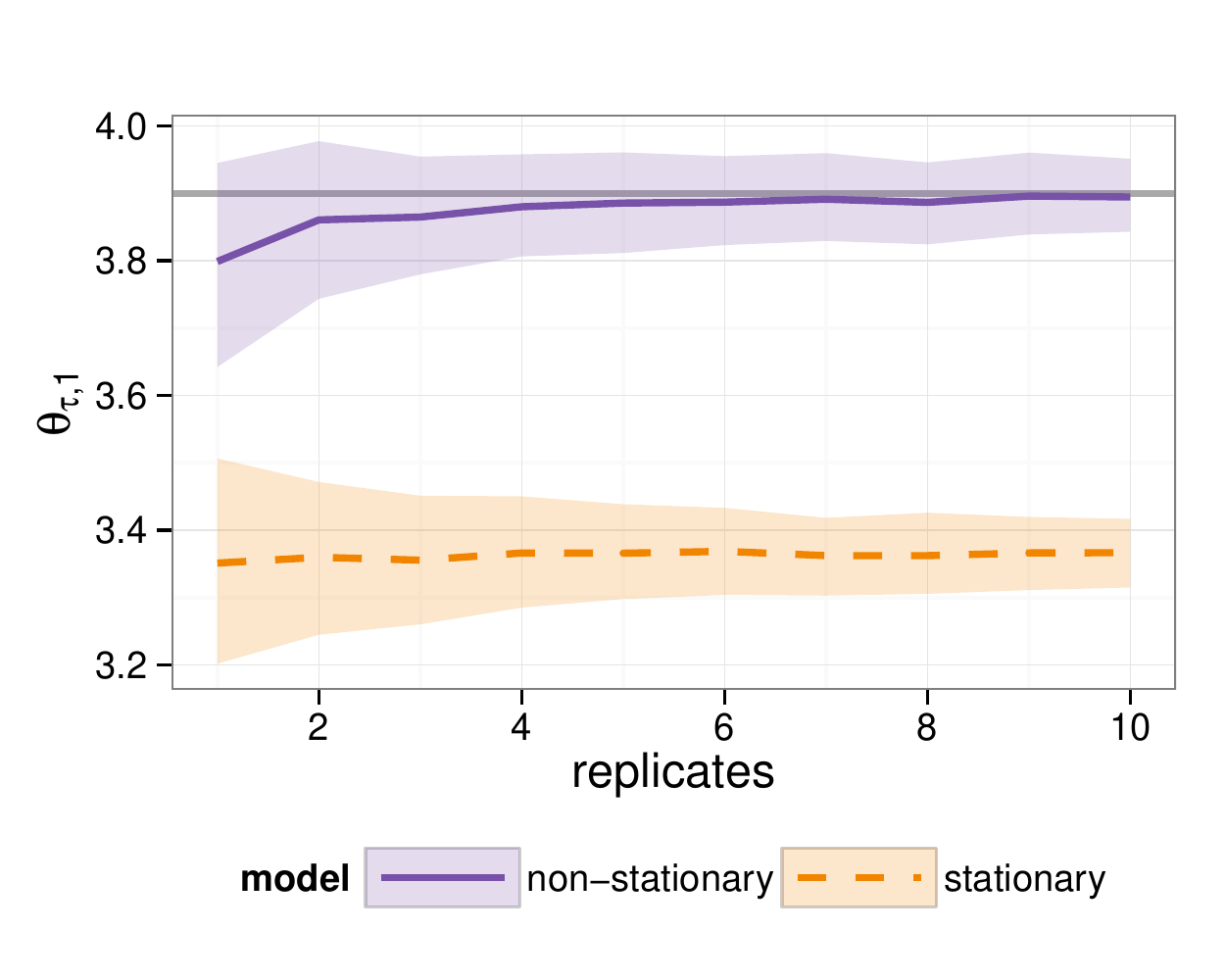}
   \caption{Posterior mean $\theta_{\tau,1}$}
   \label{fig:theta1_nonstat_data}
  \end{subfigure}
  \begin{subfigure}{0.45\textwidth}
   \includegraphics[width=\textwidth]{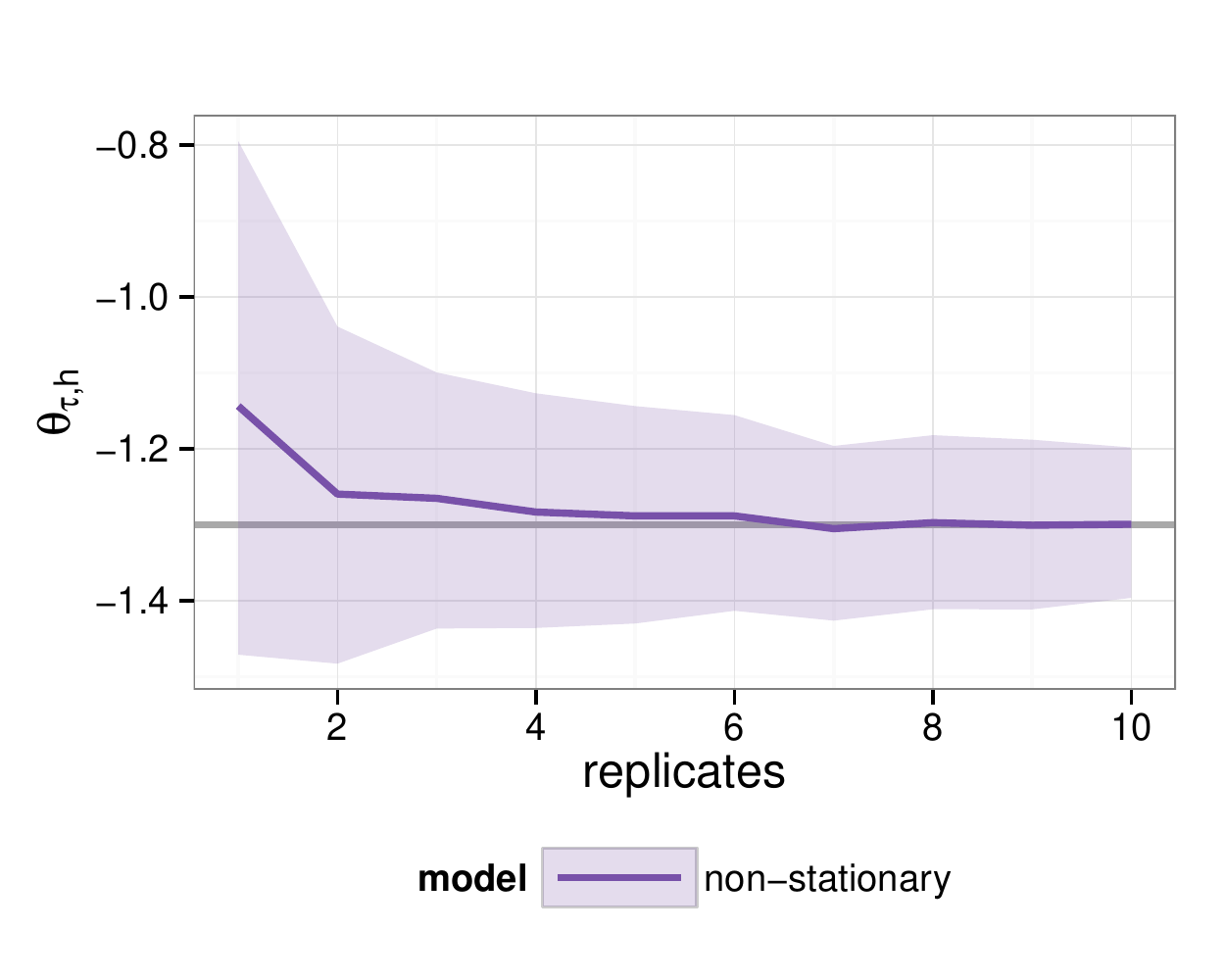}
   \caption{Posterior mean $\theta_{\tau,h}$}
   \label{fig:theta2_nonstat_data}
  \end{subfigure}

  \begin{subfigure}{0.45\textwidth}
   \includegraphics[width=\textwidth]{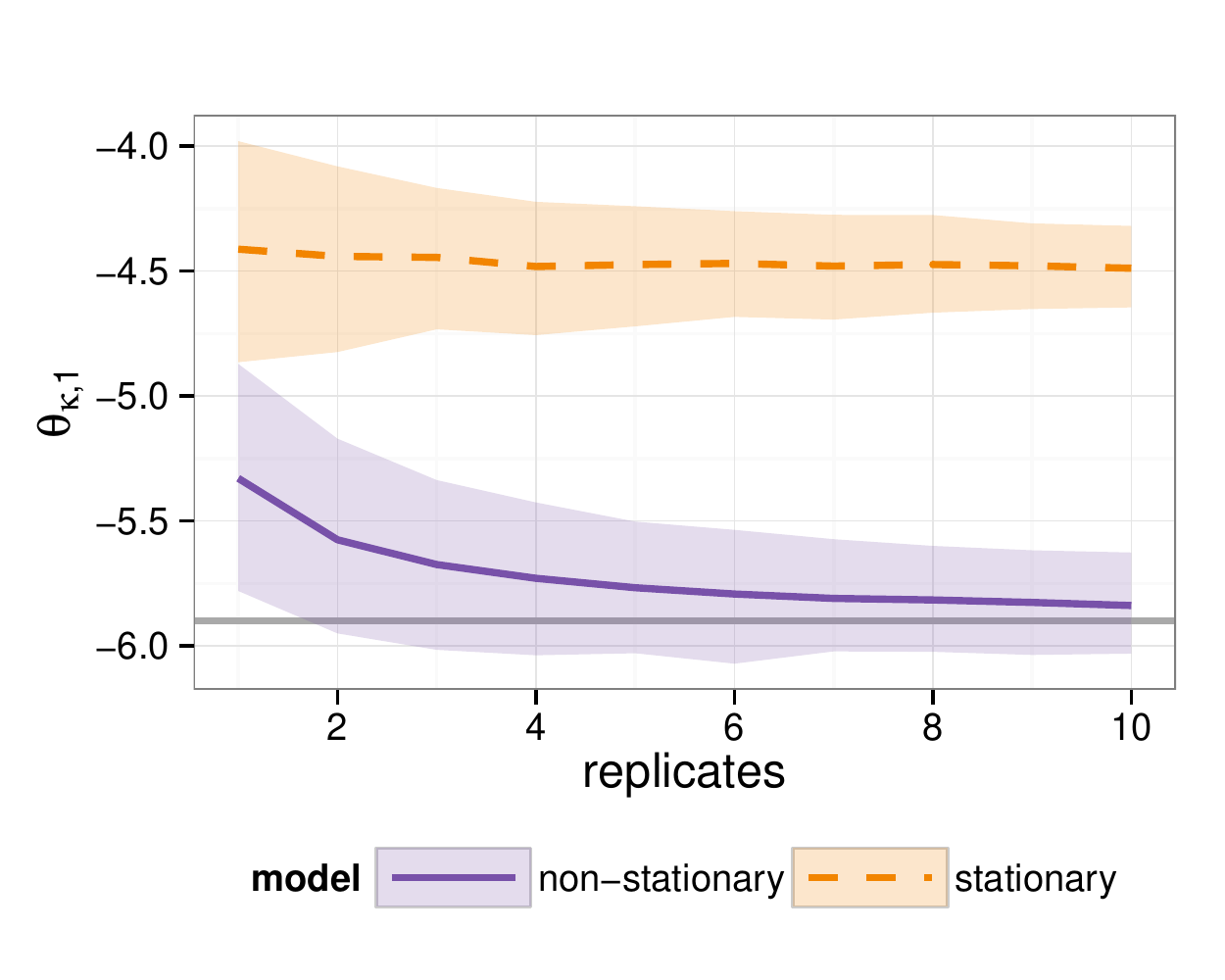}
   \caption{Posterior mean $\theta_{\kappa,1}$}
   \label{fig:theta3_nonstat_data}
  \end{subfigure}
  \begin{subfigure}{0.45\textwidth}
   \includegraphics[width=\textwidth]{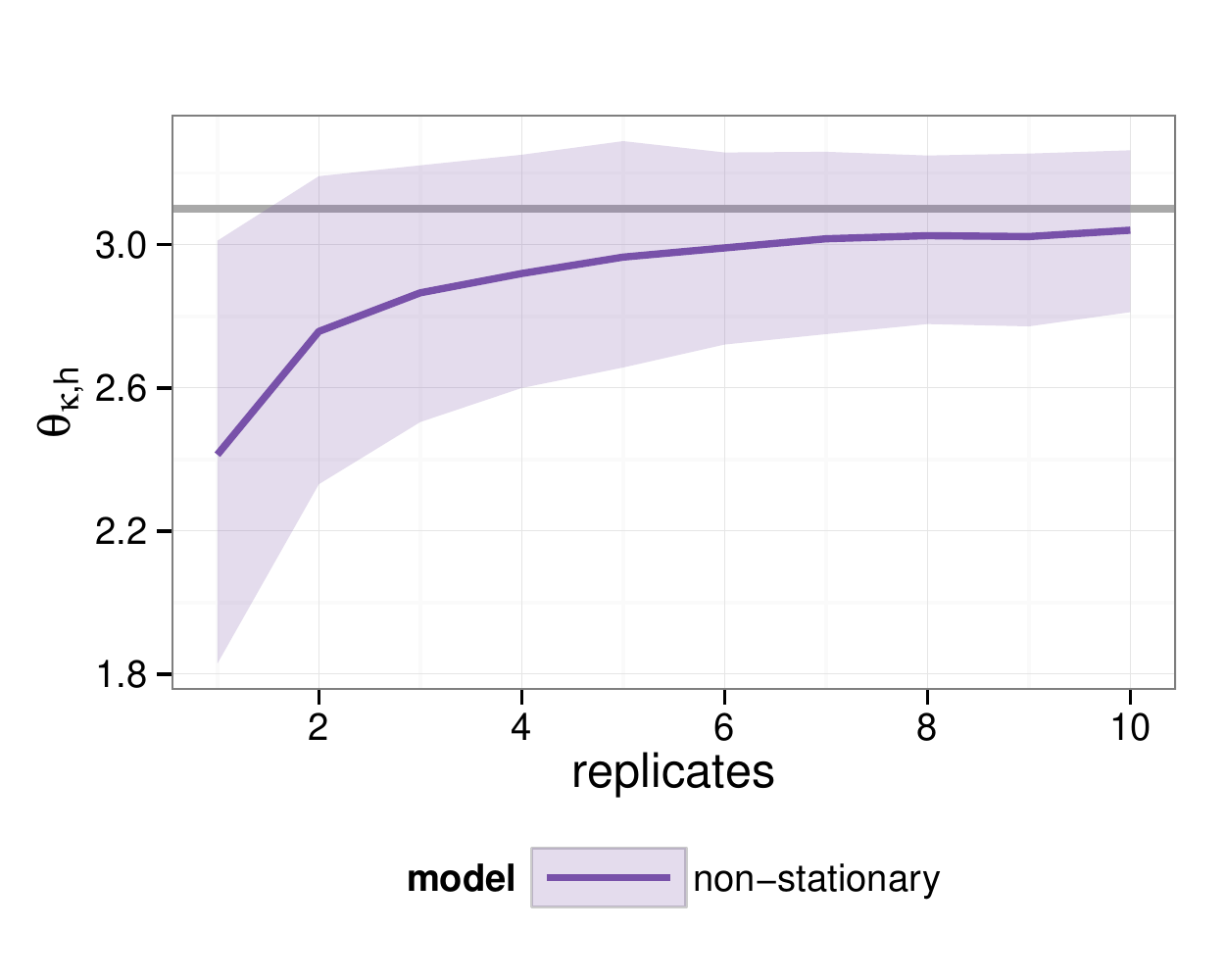}
   \caption{Posterior mean $\theta_{\kappa,h}$}
   \label{fig:theta4_nonstat_data}
  \end{subfigure}
 \caption{Stationary and non-stationary replicate models were fitted 
          to datasets sampled from the non-stationary model. Presented 
          are the posterior mean values for the spatial dependence 
          structure parameters  $\boldsymbol{\theta}_\text{S}$ and 
          $\boldsymbol{\theta}_\text{NS}$. The lines are averages 
          over all datasets, while the shaded area span the range of 
          the posterior mean values (between the 0.1- and 0.9-quantiles). 
          The true parameter values are indicated with grey lines.}
 \label{fig:mean_theta_nonstat_data}
\end{figure}

\begin{figure}[!h]
 \centering
  \begin{subfigure}{0.45\textwidth}
   \includegraphics[width=\textwidth]{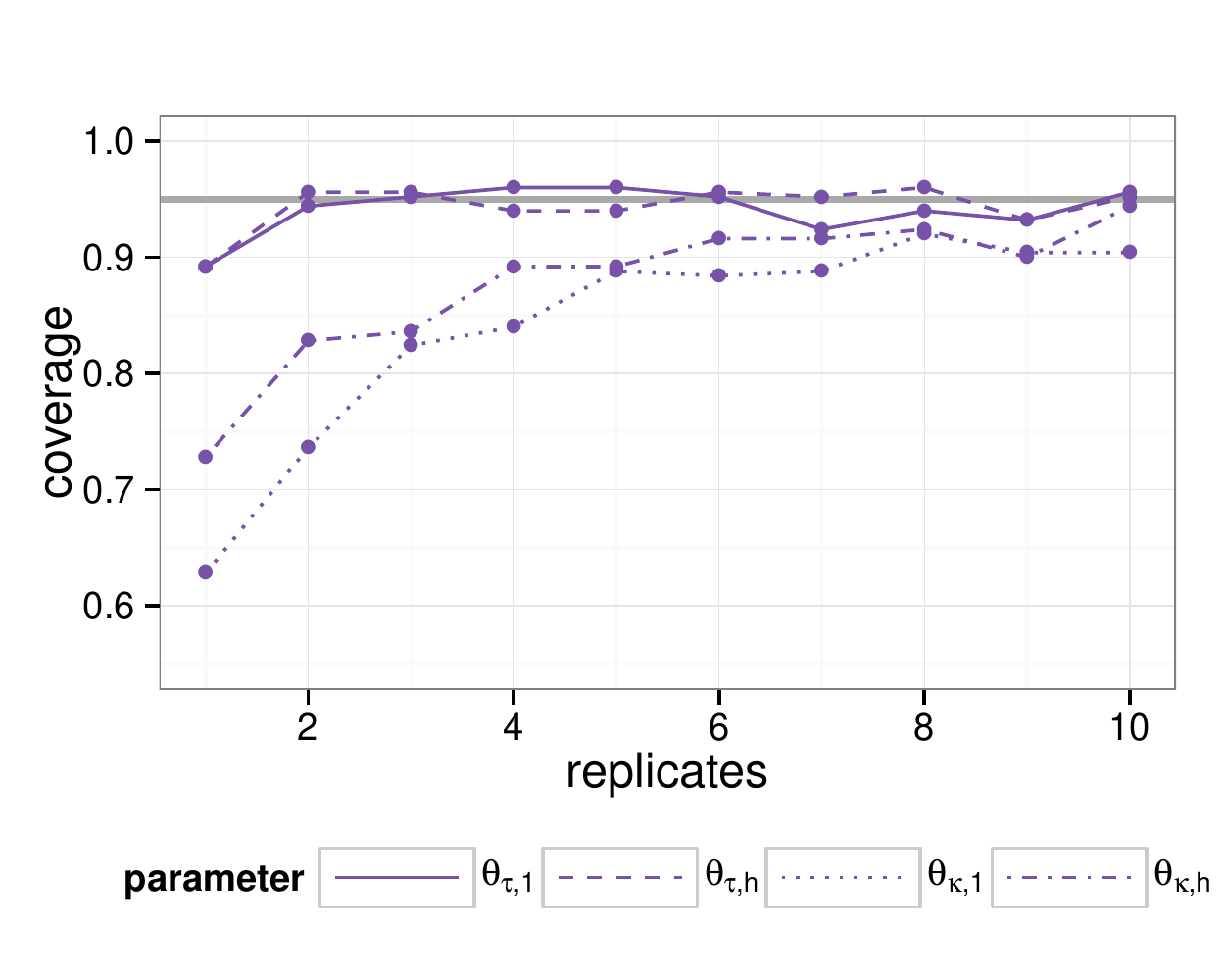}
   \caption{Coverage non-stationary model}
   \label{fig:theta_coverage_nonstat_model_nonstat_data}
  \end{subfigure}
  \begin{subfigure}{0.45\textwidth}
   \includegraphics[width=\textwidth]{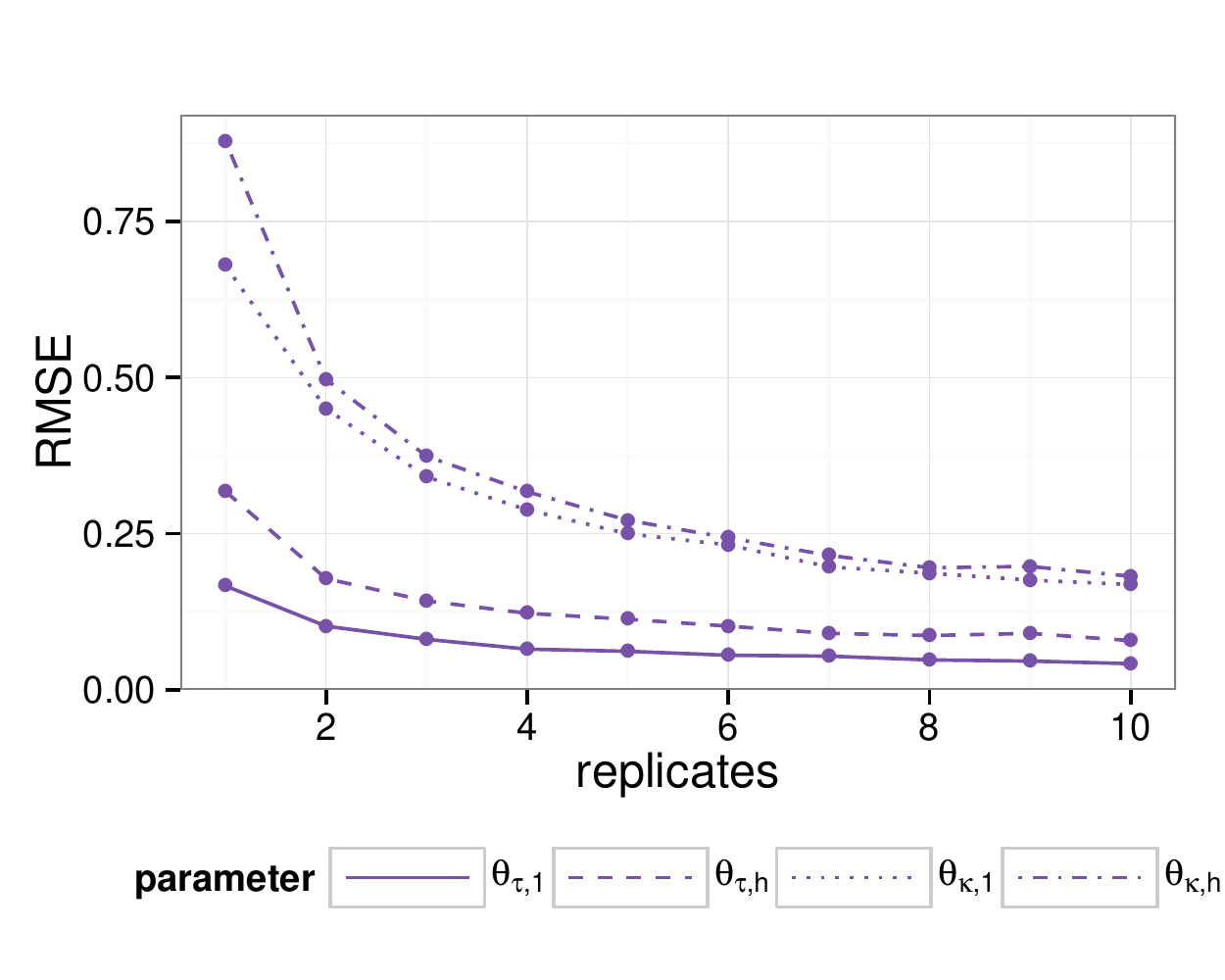}
   \caption{RMSE non-stationary model}
   \label{fig:theta_RMSE_nonstat_model_nonstat_data}
  \end{subfigure}
 \caption{Non-stationary replicate models were fitted to datasets 
          sampled from the non-stationary model. Presented are 
          95\% posterior credible interval coverage and RMSE for 
          the spatial dependence structure parameters   
          $\boldsymbol{\theta}_\text{NS}$.}
 \label{fig:coverage_RMSE_theta_nonstat_data}
\end{figure}

The parameters related to the marginal variance, $\theta_{\tau,1}$ 
and $\theta_{\tau,h}$, are better recovered than the spatial range 
parameters $\theta_{\kappa,1}$ and $\theta_{\kappa,h}$. This can 
also be seen in Figure~\ref{fig:coverage_RMSE_theta_nonstat_data}, 
which shows 95\% posterior credible interval coverage and RMSE 
for $\boldsymbol{\theta}_\text{NS}$. The coverage is as low as 
63\% for $\theta_{\kappa,1}$ when only one replicate of the 
field is used, but increases to 74\% when the number of 
replicates is two. Adding more replicates increases the coverage 
and reduces RMSE. However, the largest improvement is between 
one and two replicates. 

\begin{figure}[!h]
 \centering
  \begin{subfigure}{0.32\textwidth}
   \includegraphics[width=\textwidth]{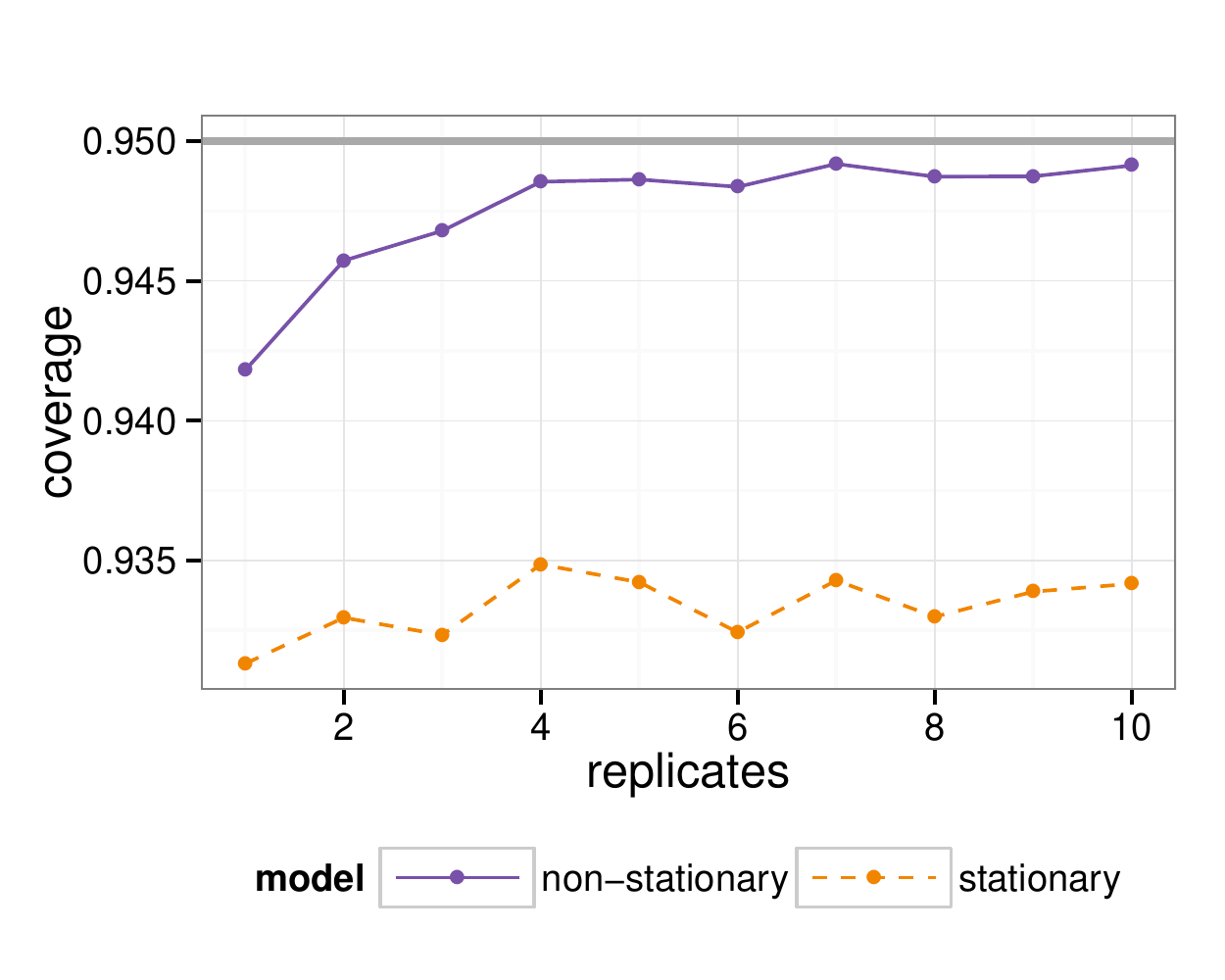}
   \caption{}
   \label{fig:coverage_repl_nonstat_data}
  \end{subfigure}
  \begin{subfigure}{0.32\textwidth}
   \includegraphics[width=\textwidth]{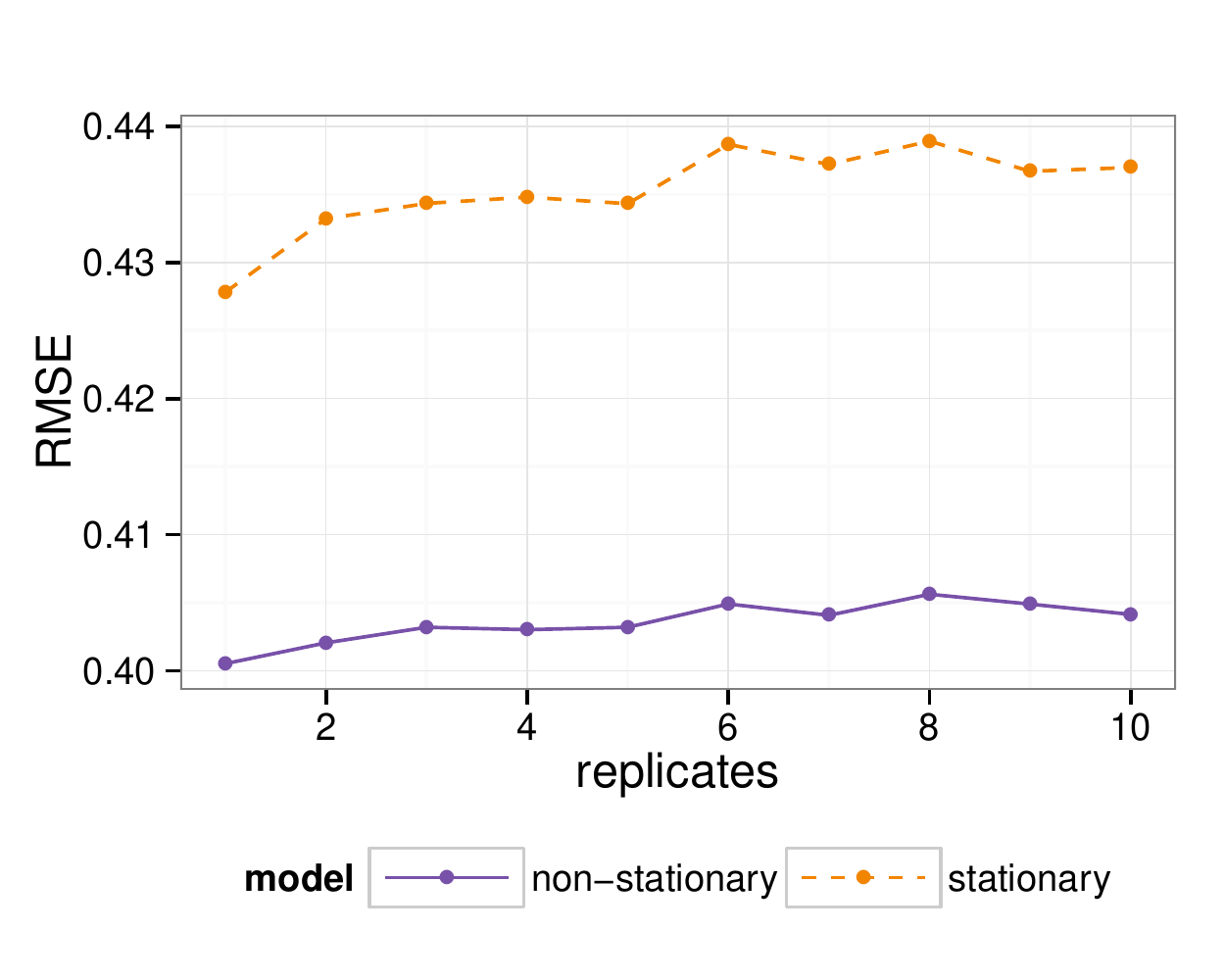}
   \caption{}
   \label{fig:rmse_repl_nonstat_data}
  \end{subfigure}
  \begin{subfigure}{0.32\textwidth}
   \includegraphics[width=\textwidth]{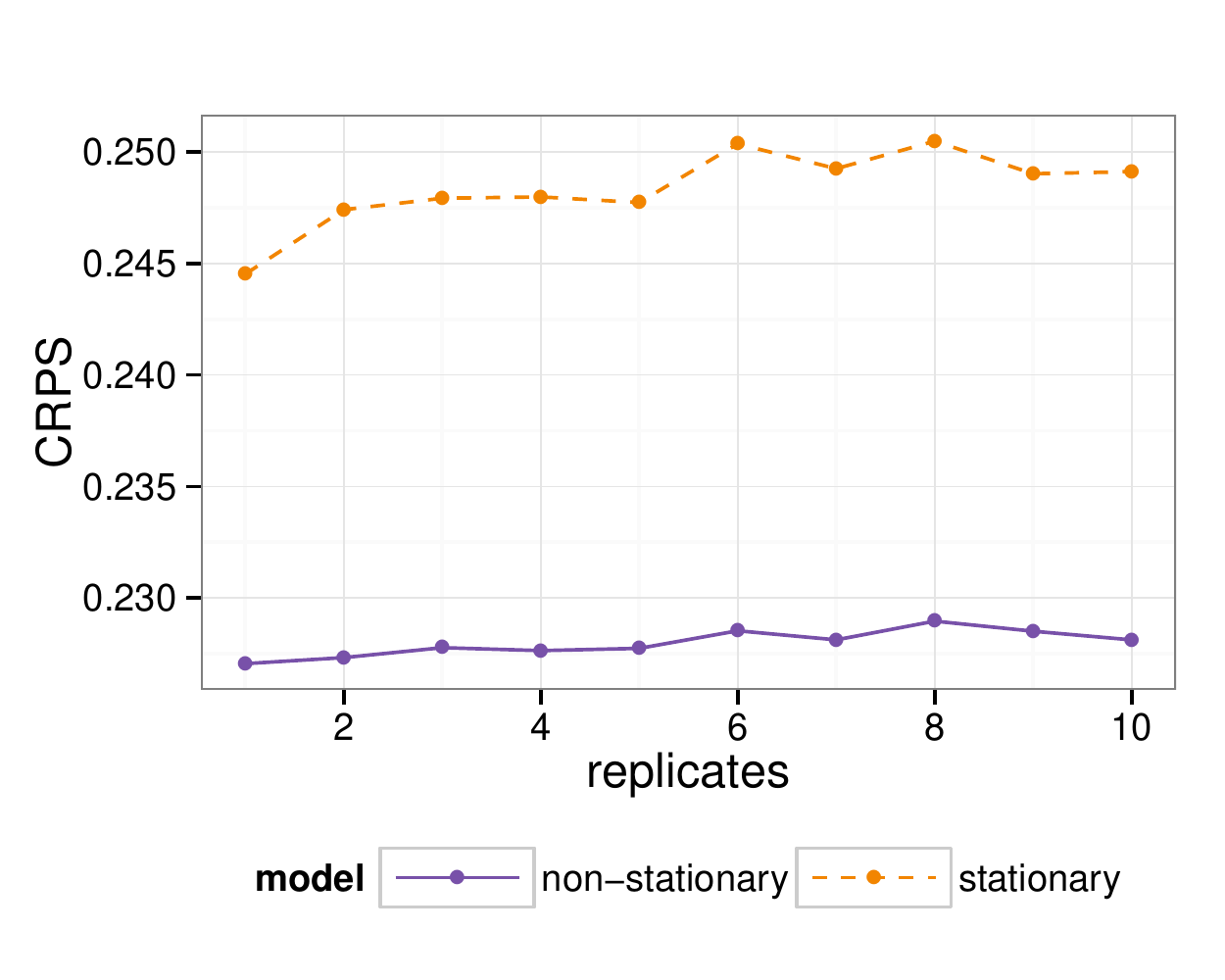}
   \caption{}
   \label{fig:crps_repl_nonstat_data}
  \end{subfigure}
 \caption{Stationary and non-stationary replicate models were fitted to datasets 
          sampled from the non-stationary replicate model. Spatial predictions at 
          the locations of the nodes of the triangulated domain are compared to 
          the sampled fields. Predictive sores; coverage, RMSE, and CRPS, 
          are averaged over all nodes.}
 \label{fig:predictions_nonstat_data}
\end{figure}

\begin{figure}[!h]
 \centering
  \begin{subfigure}{0.45\textwidth}
   \includegraphics[width=\textwidth]{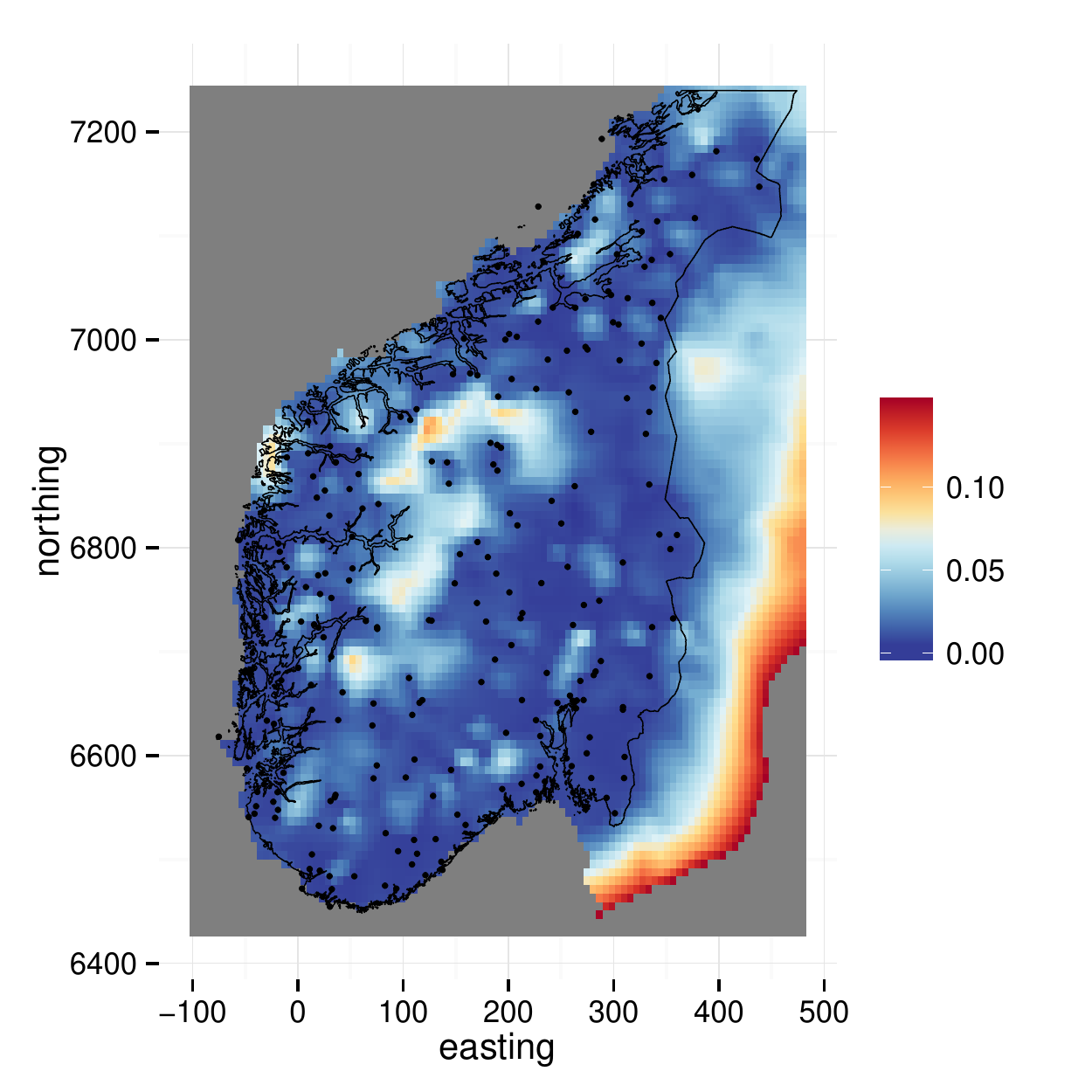}
   \caption{RMSE}
   \label{fig:rmse_diff_5_repl_nonstat_data}
  \end{subfigure}
  \begin{subfigure}{0.45\textwidth}
   \includegraphics[width=\textwidth]{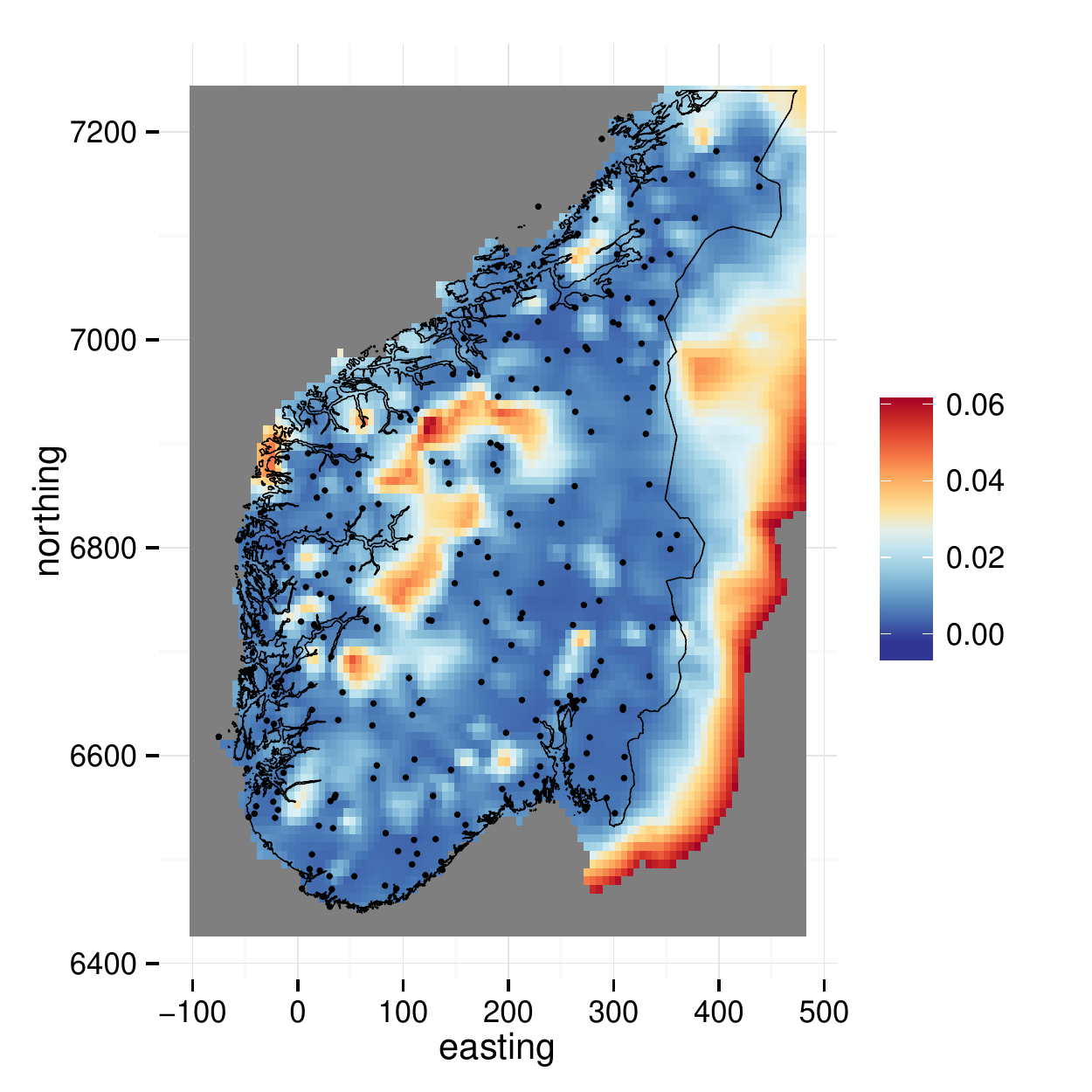}
   \caption{CRPS}
   \label{fig:crps_diff_5_repl_nonstat_data}
  \end{subfigure}
  \caption{Predictive performance for simulated data. The figures show difference in RMSE and CRPS: stationary model minus non-stationary model, 
           when the true model is non-stationary and $r=5$. The presented 
           spatial fields are pointwise averages over all replicates.}
   \label{fig:pred_perf_map}
\end{figure}

In addition to parameter estimates, we evaluate the predictive 
performance. With real data it is only possible to evaluate 
predictions at locations with observations. With simulated data 
we can investigate prediction properties at arbitrary locations 
since we are no longer restricted to know the truth only at the 
given stations. Thus, it is possible to evaluate how the models 
perform out-of-sample differently than any hold-out cross-validation 
allows. For us it is particularly interesting to investigate the 
behaviour in mountain areas with no observations. 

For each of the $250 \times r$ sampled fields, we predict 
at the locations of the nodes of the triangulation mesh conditionally on the sampled 
observations at the station locations. Thus, spatial predictions are 
compared with the simulated precipitation values in all mesh nodes.
Figure~\ref{fig:predictions_nonstat_data} 
is RMSE, CRPS and coverage for the predictions averaged over all 
nodes, and for each $r$ averaged over all replicates. 
The true model is non-stationary and both stationary and 
non-stationary models are fitted to the sampled datasets 
(observations at the 233 stations) and used for prediction. 
Figure~\ref{fig:pred_perf_map} contains spatial plots 
of the pointwise difference in RMSE and CRPS. The non-stationary 
model does a better job predicting the non-stationary precipitation 
field. Averaged over nodes, the difference between the stationary and 
non-stationary models is relatively low. However, the difference is 
considerably larger in the mountain area in the middle of Norway. 
There are also other, but smaller, mountain 
areas where the difference is enhanced
(cf. the elevation map in Figure~\ref{fig:elevation}). 

\section{Discussion}\label{sec:Discussion}

In this paper, we have studied a non-stationary 
model with elevation as a spatial explanatory variable 
in the dependence structure. 
Furthermore, 
a framework for specifying prior distributions for 
the dependence structure parameters is suggested. Our proposed 
model was used for annual precipitation observations 
from southern Norway. We used five years of annual 
precipitation observations and assumed they came 
from independent realisations (replicates) of the 
underlying process. In addition to analysis of  
real precipitation data, we explored the spatial 
replicate model in a simulation study. The SPDE based 
model enables fast simulations and inference, 
which made it feasible to run simulations 
and evaluate statistical properties of the model when 
the number of replicates increases. The non-stationary 
model was compared to the corresponding stationary model 
in the simulations, and in the data analysis.   

Key findings for the inference of parameters are that 
using replicates makes parameter estimates more precise 
and less prior sensitive, especially the dependence 
structure parameters $\boldsymbol{\theta}_\text{NS}$. 
These results are consistent for the case study and the 
simulation study. The simulation study also shows that 
having replicates improve statistical properties; the 
estimation bias is reduced and posterior credible 
interval coverage is improved when the number of 
replicates increases. The largest improvement is 
obtained changing from no replicates to two replicates. 
Also, using difference in DIC to choose between a 
stationary and non-stationary model gain power using 
replicates.

For the predictive performance, measured in average 
CRPS and RMSE, a slight improvement is achieved going 
from a stationary to a non-stationary model, but there 
are large local variations. Others 
have also reported only small improvements in predictive 
performance changing from a stationary to a non-stationary 
model \citep{Paciorek2006, Reich2011}. Using replicates 
gives a slightly poorer predictive performance for the 
observation locations we have in our case study. These 
results are consistent with the findings in the simulation 
study, where the coverage improved, but RMSE and CRPS 
increased with replicates.   

We proposed a model with elevation both in the mean 
and covariance. The model was based on physical 
understanding of the precipitation process, and an 
attempt to handle the need to extrapolate with respect 
to elevation due to the placement of the rain gauges. 
In the simulation study, we were able to explore the 
differences in predictive performance on the entire 
domain, when stationary and non-stationary models were 
fitted to observations from a non-stationary model. 
We found that the largest differences in CRPS and RMSE 
were in mountain areas (Figure~\ref{fig:pred_perf_map}). 
Evaluation by cross-validation on the observation 
locations cannot give us this insight, neither for 
the case study nor for the simulation study, as there 
are no observations at high elevation.

The result in Figure~\ref{fig:pred_perf_map} of 
course rely on the fact that the non-stationary 
replicate model we sample from is the true model. 
There are several signs that the model is, at least, 
reasonable: First, the difference in DIC between the 
stationary and non-stationary model is large (159), 
and agrees with what we expect from the simulation 
study (Figure~\ref{fig:delta_dic_nonstat}, with five 
replicates). Further, the posterior densities of 
parameters from individual models are similar, and 
posteriors of the parameters controlling the 
non-stationarity ($\theta_{\tau,h}$ and $\theta_{\kappa,h}$) 
are consistently shifted away from zero 
(Figure~\ref{fig:marginals_theta_nonstationary}). 
This suggests that it is reasonable to model different 
years of annual precipitation as realisations of the 
same process, and that this process is non-stationary 
with elevation as an important explanatory variable.

More and other explanatory variables, in both the mean 
and covariance, will probably give us a better model. 
This has been outside the scope of the paper to investigate, 
but interesting explanatory variables to explore include 
other topographical explanatory variables and using 
aggregated output from reanalysis of weather models 
\citep{Geirsson2014}. The diversity of changes in local 
predictive performance between the models (Figure~\ref{fig:crps_map}), 
illustrates the need for explanatory variables that capture the 
spatial variability, as well as good characteristics of local properties. 
Another important question is whether a non-stationary covariance 
is the best way to address spatial inhomogeneity. The measurement 
error is very likely to vary over the domain, and especially at 
wind exposed locations there is considerable measurement bias 
\citep{Wolff2012}. \citet{Fuglstad2014-precip} found that a simple 
model with a spatially varying nugget effect gave almost as good 
predictive results as a non-stationary model when they analysed 
annual precipitation in the conterminous US. To model the 
measurement error process more realistically, along with 
testing other explanatory variables, are natural 
paths to follow in the search for improved predictions 
and uncertainty estimates of precipitation in difficult and 
unmonitored terrain.

\citet{Ingebrigtsen2014} reported numerical problems, 
prior sensitivity and poor statistical properties. The 
proposed framework for specifying priors has given us a 
tool for setting priors for the dependence parameters 
controlling the non-stationarity. It is hard to have an 
intuition for these parameters. In addition, the parameters 
controlling the range ($\theta_{\kappa,1}$ and $\theta_{\kappa,h}$) 
are more challenging to identify than the parameters 
controlling the marginal standard deviation 
($\theta_{\tau,1}$ and $\theta_{\tau,h}$). Hence, the 
range parameters are more sensitive to the prior. 

The variances of the non-stationarity parameters in \citet{Ingebrigtsen2014} 
corresponds to $c_\rho = 0.4$ and $c_\sigma=0.6$, and 0.5- and 0.9-quantiles 
of the correlation range and marginal standard deviation are (154, 232) km 
and (0.282, 0.338) m. In comparison, these intervals were chosen to be 
(150, 500) km and (0.2, 2) m in this paper, and we used $c_\rho = 0.8$ and 
$c_\sigma = 1.3$, allowing for a higher relative change in nominal priors 
with respect to elevation.  
In this work, we did not experience numerical problems, and
with the new priors there is less prior sensitivity, 
and bias and coverage are improved compared to \citet{Ingebrigtsen2014}, 
also when using only one replicate. 

The simulation study showed that if a non-stationary model is 
fitted to realisations sampled from a stationary model, both 
difference in DIC and posterior distributions of the 
non-stationarity parameters will indicate that the 
observations are from a stationary model. It is therefore 
little to lose by fitting a non-stationary model with explanatory 
variables that are thought to be of importance for the dependence 
structure. It is straightforward to change from a stationary model 
to a non-stationary model within the suggested SPDE modelling 
framework using the \texttt{R-INLA} software, and the inference 
is fast.

\citet{Fuglstad2014-precip} raise the question of whether non-stationarity 
is needed in spatial models. Their conclusion is yes, but that it is important 
to understand what type of non-stationarity the process exhibits. If there is 
knowledge about the driving forces of the physical process, explanatory variables 
in the dependence structure might be a fruitful way of modelling non-stationarity. 
Our analysis of the southern Norway data illustrates that priors for these type of 
models should be chosen with care, and that the dataset ideally should 
consist of more than one realisation of the process. 

\section*{Acknowledgements}

This work was supported by the project Spatio-temporal modelling and approximate Bayesian inference (196670/V30) 
funded by the Research Council of Norway.


\begin{thebibliography}{38}
\expandafter\ifx\csname natexlab\endcsname\relax\def\natexlab#1{#1}\fi
\expandafter\ifx\csname url\endcsname\relax
  \def\url#1{\texttt{#1}}\fi
\expandafter\ifx\csname urlprefix\endcsname\relax\def\urlprefix{URL }\fi


\bibitem[Banerjee et~al., 2004]{Banerjee2004}
Banerjee, S., Carlin, B.~P., and Gelfand, A.~E. (2004).
\newblock {\em Hierarchical Modeling and Analysis for Spatial Data}, volume 101
  of {\em Monographs on Statistics and Applied Probability}.
\newblock Chapman \& Hall/CRC, Boca Raton, Florida.

\bibitem[Bolin and Lindgren, 2011]{Bolin2011}
Bolin, D. and Lindgren, F. (2011).
\newblock Spatial models generated by nested stochastic partial differential
  equations, with an application to global ozone mapping.
\newblock {\em The Annals of Applied Statistics}, 5(1):523--550.

\bibitem[Brenner and Scott, 2007]{FEM}
Brenner, S.~C. and Scott, L.~R. (2007).
\newblock {\em The Mathematical Theory of Finite Element Methods}, volume~15 of
  {\em Texts in Applied Mathematics}.
\newblock Springer, New York, 3 edition.

\bibitem[Cressie and Wikle, 2011]{Cressie2011}
Cressie, N. and Wikle, C.~K. (2011).
\newblock {\em Statistics for spatio-temporal data}.
\newblock Wiley series in probability and statistics. Wiley, Hoboken, New
  Jersey.

\bibitem[Dyrrdal et~al., 2014]{Dyrrdal2014}
Dyrrdal, A.~V., Lenkoski, A., Thorarinsdottir, T.~L., and Stordal, F. (2014).
\newblock Bayesian hierarchical modeling of extreme hourly precipitation in
  {N}orway.
\newblock {\em Environmetrics}.
\newblock To appear.

\bibitem[Fuentes, 2002]{Fuentes2002}
Fuentes, M. (2002).
\newblock Spectral methods for nonstationary spatial processes.
\newblock {\em Biometrika}, 89(1):197--210.

\bibitem[Fuglstad et~al., 2014a]{Fuglstad2014-precip}
Fuglstad, G.-A., Lindgren, F., Simpson, D., and Rue, H. (2014a).
\newblock Do we need non-stationarity in spatial models?
\newblock arXiv:1409.0743.

\bibitem[Fuglstad et~al., 2014b]{Fuglstad2014}
Fuglstad, G.-A., Lindgren, F., Simpson, D., and Rue, H. (2014b).
\newblock Exploring a new class of non-stationary spatial {G}aussian random
  fields with varying local anisotropy.
\newblock {\em Statistica Sinica}.
\newblock To appear.

\bibitem[Geirsson et~al., 2014]{Geirsson2014}
Geirsson, {\'O}., Hrafnkelsson, B., and Simpson, D. (2014).
\newblock Computationally efficient spatial modeling of annual maximum 24 hour
  precipitation. {A}n application to data from {I}celand.
\newblock ar{X}iv:1405.6947.

\bibitem[Gelman, 2006]{Gelman2006}
Gelman, A. (2006).
\newblock Prior distributions for variance parameters in hierarchical models
  (comment on article by {B}rowne and {D}raper).
\newblock {\em Bayesian Analysis}, 1(3):515--534.

\bibitem[{GLOBE Task Team}, 1999]{GLOBE}
{GLOBE Task Team} (1999).
\newblock Globe: The {G}lobal {L}and {O}ne-kilometer {B}ase {E}levation
  {D}igital {E}levation {M}odel, {V}ersion 1.0.
\newblock http://www.ngdc.noaa.gov/mgg/topo/globe.html.

\bibitem[Gneiting and Raftery, 2007]{crps_gneiting}
Gneiting, T. and Raftery, A.~E. (2007).
\newblock Strictly proper scoring rules, prediction, and estimation.
\newblock {\em Journal of the American Statistical Association},
  102(477):359--378.

\bibitem[Hanks et~al., 2015]{Hanks2015}
Hanks, E.~M., Schliep, E.~M., Hooten, M.~B., and Hoeting, J.~A. (2015).
\newblock Restricted spatial regression in practice: geostatistical models, confounding, and robustness under model misspecification. 
\newblock {\em Environmetrics}. 
\newblock http://dx.doi.org/10.1002/env.2331.

\bibitem[Higdon et~al., 1999]{Higdon1999}
Higdon, D., Swall, J., and Kern, J. (1999).
\newblock Non-stationary spatial modeling.
\newblock In Bernardo, J.~M., Berger, J.~O., Dawid, A.~P., and Smith, A.~F.~M.,
  editors, {\em Bayesian Statistics 6. Proceedings of the Sixth Valencia
  International Meeting}, pages 761--768. Oxford University Press.

\bibitem[Holand et~al., 2013]{Holand2013}
Holand, A.~M., Steinsland, I., Martino, S., and Jensen, H. (2013).
\newblock Animal models and integrated nested {L}aplace approximations.
\newblock {\em G3: Genes, Genomes, Genetics}, 3(8):1241--1251.

\bibitem[Ingebrigtsen et~al., 2014]{Ingebrigtsen2014}
Ingebrigtsen, R., Lindgren, F., and Steinsland, I. (2014).
\newblock Spatial models with explanatory variables in the dependence
  structure.
\newblock {\em Spatial Statistics}, 8:20--38.

\bibitem[Kaufman and Shaby, 2013]{Kaufman2013}
Kaufman, C.~G. and Shaby, B.~A. (2013).
\newblock The role of the range parameter for estimation and prediction in
  geostatistics.
\newblock {\em Biometrika}, 100(2):473--484.

\bibitem[Lele et~al., 2010]{Lele2010}
Lele, S.~R., Nadeem, K., and Schmuland, B. (2010).
\newblock Estimability and likelihood inference for {G}eneralized {L}inear
  {M}ixed {M}odels using data cloning.
\newblock {\em Journal of the American Statistical Association},
  105(492):1617--1625.

\bibitem[Lindgren and Rue, 2014]{Lindgren2014}
Lindgren, F. and Rue, H. (2014).
\newblock Bayesian spatial modelling with {R-INLA}.
\newblock {\em Journal of Statistical Software}.
\newblock To appear.

\bibitem[Lindgren et~al., 2011]{Lindgren2011}
Lindgren, F., Rue, H., and Lindstr\"{o}m, J. (2011).
\newblock An explicit link between {G}aussian fields and {G}aussian {M}arkov
  random fields: the stochastic partial differential equation approach (with
  discussion).
\newblock {\em Journal of the Royal Statistical Society: Series B (Statistical
  Methodology)}, 73(4):423--498.

\bibitem[{{NCAR} - {R}esearch {A}pplications {L}aboratory}, 2014]{verification}
{{NCAR} - {R}esearch {A}pplications {L}aboratory} (2014).
\newblock {\em verification: {W}eather {F}orecast {V}erification {U}tilities.}
\newblock R package version 1.40.

\bibitem[Neto et~al., 2014]{Neto2014}
Neto, J.~H.~V., Schmidt, A.~M., and Guttorp, P. (2014).
\newblock Accounting for spatially varying directional effects in spatial
  covariance structures.
\newblock {\em Journal of the Royal Statistical Society: Series C (Applied
  Statistics)}, 63(1):103--122.

\bibitem[Orskaug et~al., 2011]{Orskaug2011}
Orskaug, E., Scheel, I., Frigessi, A., Guttorp, P., Haugen, J.~E., Tveito,
  O.~E., and Haug, O. (2011).
\newblock Evaluation of a dynamic downscaling of precipitation over the
  {N}orwegian mainland.
\newblock {\em Tellus, Series A: Dynamic Meteorology and Oceanography},
  63A(4):746--756.

\bibitem[Paciorek and Schervish, 2006]{Paciorek2006}
Paciorek, C.~J. and Schervish, M.~J. (2006).
\newblock Spatial modelling using a new class of nonstationary covariance
  functions.
\newblock {\em Environmetrics}, 17:483--506.

\bibitem[{R Core Team}, 2013]{Rlanguage}
{R Core Team} (2013).
\newblock {\em R: A Language and Environment for Statistical Computing}.
\newblock R Foundation for Statistical Computing, Vienna, Austria.
\newblock http://www.R-project.org/.

\bibitem[Reich et~al., 2011]{Reich2011}
Reich, B.~J., Eidsvik, J., Guindani, M., Nail, A.~J., and Schmidt, A.~M.
  (2011).
\newblock A class of covariate-dependent spatiotemporal covariance functions
  for the analysis of daily ozone concentration.
\newblock {\em The Annals of Applied Statistics}, 5(4):2425--2447.

\bibitem[Rue and Held, 2005]{GMRF}
Rue, H. and Held, L. (2005).
\newblock {\em Gaussian Markov Random Fields: theory and applications}, volume
  104 of {\em Monographs on Statistics and Applied Probability}.
\newblock Chapman \& Hall/CRC, Boca Raton, Florida.

\bibitem[Rue et~al., 2009]{Rue2009}
Rue, H., Martino, S., and Chopin, N. (2009).
\newblock Approximate {B}ayesian inference for latent {G}aussian models by
  using integrated nested {L}aplace approximations (with discussion).
\newblock {\em Journal of the Royal Statistical Society: Series B (Statistical
  Methodology)}, 71(2):319--392.

\bibitem[Sampson and Guttorp, 1992]{Sampson1992}
Sampson, P.~D. and Guttorp, P. (1992).
\newblock Nonparametric estimation of nonstationary spatial covariance
  structure.
\newblock {\em Journal of the American Statistical Association},
  87(417):108--119.

\bibitem[Schmidt et~al., 2011]{Schmidt2011}
Schmidt, A.~M., Guttorp, P., and O'Hagan, A. (2011).
\newblock Considering covariates in the covariance structure of spatial
  processes.
\newblock {\em Environmetrics}, 22:487--500.

\bibitem[{Simpson} et~al., 2014]{Simpson2014}
{Simpson}, D.~P., {Martins}, T.~G., {Riebler}, A., {Fuglstad}, G.-A., {Rue},
  H., and {S{\o}rbye}, S.~H. (2014).
\newblock Penalising model component complexity: {A} principled, practical
  approach to constructing priors.
\newblock arXiv:1403.4630.

\bibitem[S{\o}rbye and Rue, 2014]{Sorbye2014}
S{\o}rbye, S.~H. and Rue, H. (2014).
\newblock Scaling intrinsic {G}aussian {M}arkov random field priors in spatial
  modelling.
\newblock {\em Spatial Statistics}, 8:39--51.

\bibitem[Spiegelhalter et~al., 2002]{DIC}
Spiegelhalter, D.~J., Best, N.~G., Carlin, B.~P., and {Van Der Linde}, A.
  (2002).
\newblock Bayesian measures of model complexity and fit.
\newblock {\em Journal of the Royal Statistical Society: Series B (Statistical
  Methodology)}, 64(4):583--639.

\bibitem[Spiegelhalter et~al., 2014]{DIC-2014}
Spiegelhalter, D.~J., Best, N.~G., Carlin, B.~P., and {Van Der Linde}, A.
  (2014).
\newblock The deviance information criterion: 12 years on.
\newblock {\em Journal of the Royal Statistical Society: Series B (Statistical
  Methodology)}, 76(3):485--493.

\bibitem[Whittle, 1954]{Whittle1954}
Whittle, P. (1954).
\newblock On stationary processes in the plane.
\newblock {\em Biometrika}, 41(3/4):434--449.

\bibitem[Whittle, 1963]{Whittle1963}
Whittle, P. (1963).
\newblock Stochastic processes in several dimensions.
\newblock {\em Bulletin of the International Statistical Institute},
  40:974--994.

\bibitem[Wolff et~al., 2012]{Wolff2012}
Wolff, M., Isaksen, K., Br{\ae}kkan, R., Alfnes, E., {Petersen-{\O}verleir},
  A., and Ruud, E. (2012).
\newblock Measurements of wind-induced loss of solid precipitation: description
  of a {N}orwegian field study.
\newblock {\em Hydrology Research}.

\bibitem[Zhang, 2004]{Zhang2004}
Zhang, H. (2004).
\newblock Inconsistent estimation and asymptotically equal interpolations in
  model-based geostatistics.
\newblock {\em Journal of the American Statistical Association},
  99(465):250--261.

\end{thebibliography}


\bibliographystyle{elsarticle-harv}

\clearpage
\begin{appendices}

\section{Figures}\label{app:Figures}

This appendix contains figures included for completeness of the results. 
\ref{app:Case_study} for the real annual precipitation dataset, 
and \ref{app:Non-stationary_datasets} and \ref{app:Stationary_datasets} 
for simulated datasets.  

\subsection{Southern Norway data}\label{app:Case_study}

\begin{figure}[!h]
 \centering
  \begin{subfigure}{0.32\textwidth}
   \includegraphics[width=\textwidth]{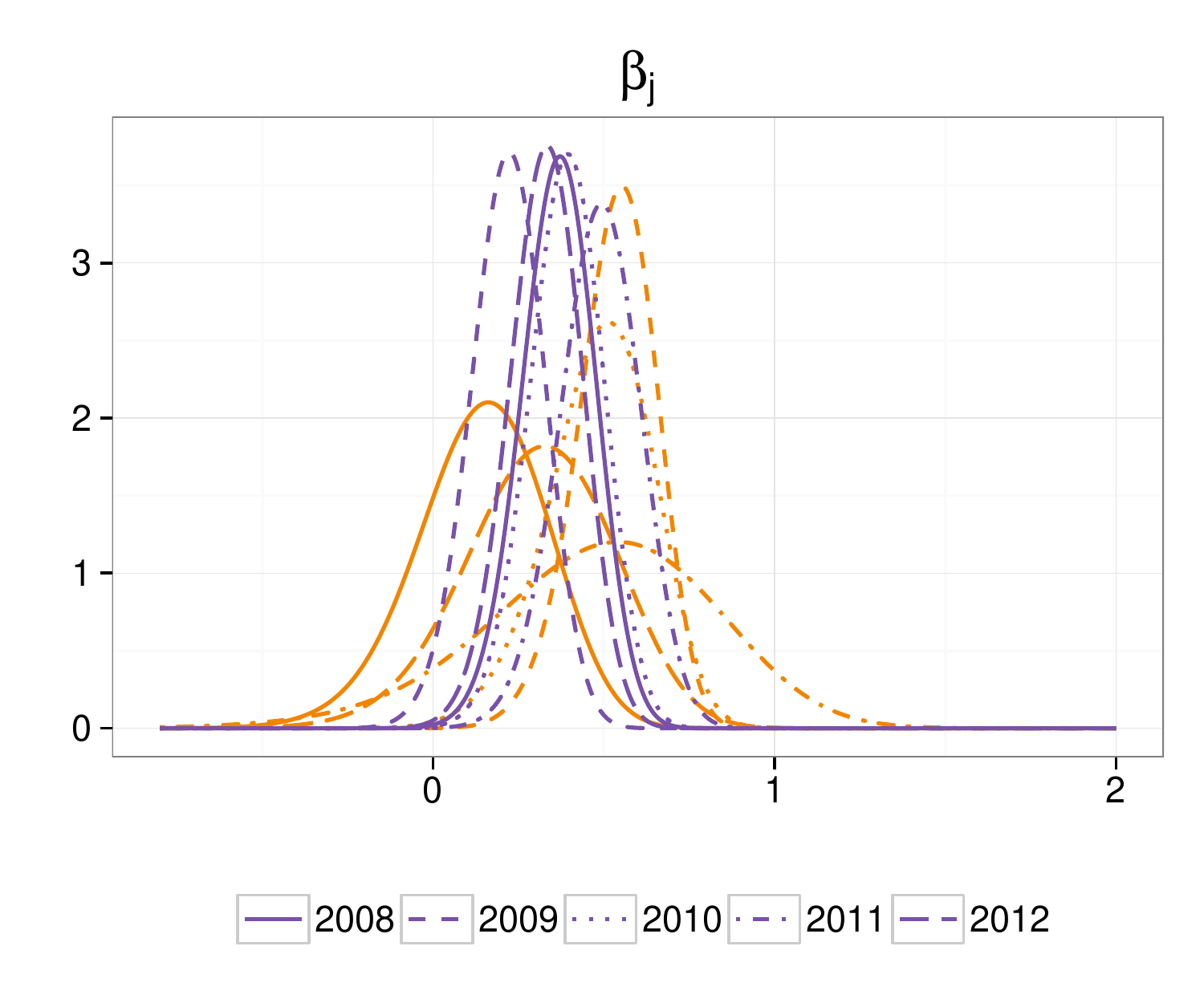}
   \caption{}
   \label{fig:posterior_intercepts_nonstat}
  \end{subfigure}
  \begin{subfigure}{0.32\textwidth}
   \includegraphics[width=\textwidth]{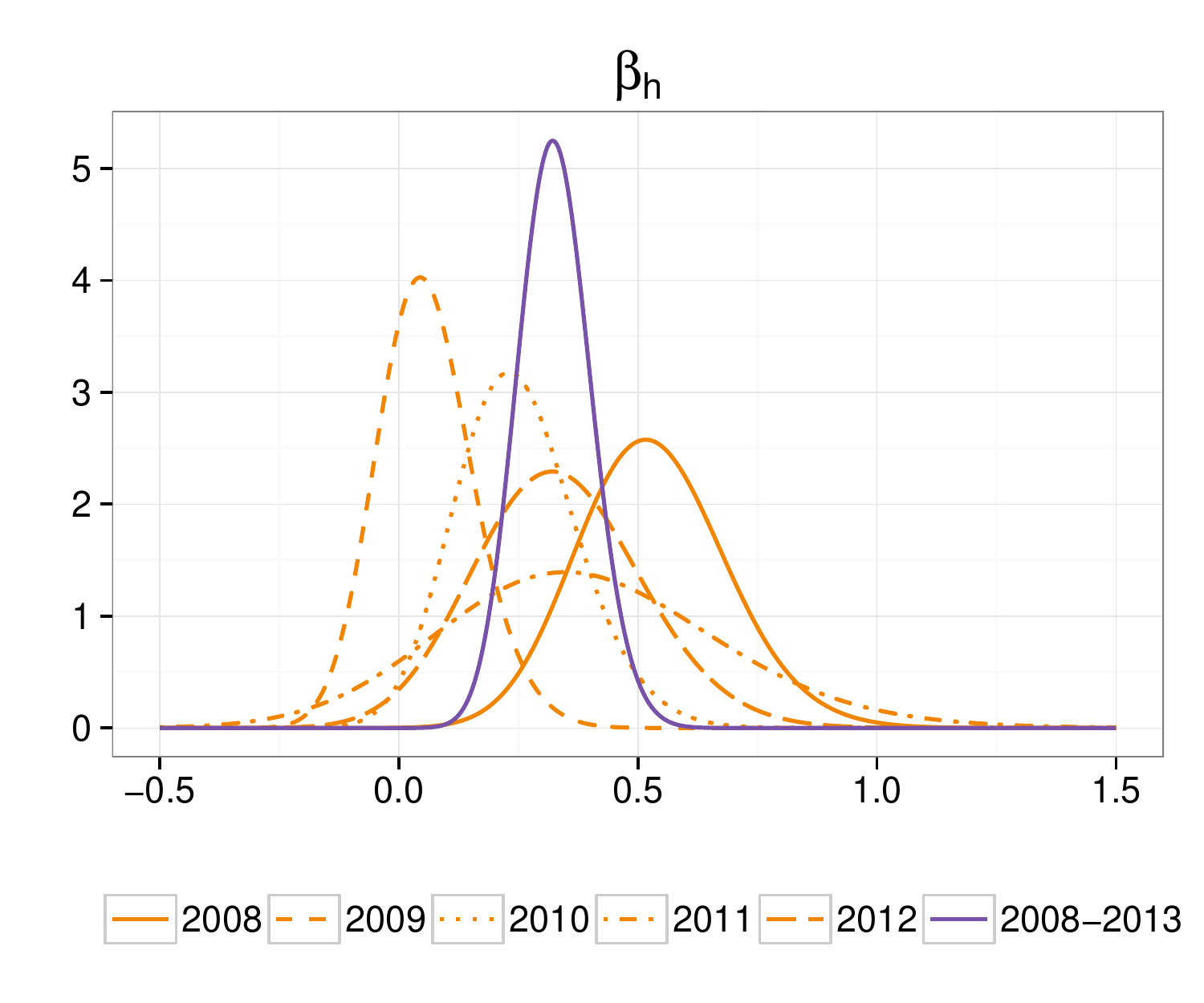}
   \caption{}
   \label{fig:posterior_elevation_nonstat}
  \end{subfigure}
  \begin{subfigure}{0.32\textwidth}
   \includegraphics[width=\textwidth]{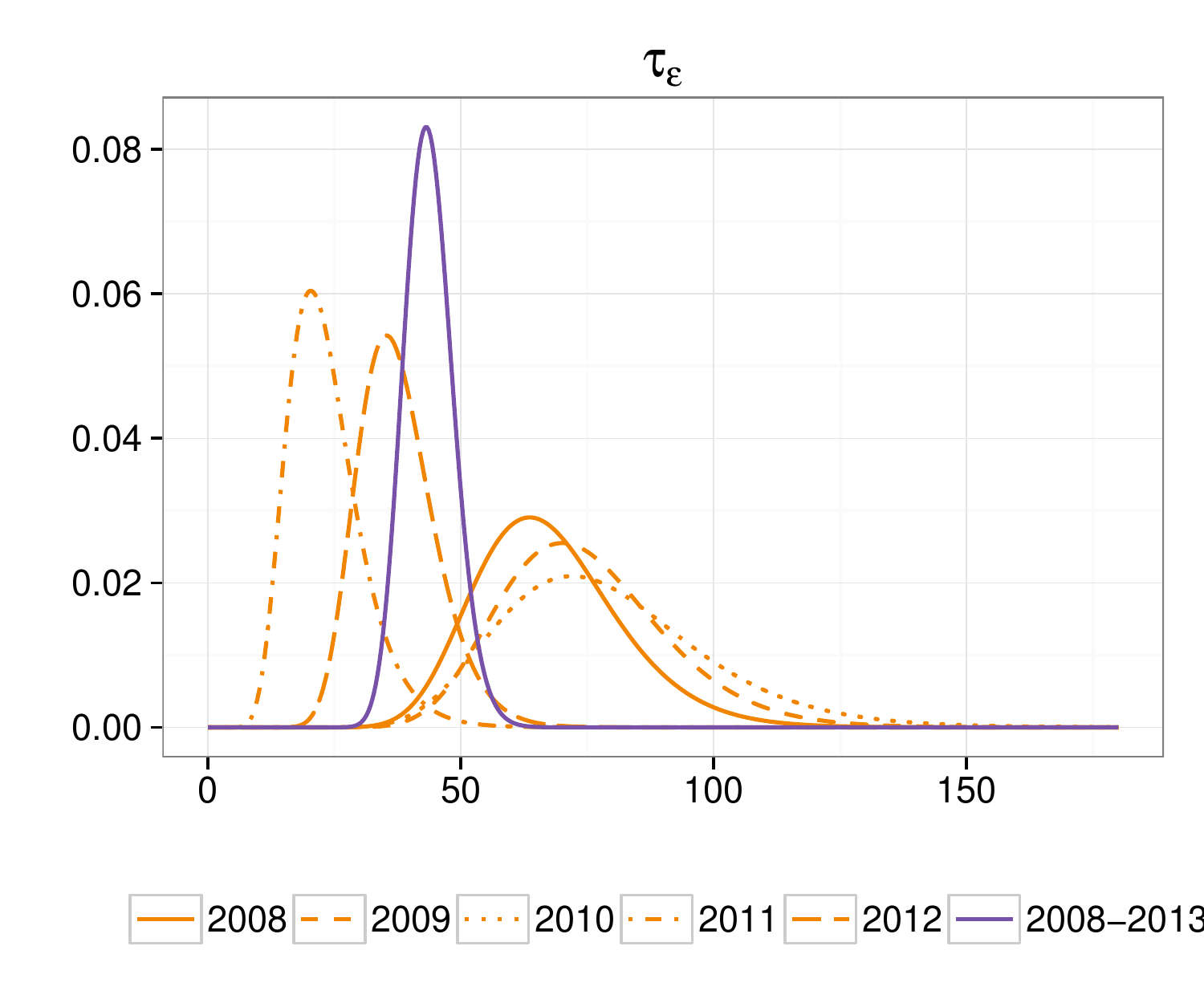}
   \caption{}
   \label{fig:posterior_tau_nonstat}
  \end{subfigure}
  \caption{Posterior marginal densities for $\beta_j, j = 1, \dots, 5$, $\beta_h$, 
           and $\tau_\epsilon$ with the non-stationary precipitation model. 
           The individual model was fitted to each annual dataset and the replicate 
           model was fitted to the five years together.  
          }
 \label{fig:marginals_parameters_nonstat}
\end{figure}

\subsection{Simulation study: non-stationary datasets}\label{app:Non-stationary_datasets}

This section contains results from the simulation study described in Section~\ref{sec:Simulation_study}. 
It only includes results for the parameters $\beta_0$, $\beta_h$, and $\tau_\epsilon$, while 
results for the dependence structure parameters are in Section~\ref{sec:Simulation_study_results}. 
The observations are sampled from the non-stationary model.

\begin{figure}[!h]
 \centering
  \begin{subfigure}{0.32\textwidth}
   \includegraphics[width=\textwidth]{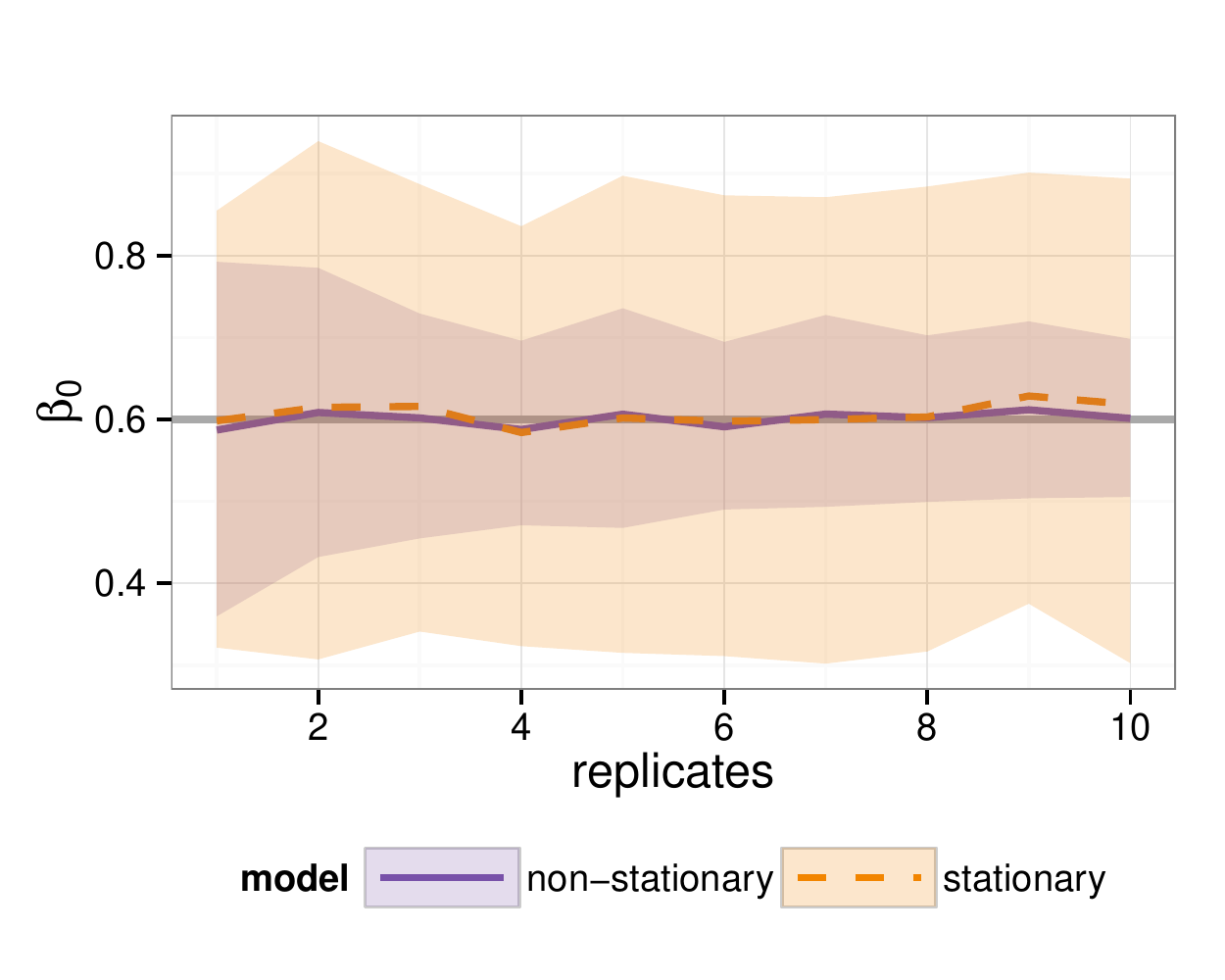}
   \caption{Posterior mean $\beta_0$}
   \label{fig:intercept_nonstat_data}
  \end{subfigure}
  \begin{subfigure}{0.32\textwidth}
   \includegraphics[width=\textwidth]{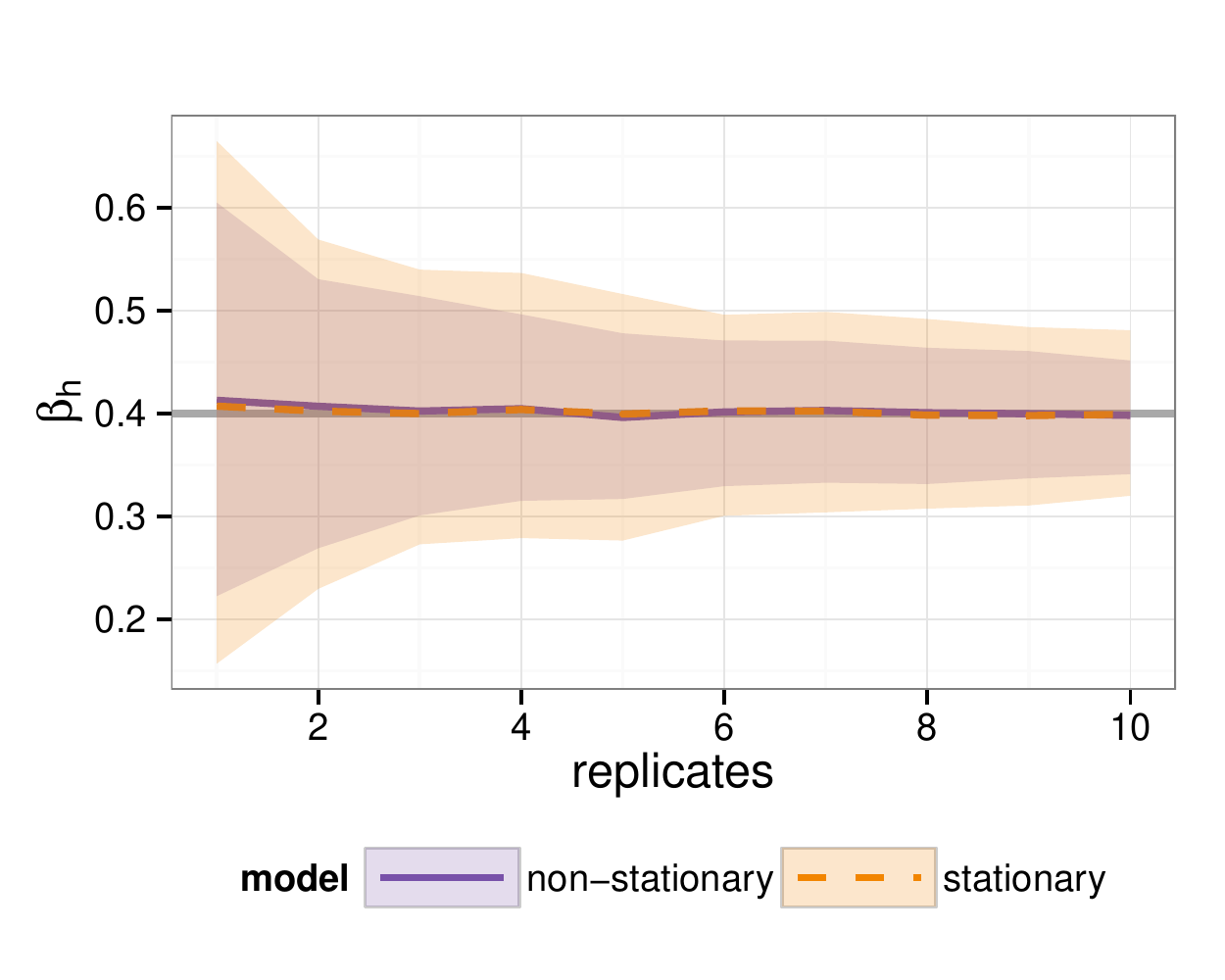}
   \caption{Posterior mean $\beta_h$}
   \label{fig:elevation_nonstat_data}
  \end{subfigure}
  \begin{subfigure}{0.32\textwidth}
   \includegraphics[width=\textwidth]{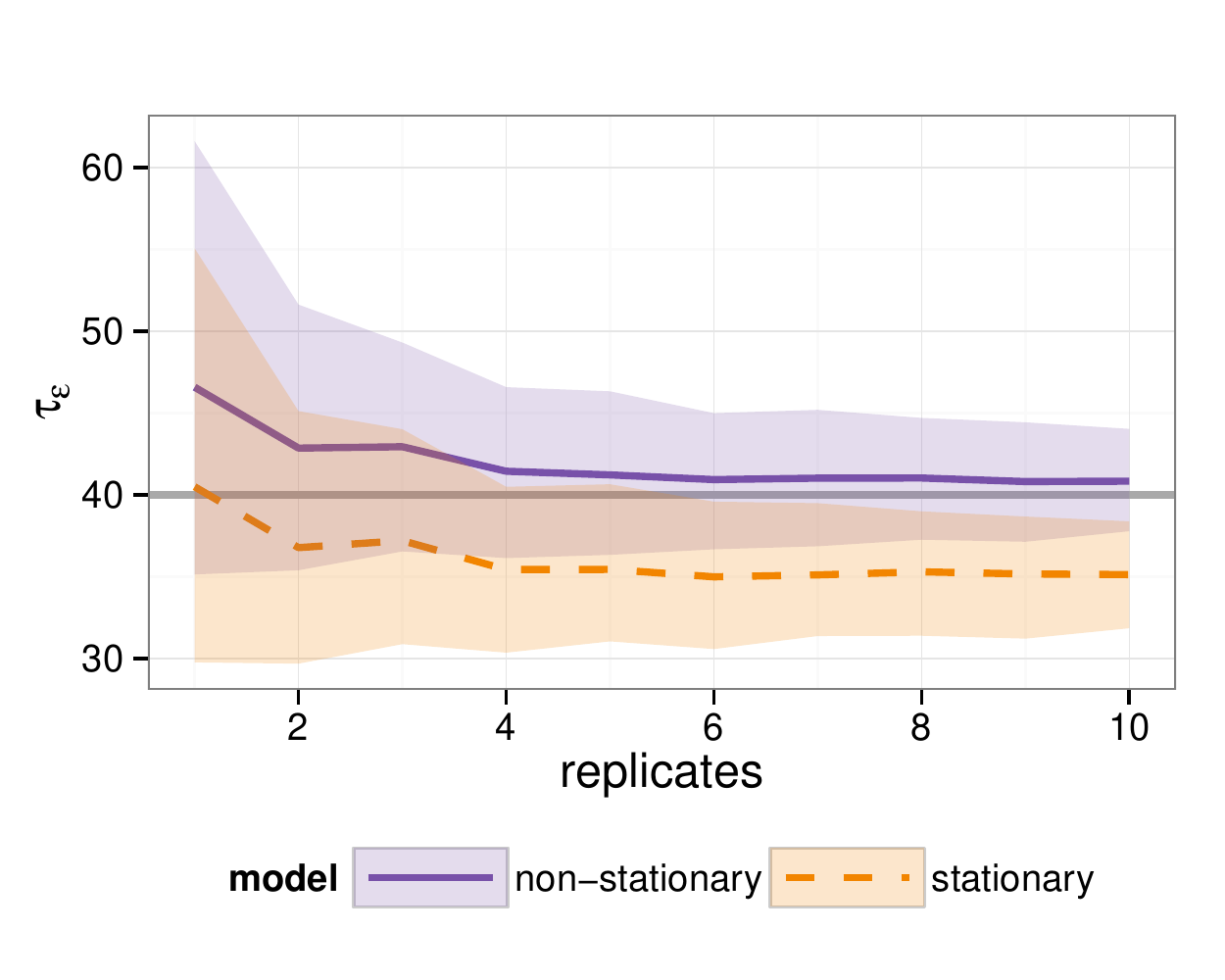}
   \caption{Posterior mean $\tau_\epsilon$}
   \label{fig:tau_nonstat_data}
  \end{subfigure}
 \caption{Stationary and non-stationary replicate models were fitted 
          to datasets sampled from the non-stationary model. Presented 
          are the posterior mean values for $\beta_0$, $\beta_h$, and 
          $\tau_\epsilon$. The lines are averages over all datasets, 
          while the shaded area span the range of the posterior mean 
          values (between the 0.1- and 0.9-quantiles). The true parameter values 
          are indicated with grey lines.}
 \label{fig:mean_betas_tau_nonstat_data}
\end{figure}

\begin{figure}
 \centering
  \begin{subfigure}{0.45\textwidth}
   \includegraphics[width=\textwidth]{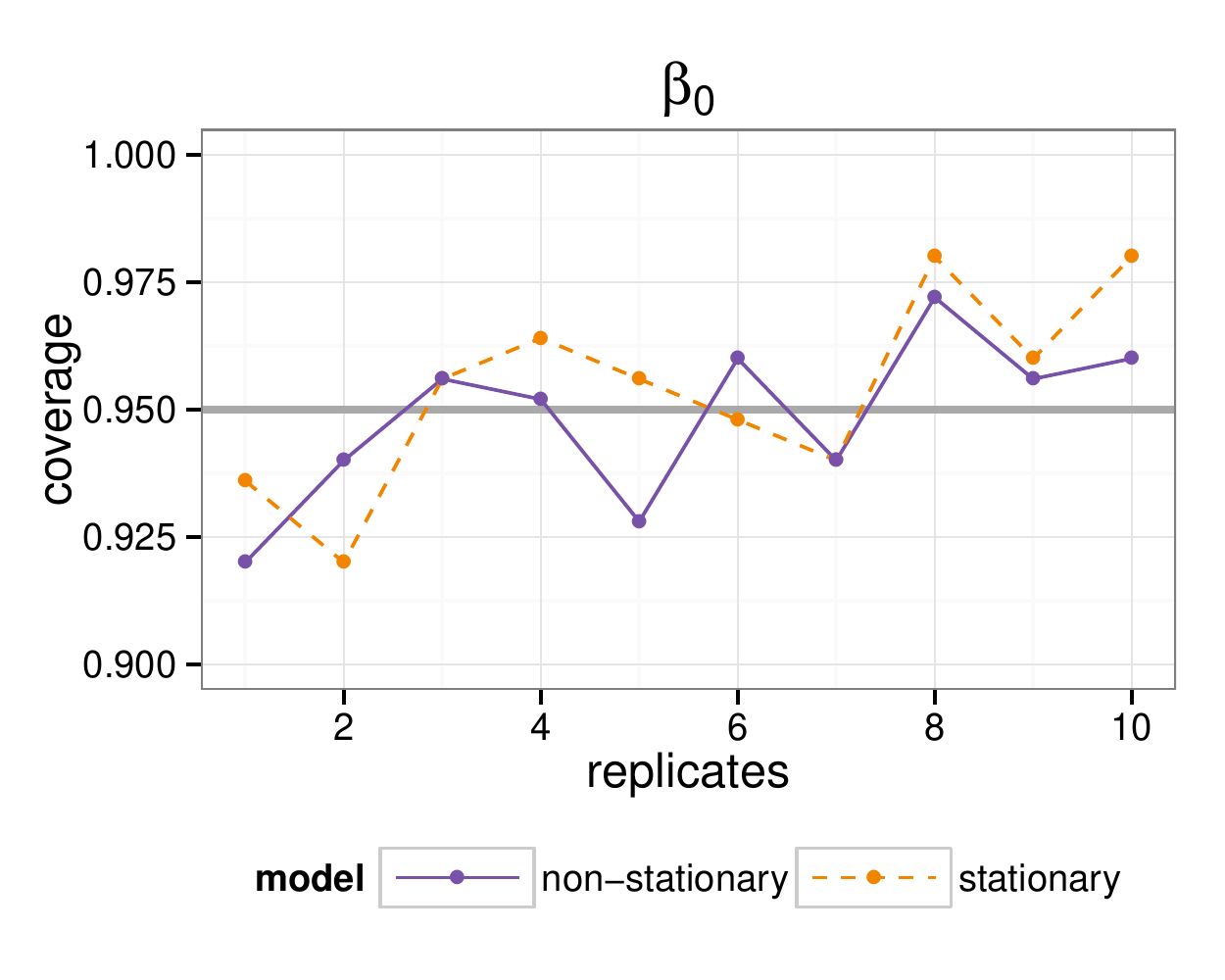}
   \caption{Coverage $\beta_0$}
   \label{fig:intercept_coverage_nonstat_data}
  \end{subfigure}%
  \begin{subfigure}{0.45\textwidth}
   \includegraphics[width=\textwidth]{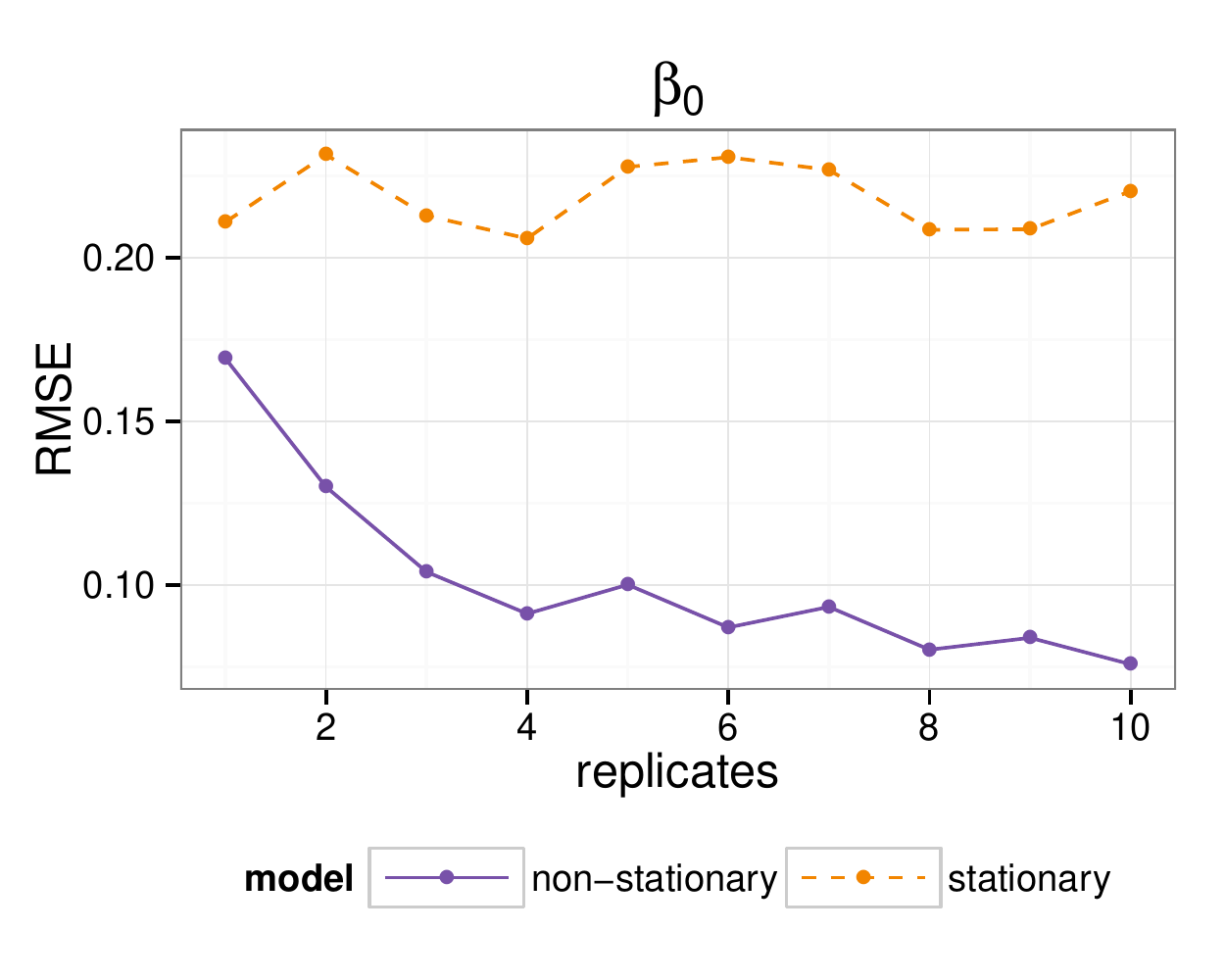}
   \caption{RMSE $\beta_0$}
   \label{fig:intercept_RMSE_nonstat_data}
  \end{subfigure}

  \begin{subfigure}{0.45\textwidth}
   \includegraphics[width=\textwidth]{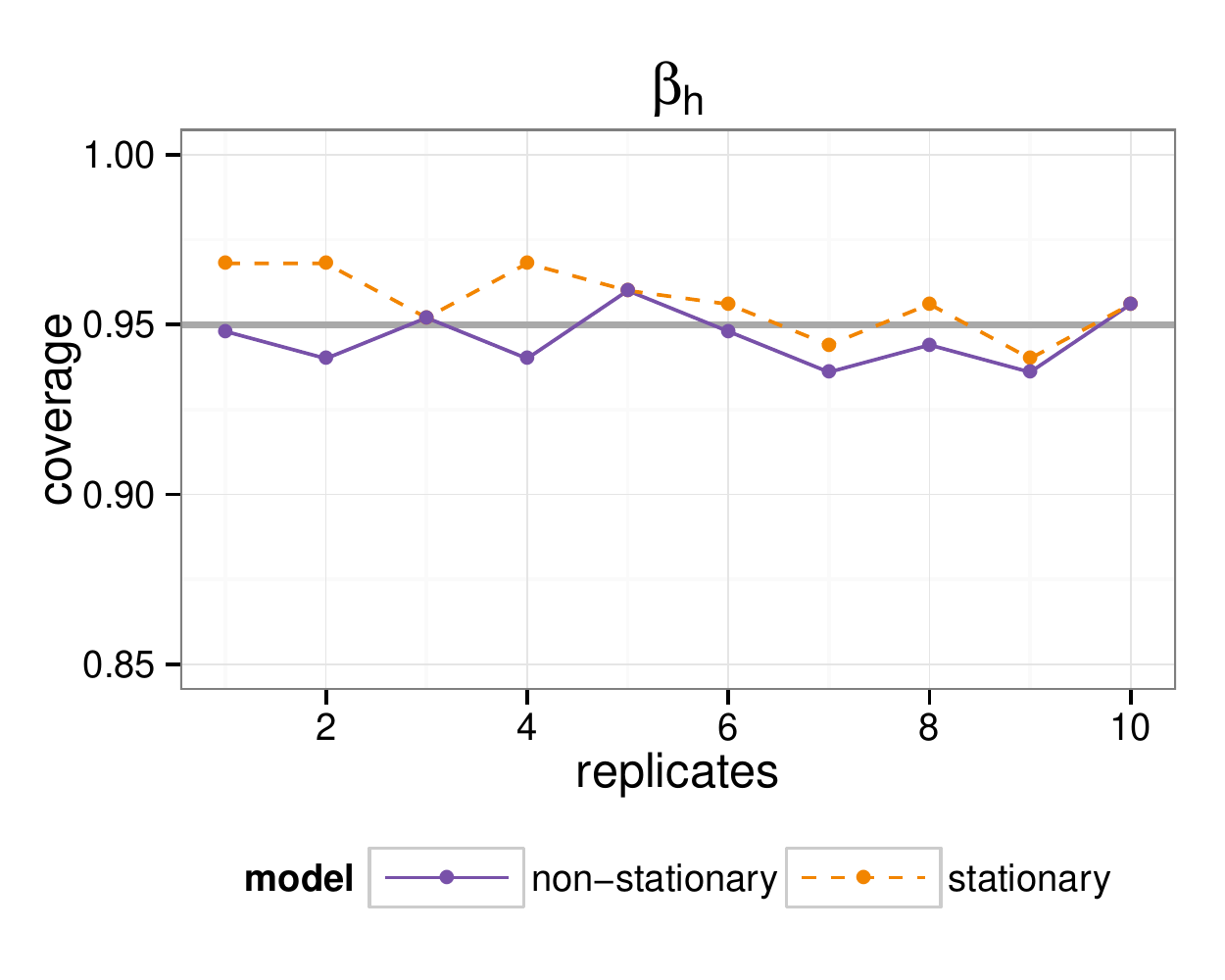}
   \caption{Coverage $\beta_h$}
   \label{fig:elevation_coverage_nonstat_data}
  \end{subfigure}%
  \begin{subfigure}{0.45\textwidth}
   \includegraphics[width=\textwidth]{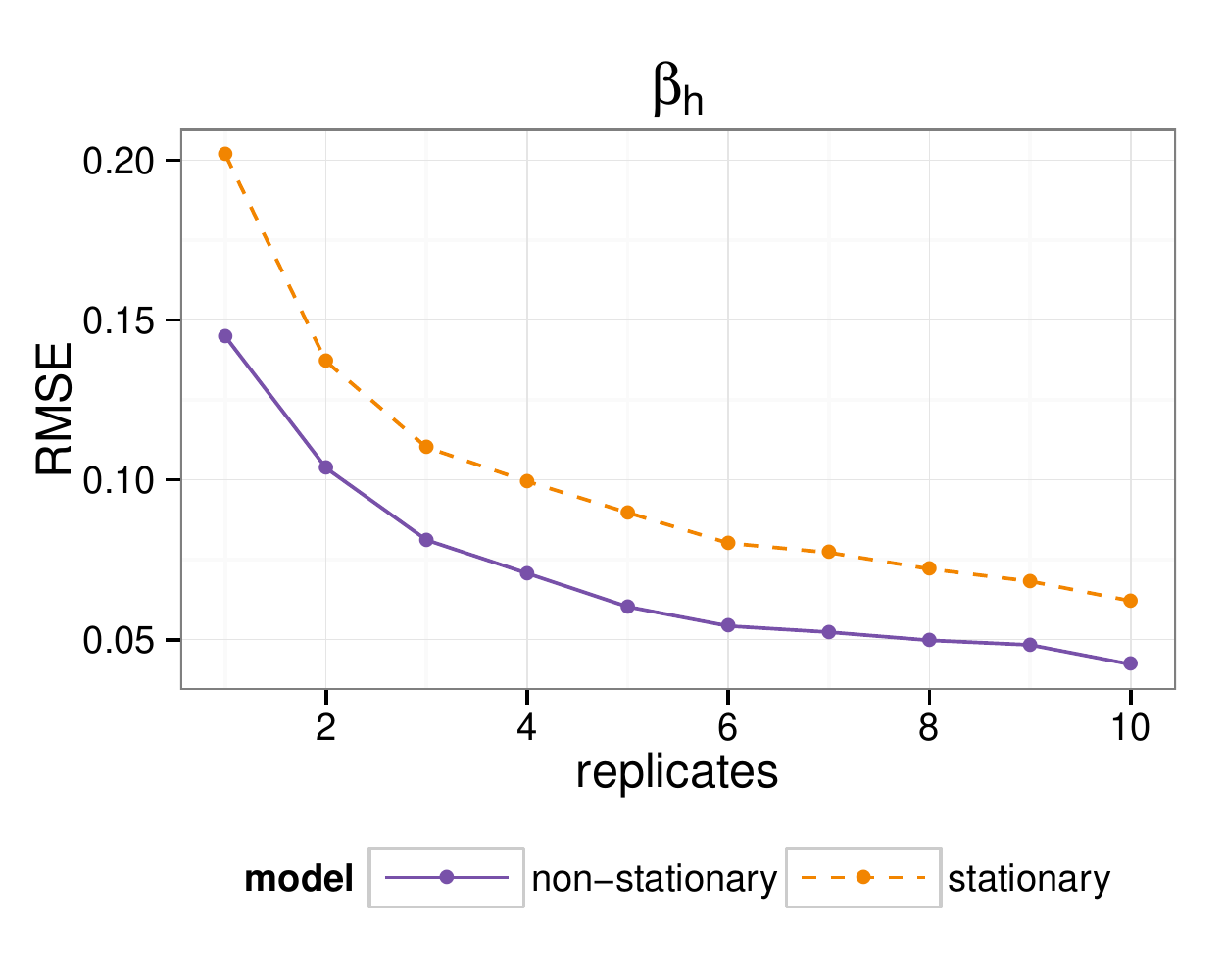}
   \caption{RMSE $\beta_h$}
   \label{fig:elevation_RMSE_nonstat_data}
  \end{subfigure}

  \begin{subfigure}{0.45\textwidth}
   \includegraphics[width=\textwidth]{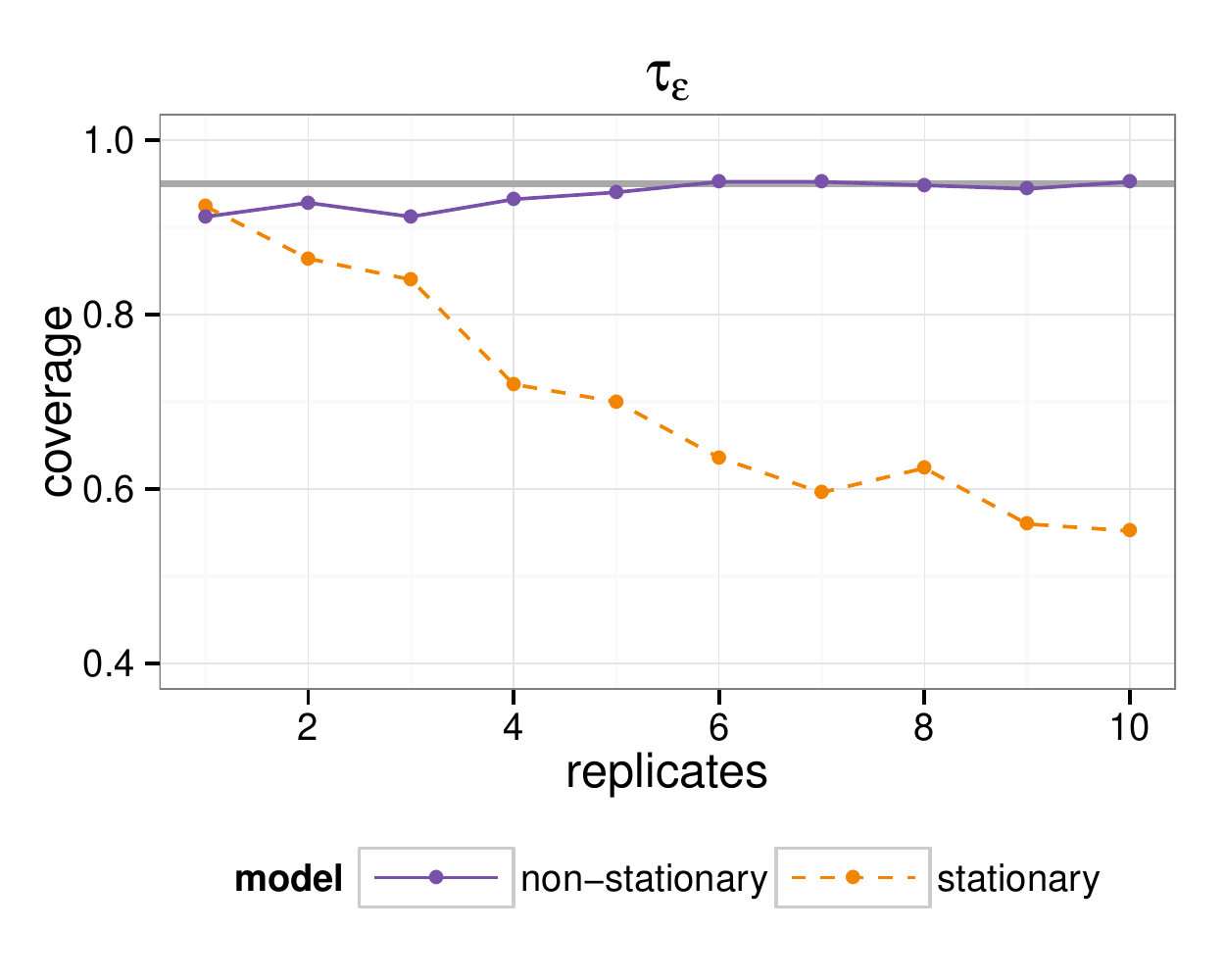}
   \caption{Coverage $\tau_\epsilon$}
   \label{fig:tau_coverage_nonstat_data}
  \end{subfigure}
  \begin{subfigure}{0.45\textwidth}
   \includegraphics[width=\textwidth]{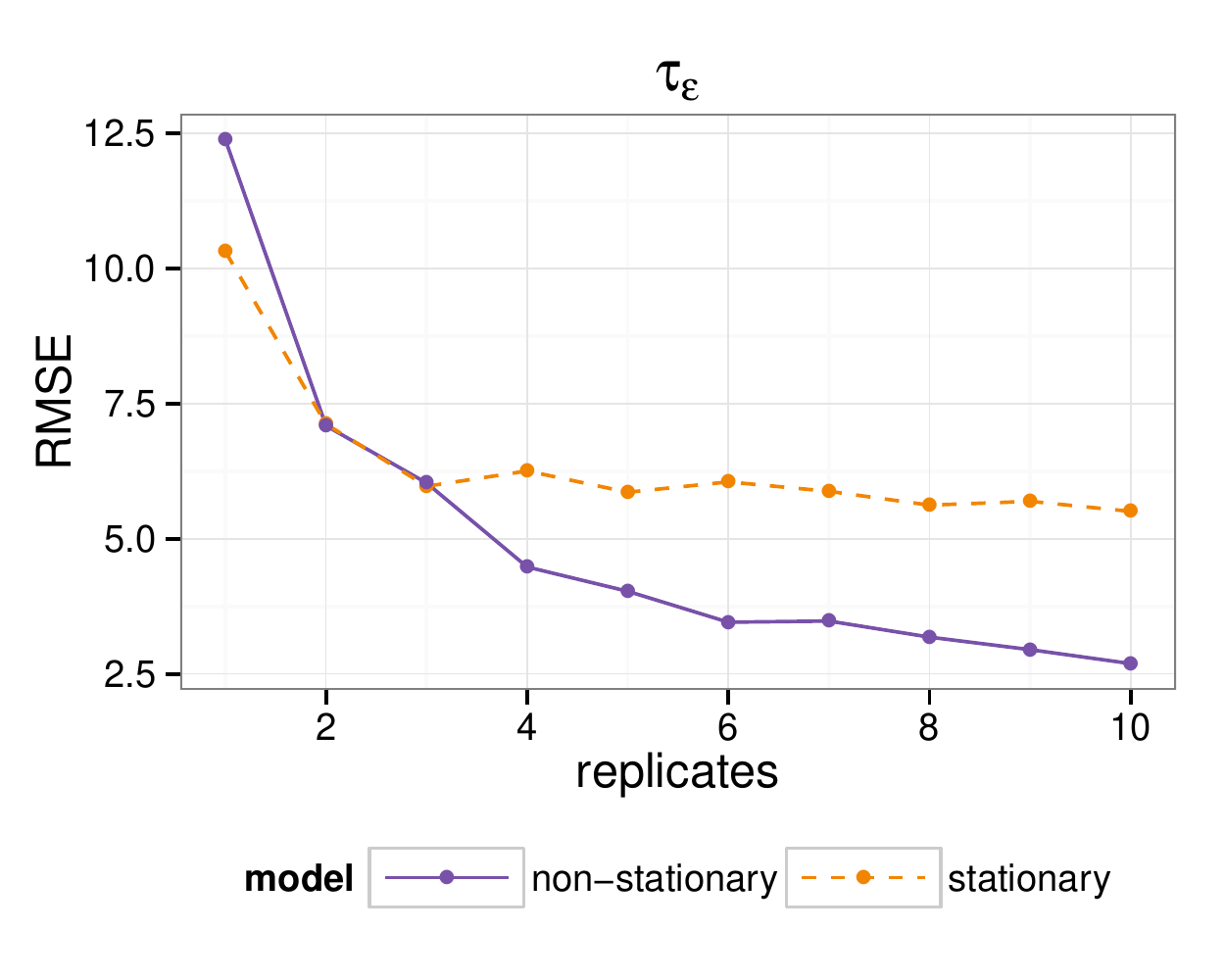}
   \caption{RMSE $\tau_\epsilon$}
   \label{fig:tau_RMSE_nonstat_data}
  \end{subfigure}
 \caption{Stationary and non-stationary replicate models were fitted 
          to datasets sampled from the non-stationary model. Presented 
          are 95\% posterior credible interval coverage and RMSE for 
          the parameters $\beta_0$, $\beta_h$, and $\tau_\epsilon$.}
 \label{fig:coverage_RMSE_betas_tau_nonstat_data}
\end{figure}

\clearpage

\subsection{Simulation study: stationary datasets}\label{app:Stationary_datasets}

This section contains results from the simulation study described 
in Section~\ref{sec:Simulation_study}. Results for all model parameters 
are included, and the observations are sampled from a stationary model.

\begin{figure}[!h]
 \centering
  \begin{subfigure}{0.45\textwidth}
   \includegraphics[width=\textwidth]{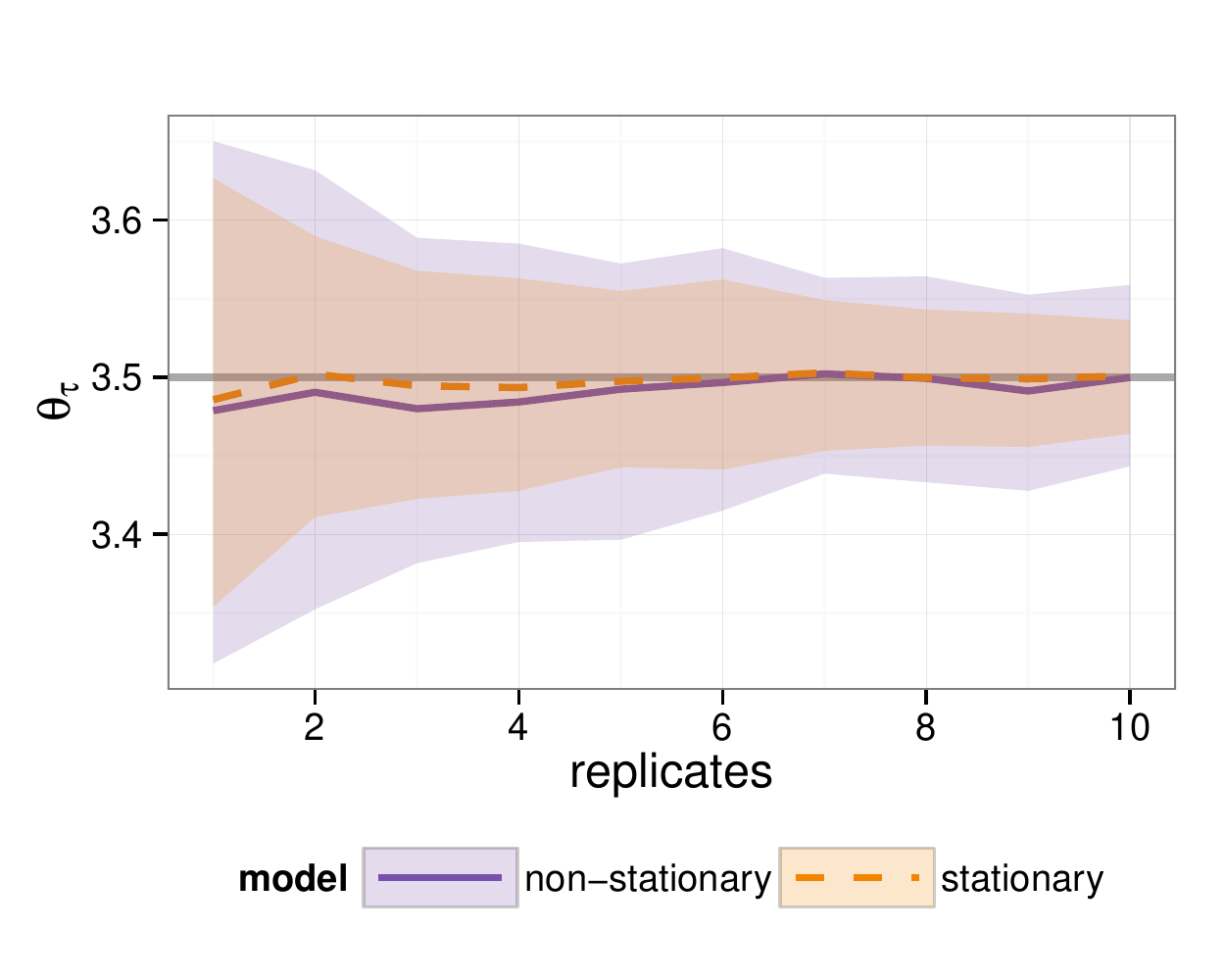}
   \caption{Posterior mean $\theta_{\tau,1}$}
   \label{fig:theta1_stat_data}
  \end{subfigure}
  \begin{subfigure}{0.45\textwidth}
   \includegraphics[width=\textwidth]{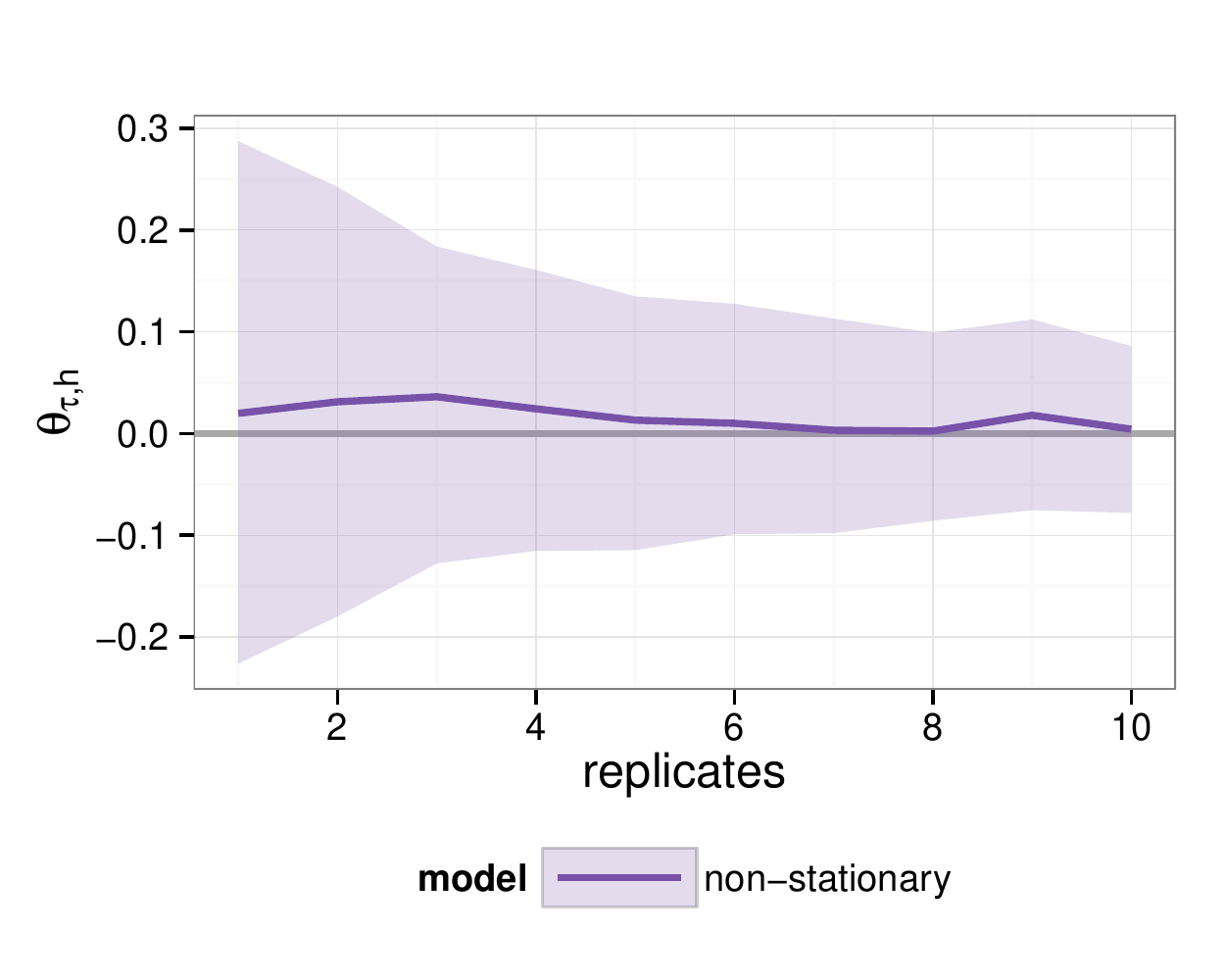}
   \caption{Posterior mean $\theta_{\tau,h}$}
   \label{fig:theta2_stat_data}
  \end{subfigure}

  \begin{subfigure}{0.45\textwidth}
   \includegraphics[width=\textwidth]{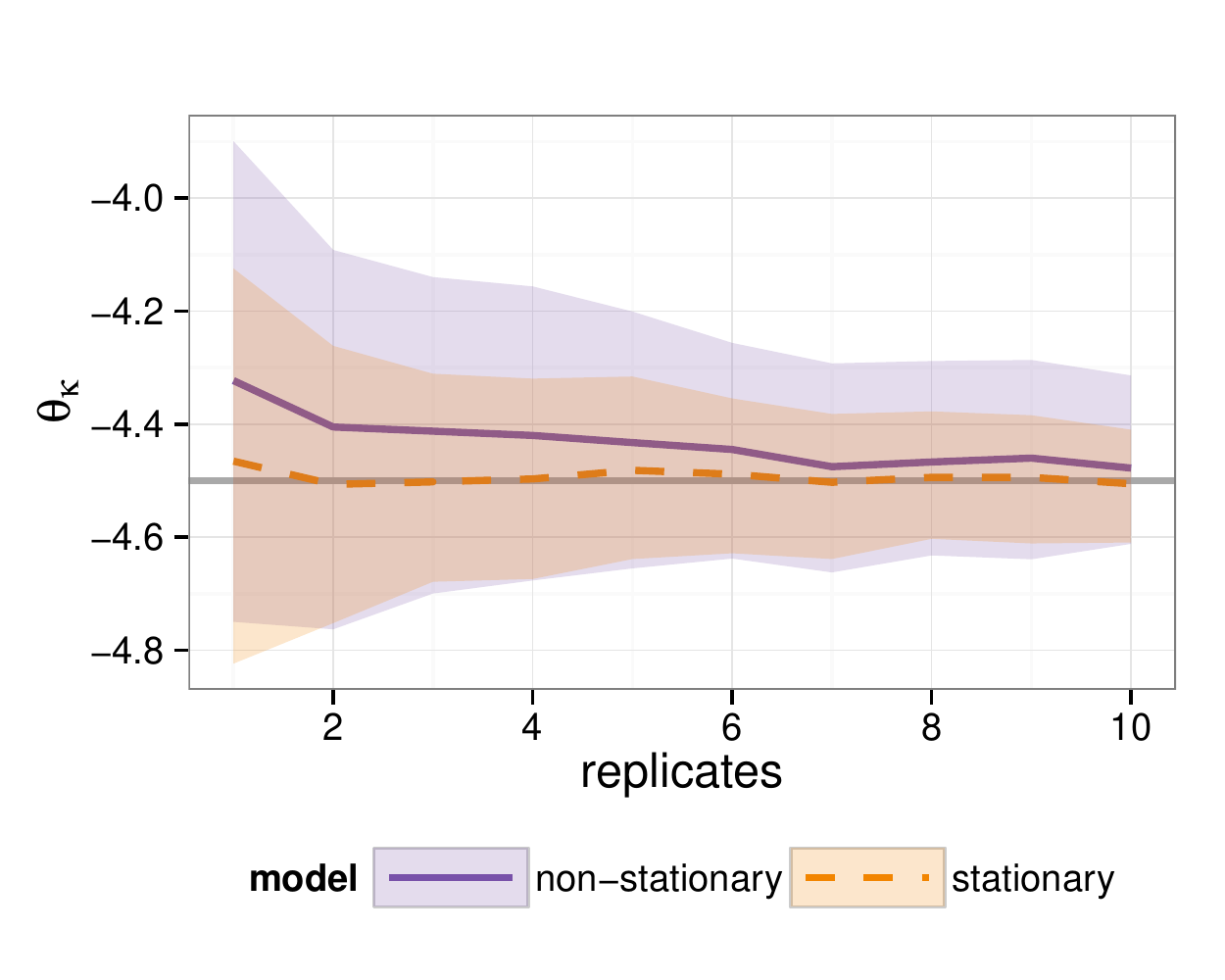}
   \caption{Posterior mean $\theta_{\kappa,1}$}
   \label{fig:theta3_stat_data}
  \end{subfigure}
  \begin{subfigure}{0.45\textwidth}
   \includegraphics[width=\textwidth]{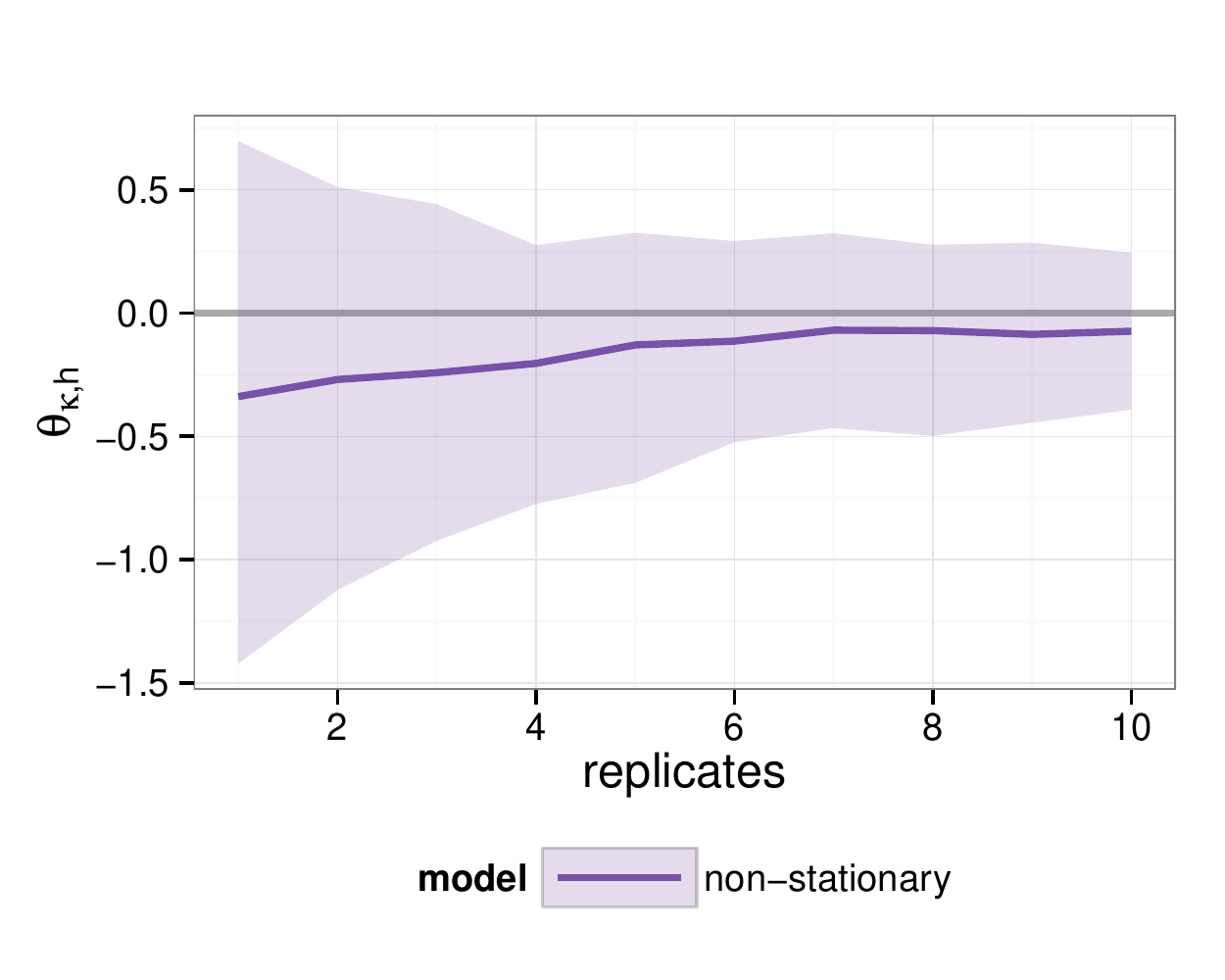}
   \caption{Posterior mean $\theta_{\kappa,h}$}
   \label{fig:theta4_stat_data}
  \end{subfigure}
 \caption{Stationary and non-stationary replicate models were fitted 
          to datasets sampled from the stationary model. Presented 
          are the posterior mean values for the spatial dependence 
          structure parameters $\boldsymbol{\theta}_\text{S}$ and 
          $\boldsymbol{\theta}_\text{NS}$. The 
          lines are averages over all datasets, while the shaded area 
          span the range of the posterior mean values (between the
          0.1- and 0.9-quantiles). The true parameter values are indicated with grey 
          lines.}
 \label{fig:mean_theta_stat_data}
\end{figure}

\begin{figure}
 \centering
  \begin{subfigure}{0.32\textwidth}
   \includegraphics[width=\textwidth]{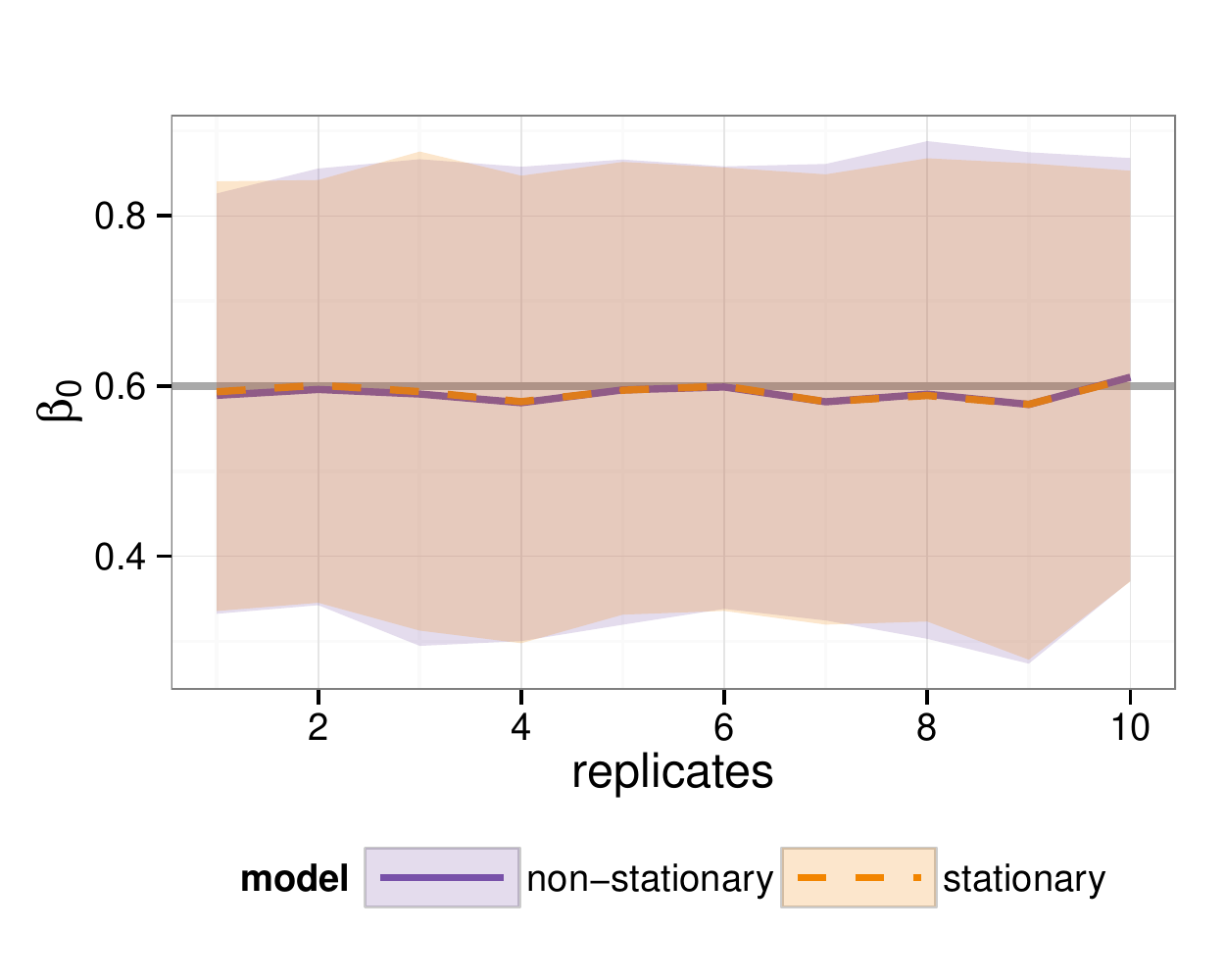}
   \caption{Posterior mean $\beta_0$}
   \label{fig:intercept_stat_data}
  \end{subfigure}
  \begin{subfigure}{0.32\textwidth}
   \includegraphics[width=\textwidth]{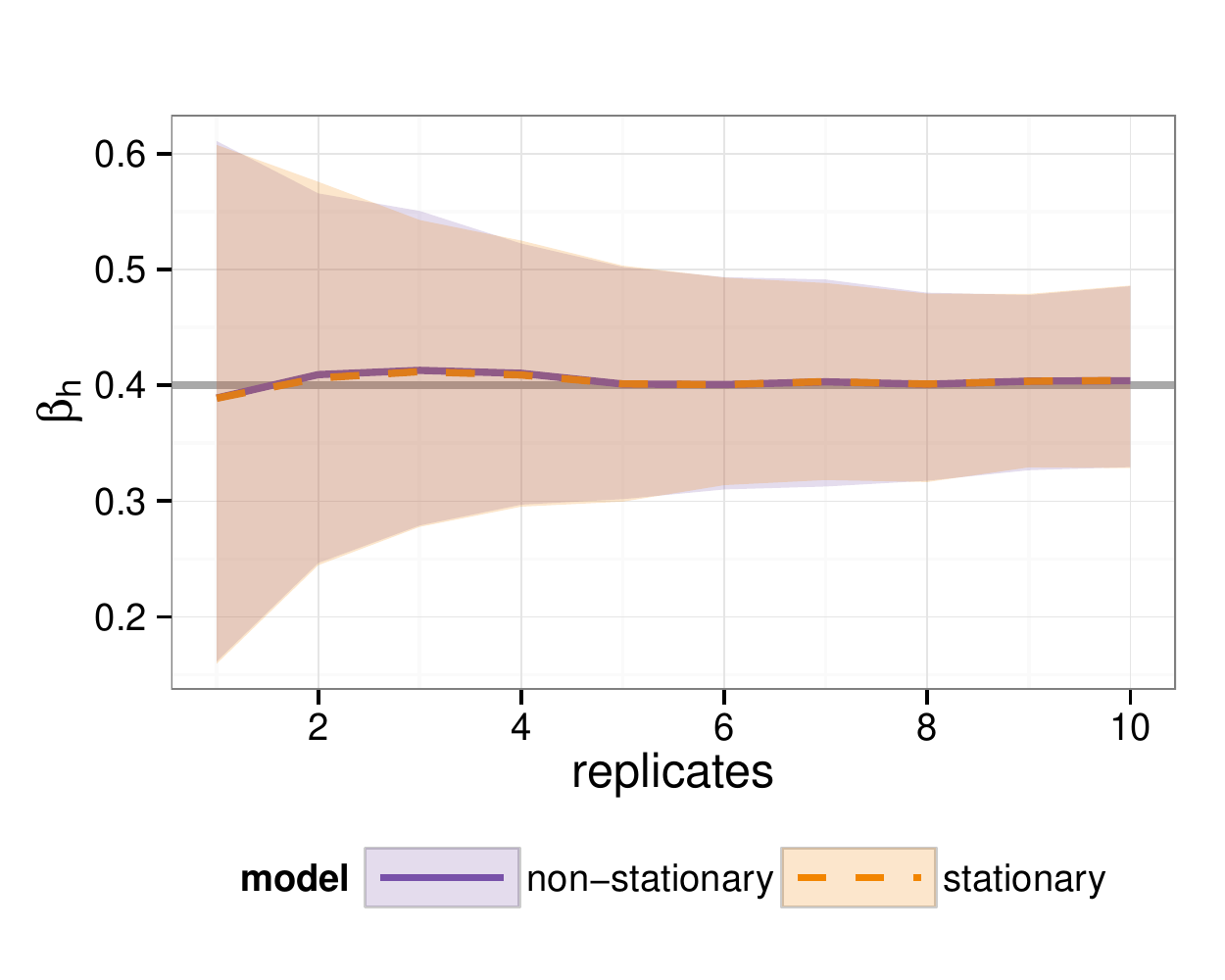}
   \caption{Posterior mean $\beta_h$}
   \label{fig:elevation_stat_data}
  \end{subfigure}
  \begin{subfigure}{0.32\textwidth}
   \includegraphics[width=\textwidth]{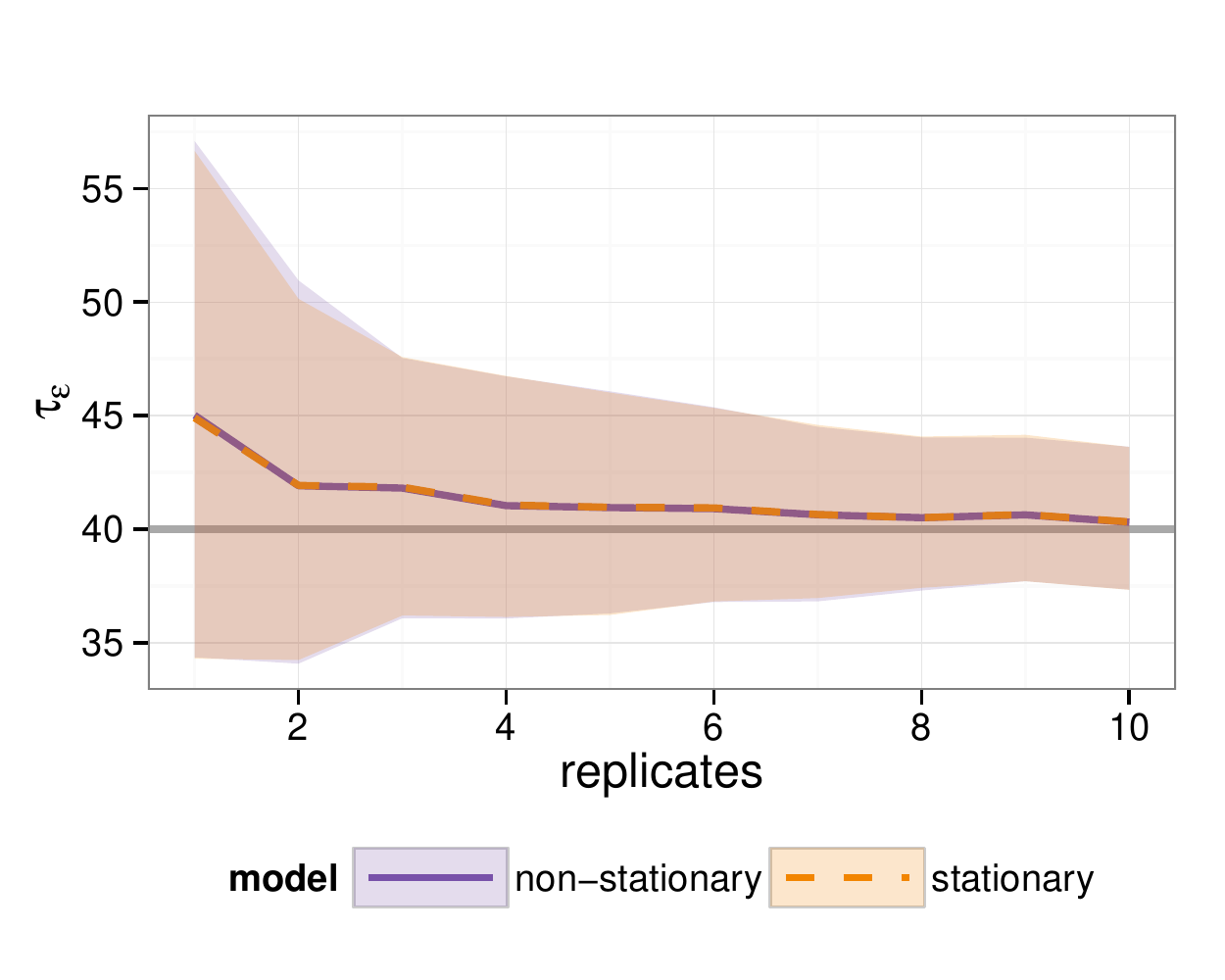}
   \caption{Posterior mean $\tau_\epsilon$}
   \label{fig:tau_stat_data}
  \end{subfigure}
 \caption{Stationary and non-stationary replicate models were fitted 
          to datasets sampled from the stationary model. Presented 
          are the posterior mean values for $\beta_0$, $\beta_h$, and 
          $\tau_\epsilon$. The lines are averages over all datasets, 
          while the shaded area span the range of the posterior mean 
          values (between the 0.1- and 0.9-quantiles). The true parameter values 
          are indicated with grey lines.}
 \label{fig:mean_betas_tau_stat_data}
\end{figure}

\begin{figure}
 \centering
  \begin{subfigure}{0.45\textwidth}
   \includegraphics[width=\textwidth]{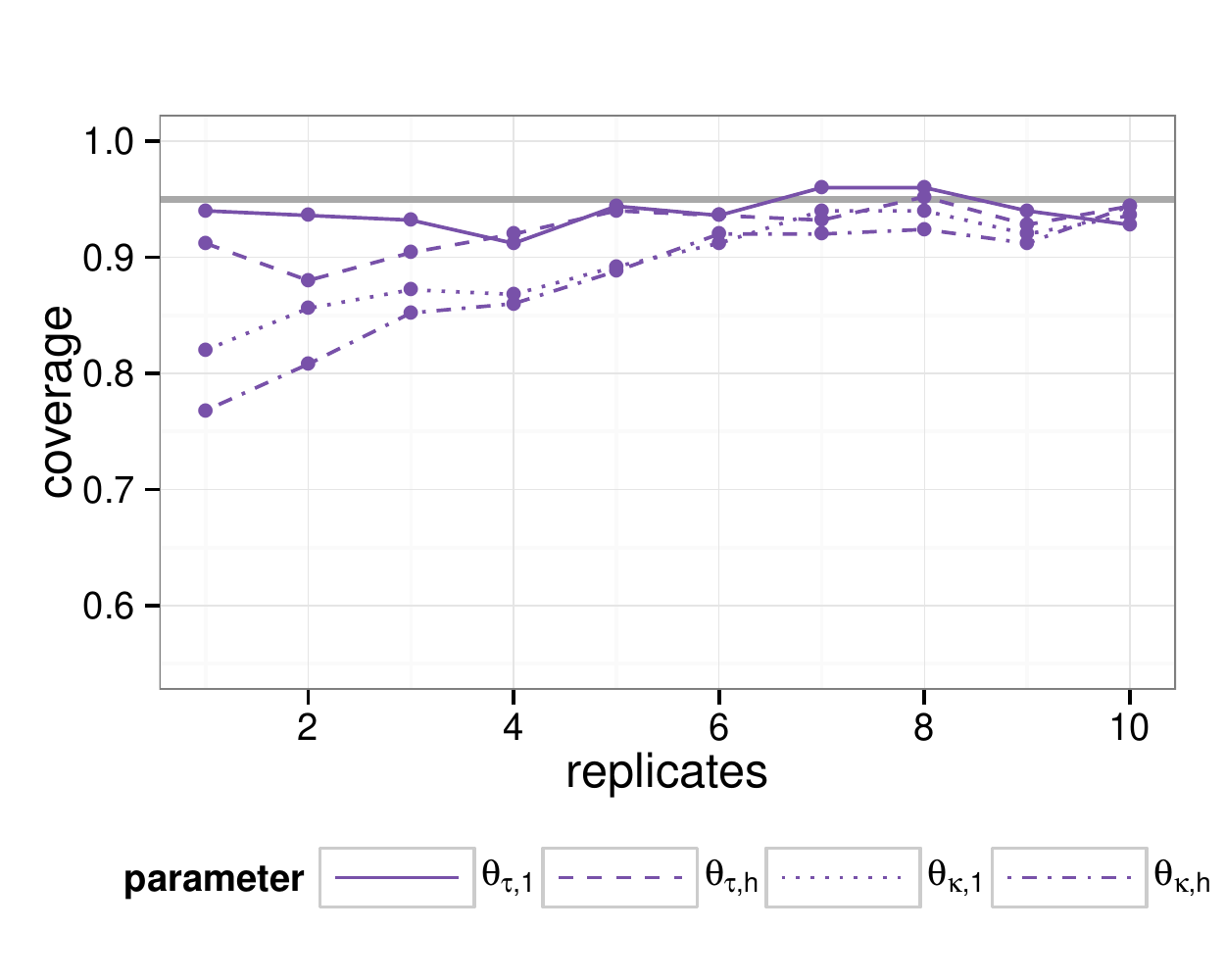}
   \caption{Coverage non-stationary model}
   \label{fig:theta_coverage_nonstat_model_stat_data}
  \end{subfigure}
  \begin{subfigure}{0.45\textwidth}
   \includegraphics[width=\textwidth]{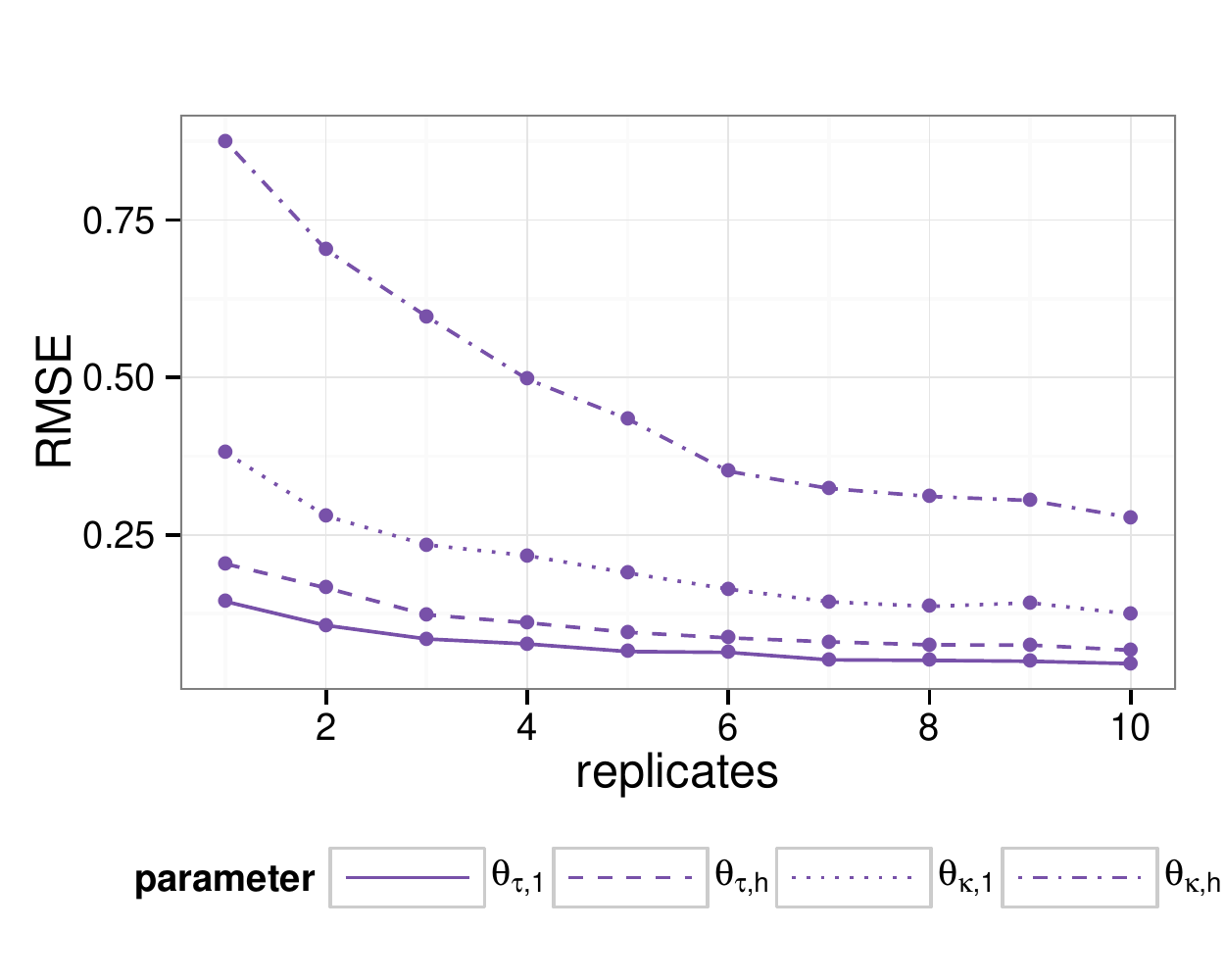}
   \caption{RMSE non-stationary model}
   \label{fig:theta_RMSE_nonstat_model_stat_data}
  \end{subfigure}

  \begin{subfigure}{0.45\textwidth}
   \includegraphics[width=\textwidth]{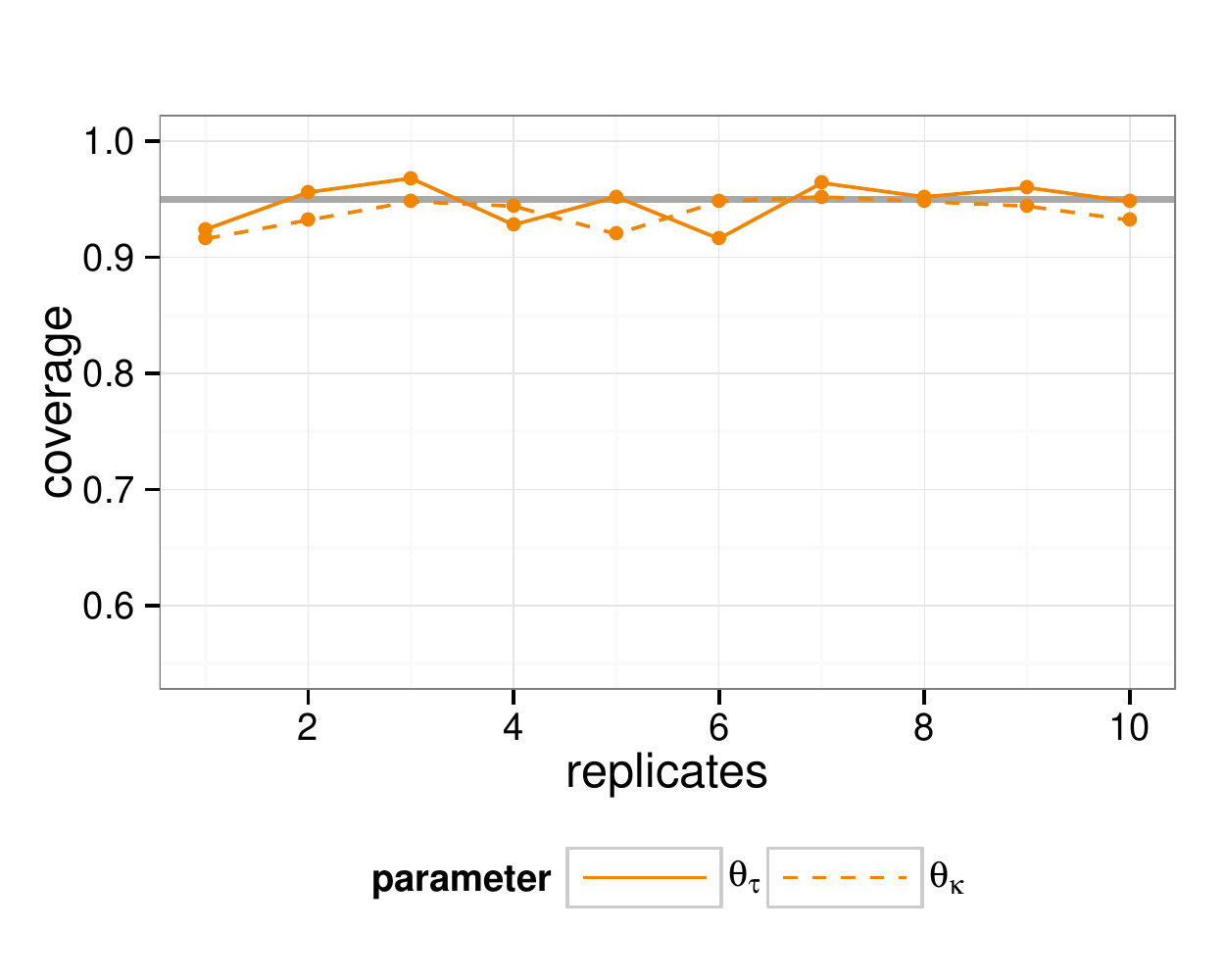}
   \caption{Coverage stationary model}
   \label{fig:theta_coverage_stat_model_stat_data}
  \end{subfigure}
  \begin{subfigure}{0.45\textwidth}
   \includegraphics[width=\textwidth]{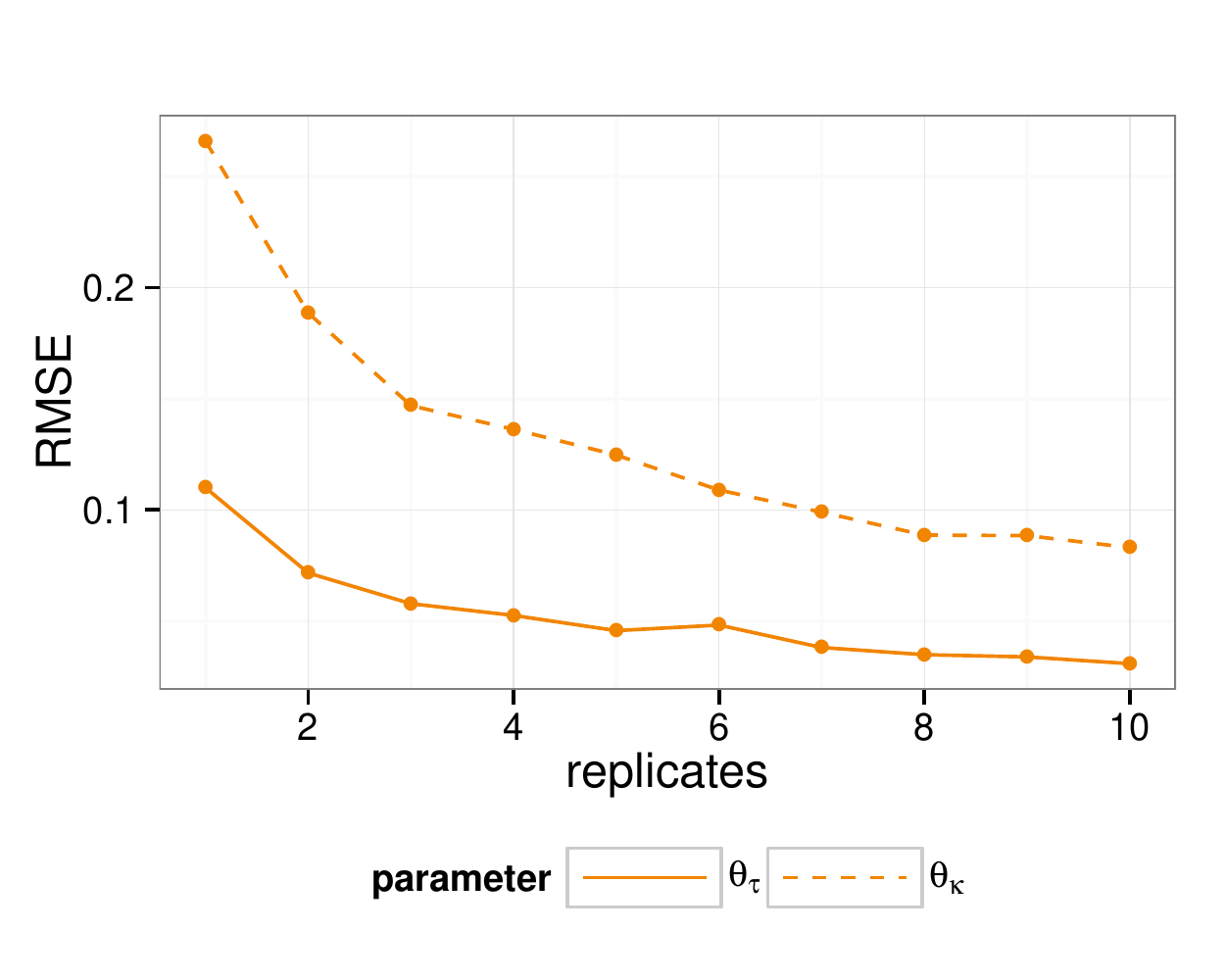}
   \caption{RMSE stationary model}
   \label{fig:theta_RMSE_stat_model_stat_data}
  \end{subfigure}
 \caption{Stationary and non-stationary replicate models were fitted to datasets 
          sampled from the stationary model. Presented are 95\% posterior credible 
          interval coverage and RMSE for the spatial dependence structure parameters   
          $\boldsymbol{\theta}_\text{S}$ and $\boldsymbol{\theta}_\text{NS}$.}
 \label{fig:coverage_RMSE_theta_stat_data}
\end{figure}

\begin{figure}
 \centering
  \begin{subfigure}{0.45\textwidth}
   \includegraphics[width=\textwidth]{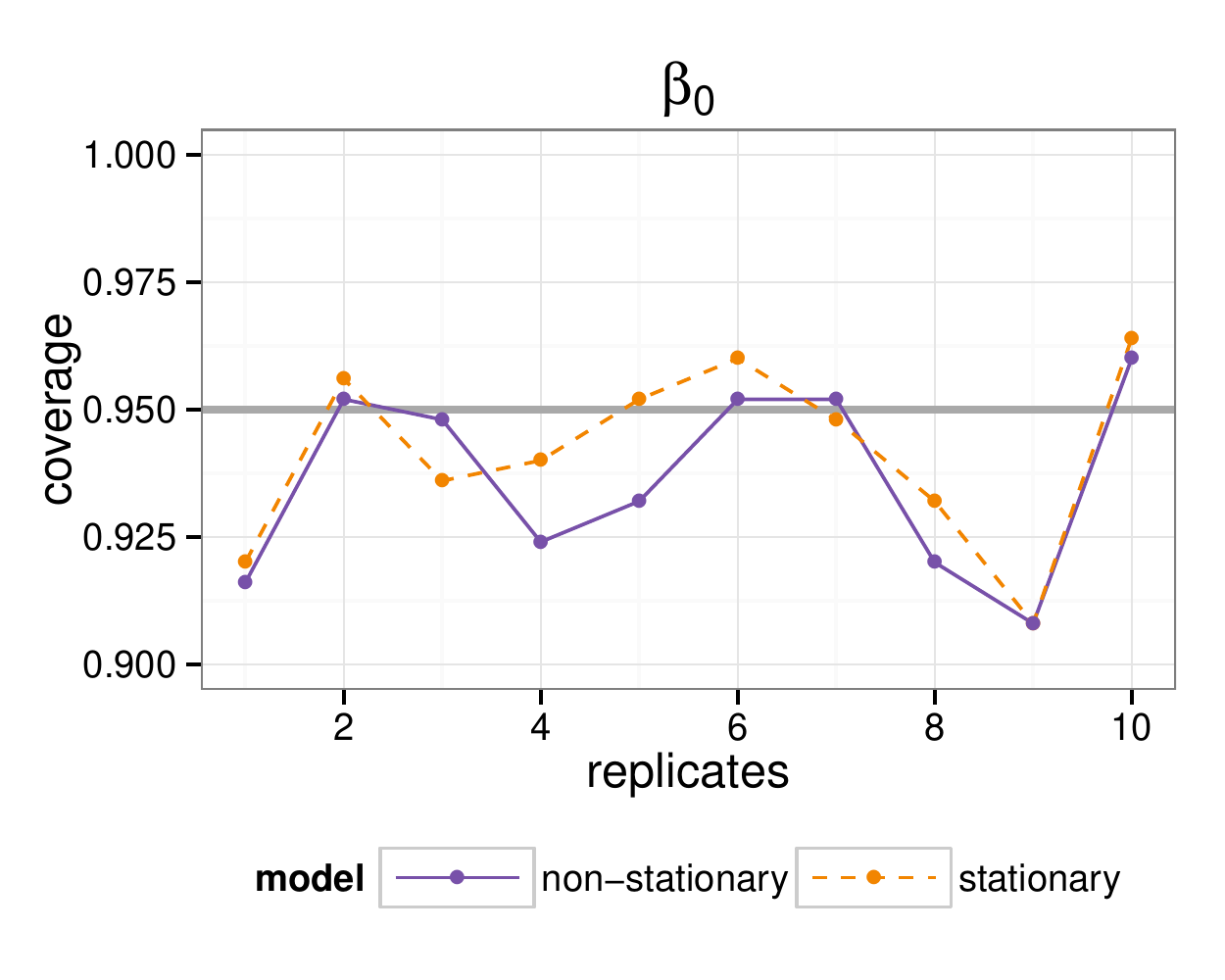}
   \caption{Coverage $\beta_0$}
   \label{fig:intercept_coverage_stat_data}
  \end{subfigure}%
  \begin{subfigure}{0.45\textwidth}
   \includegraphics[width=\textwidth]{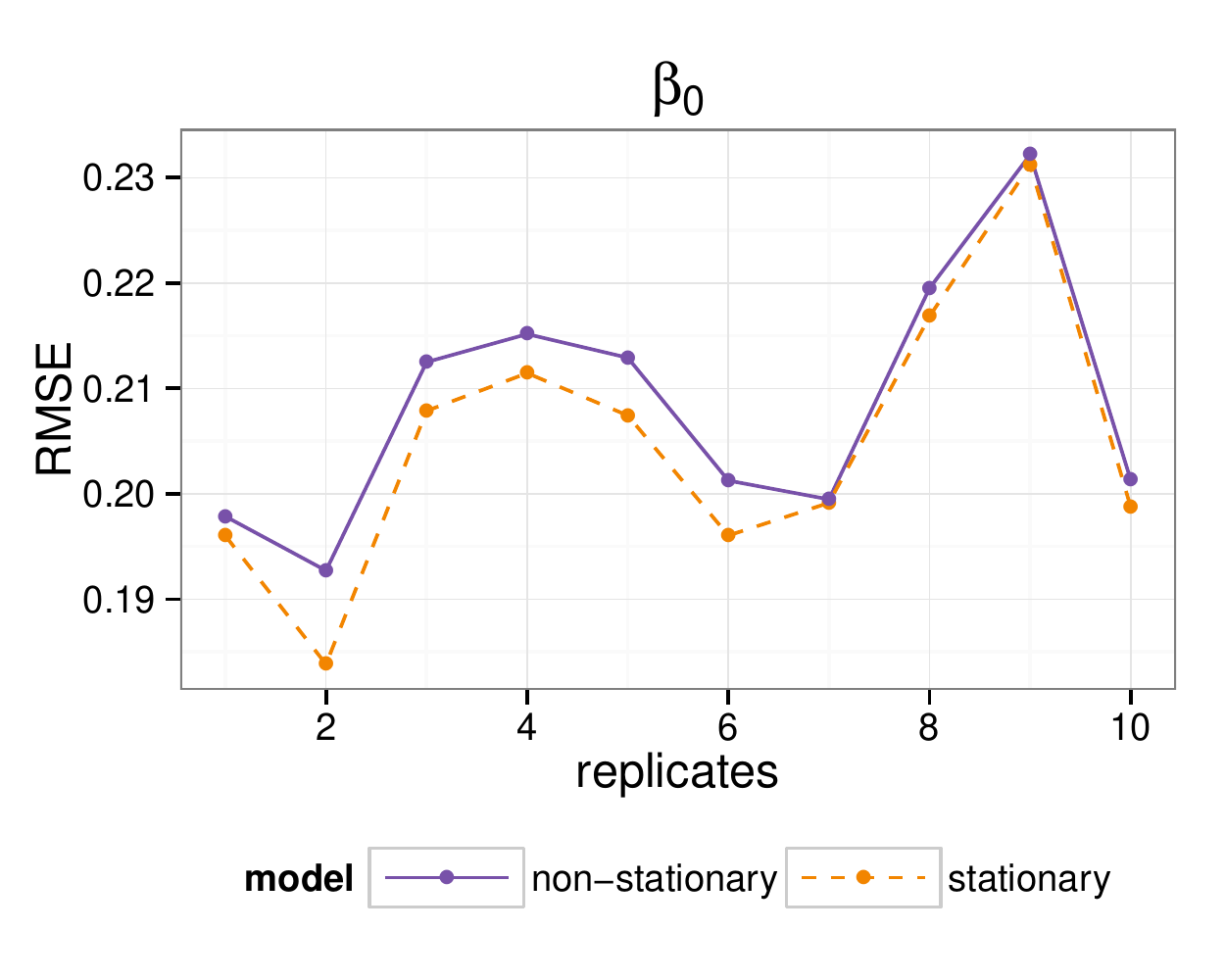}
   \caption{RMSE $\beta_0$}
   \label{fig:intercept_RMSE_stat_data}
  \end{subfigure}

  \begin{subfigure}{0.45\textwidth}
   \includegraphics[width=\textwidth]{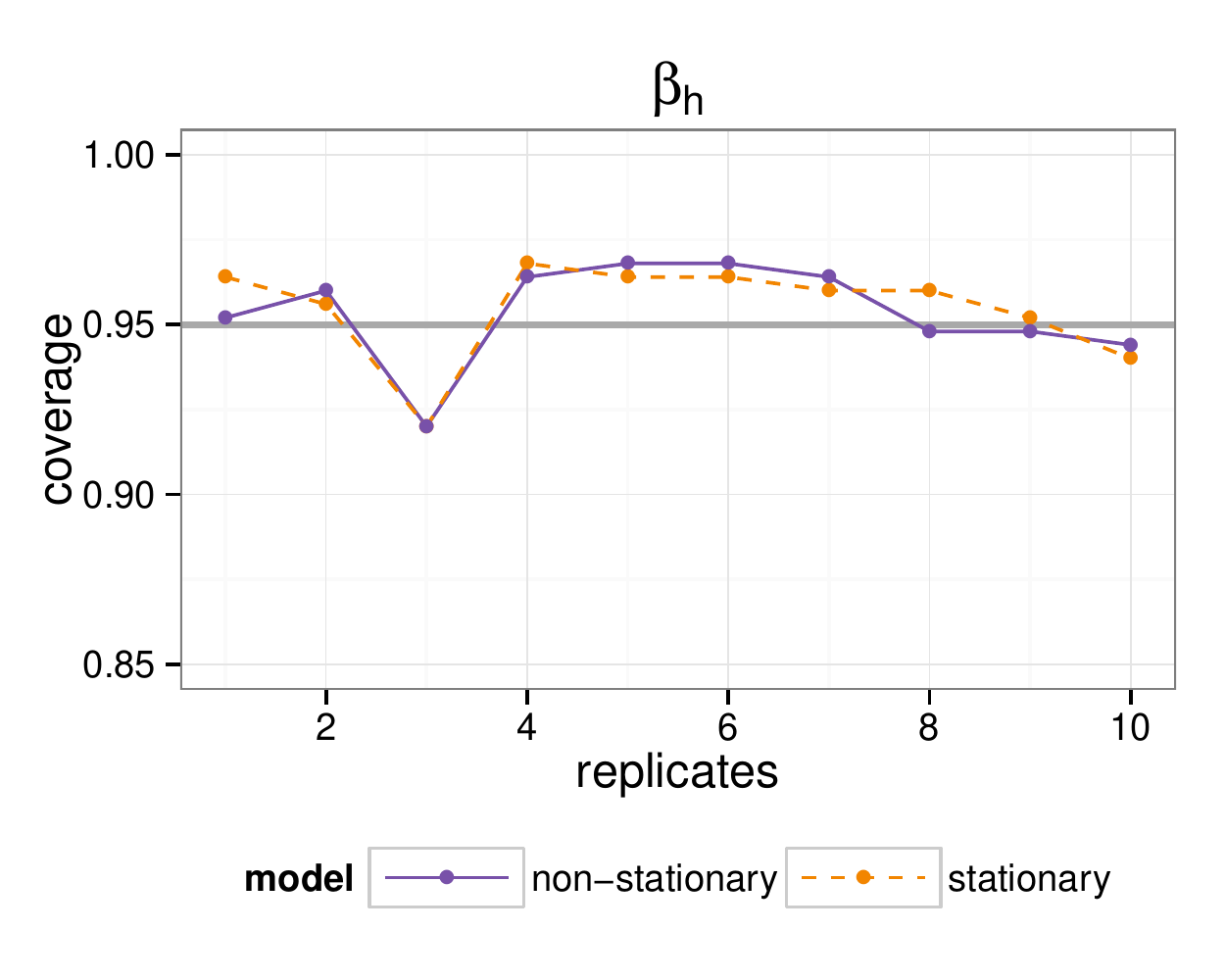}
   \caption{Coverage $\beta_h$}
   \label{fig:elevation_coverage_stat_data}
  \end{subfigure}%
  \begin{subfigure}{0.45\textwidth}
   \includegraphics[width=\textwidth]{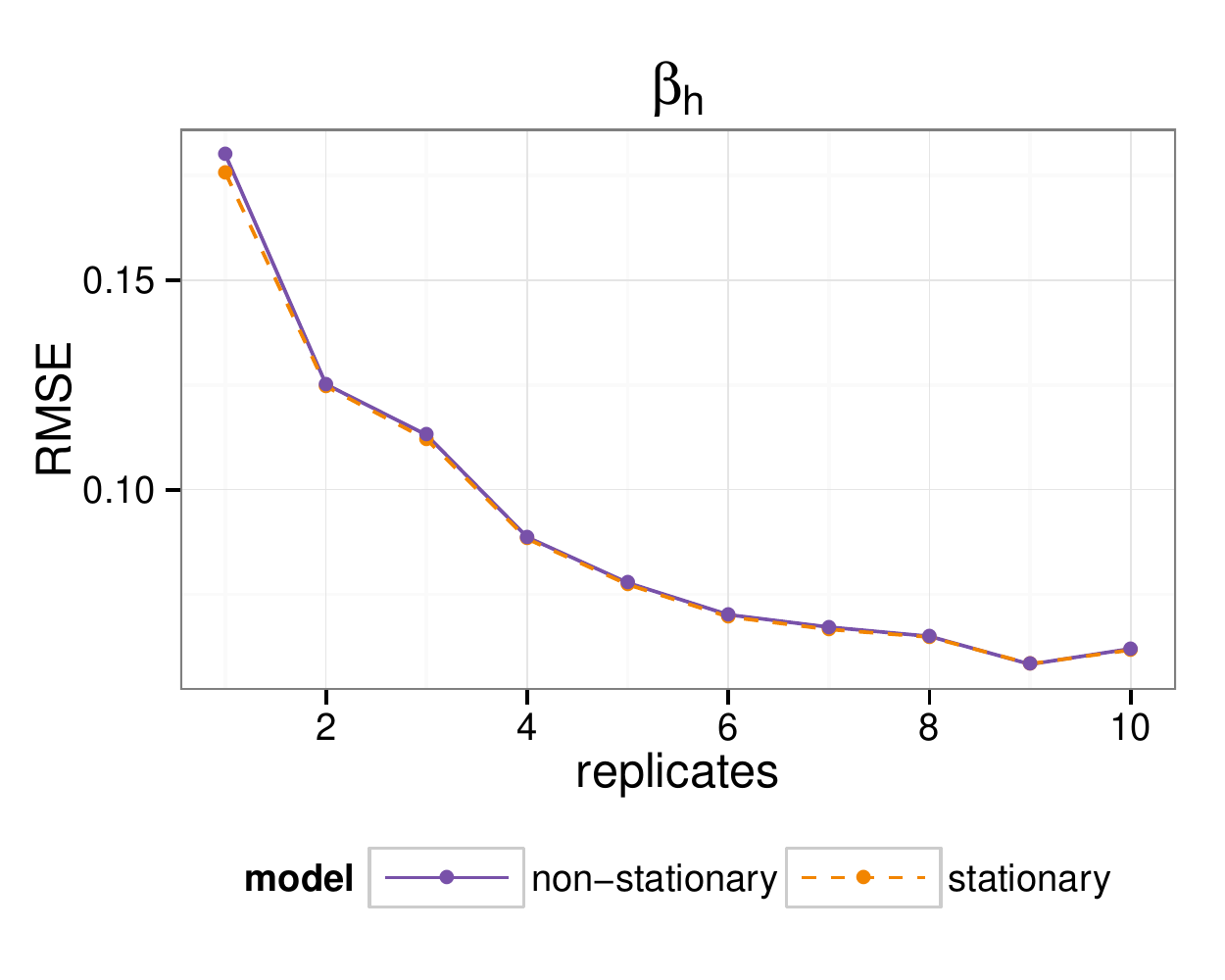}
   \caption{RMSE $\beta_h$}
   \label{fig:elevation_RMSE_stat_data}
  \end{subfigure}

  \begin{subfigure}{0.45\textwidth}
   \includegraphics[width=\textwidth]{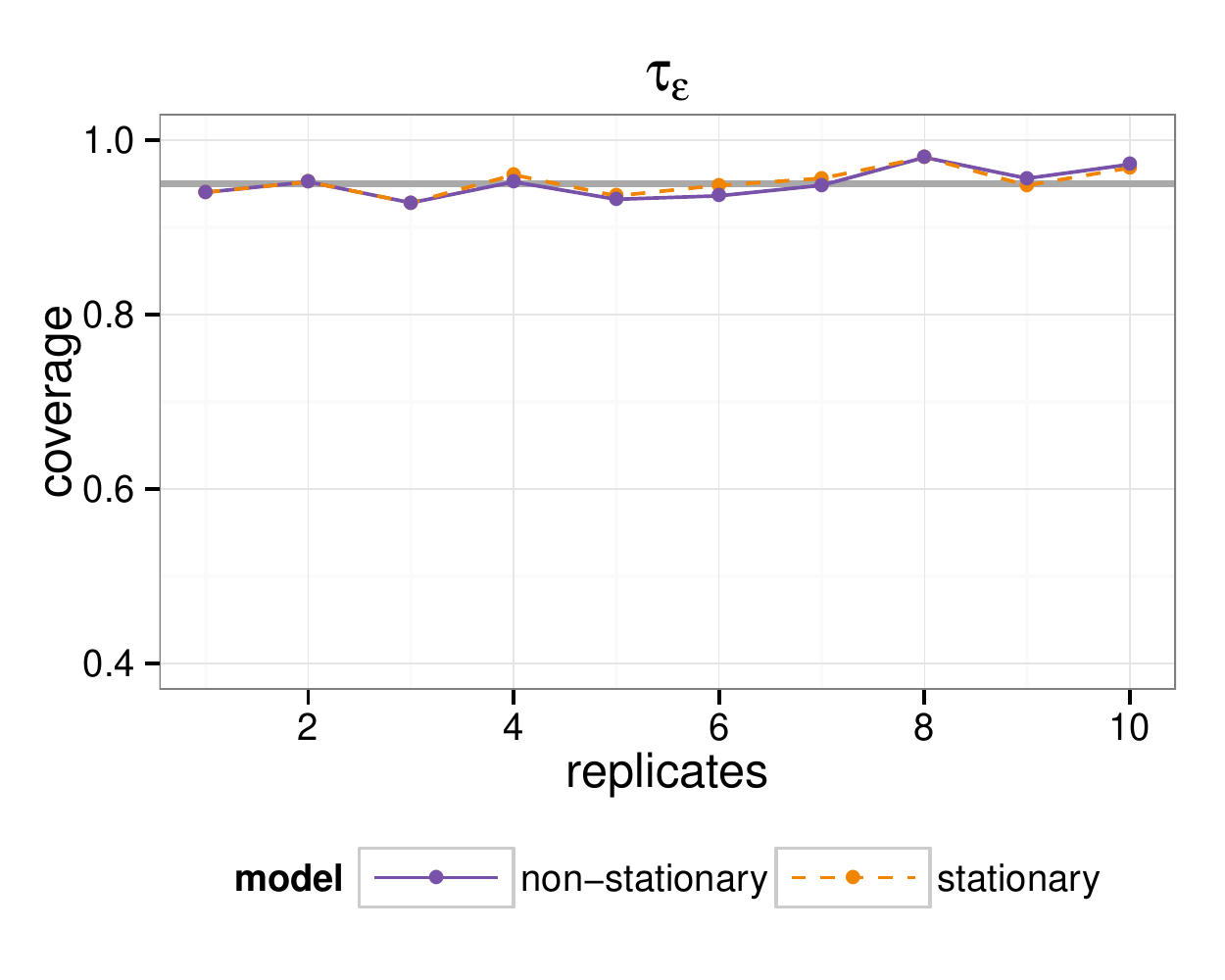}
   \caption{Coverage $\tau_\epsilon$}
   \label{fig:tau_coverage_stat_data}
  \end{subfigure}
  \begin{subfigure}{0.45\textwidth}
   \includegraphics[width=\textwidth]{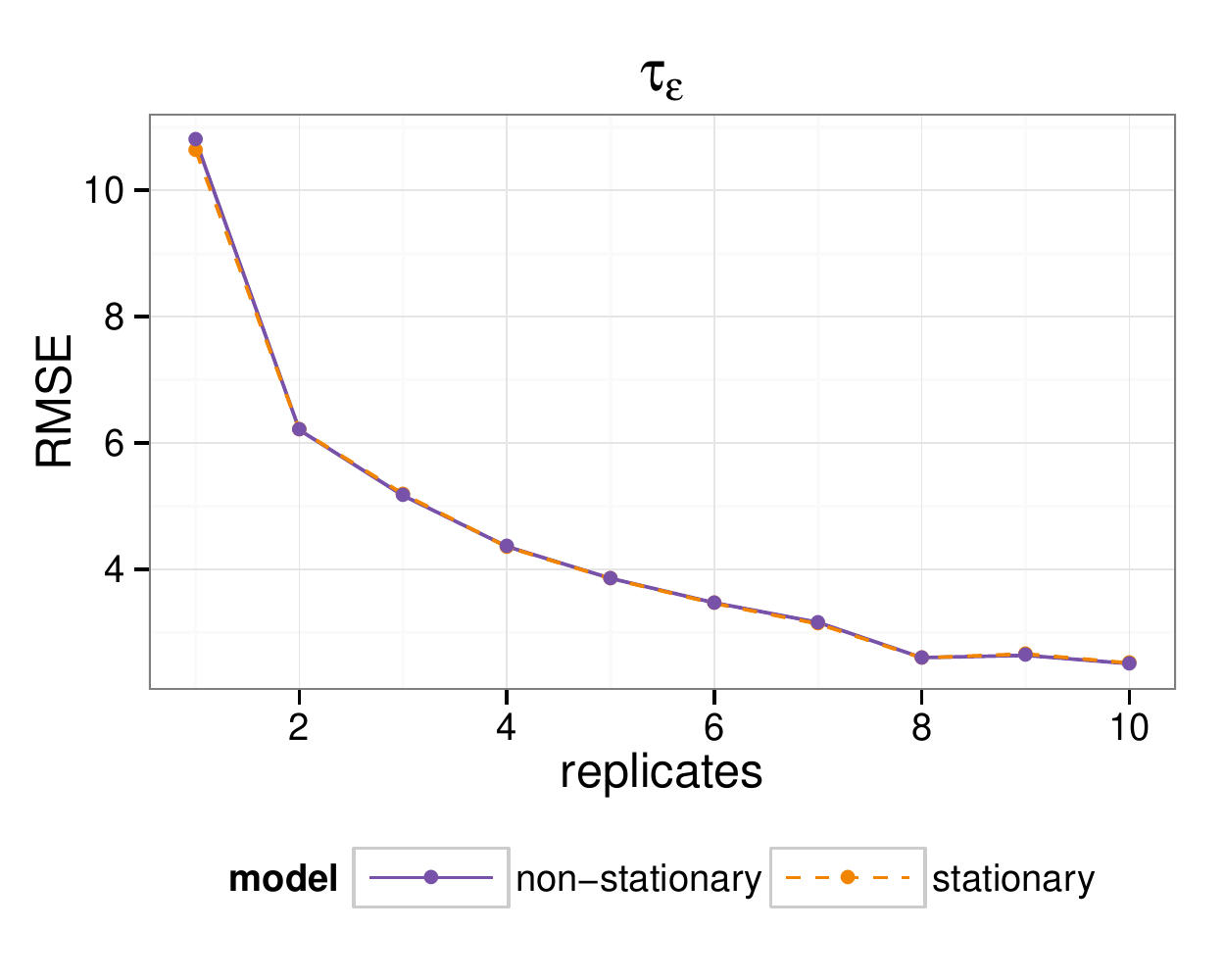}
   \caption{RMSE $\tau_\epsilon$}
   \label{fig:tau_RMSE_stat_data}
  \end{subfigure}
 \caption{Stationary and non-stationary replicate models were fitted 
          to datasets sampled from the stationary model. Presented 
          are 95\% posterior credible interval coverage and RMSE for 
          the parameters $\beta_0$, $\beta_h$, and $\tau_\epsilon$}
 \label{fig:coverage_RMSE_betas_tau_stat_data}
\end{figure}

\clearpage
\section{Prior sensitivity}\label{app:Prior_sensitivity}

In this appendix, the effect of using different coefficients of 
variation is investigated. The priors defined in Section~\ref{sec:prior} 
are used, but with different values for $c_\rho$ and $c_\sigma$, 
thus different variances for the non-stationarity parameters 
$\theta_{\tau, h}$ and $\theta_{\kappa, h}$. 
The stationary priors (and non-stationary prior at sea level) are given 
by the 0.5- and 0.9-quantiles: 150 km and 500 km for the range, and 0.2 m 
and 2 m for the marginal standard deviation. This yields $\mu_\kappa = -3.97$, 
$\sigma_\kappa^2 = 0.88$, $\mu_\tau = 4.31$, and $\sigma_\tau^2 = 2.35$.    
The reference elevation $h_0$ is 0.4 km. Six different non-stationary priors 
are compared, these are summarised in Table~\ref{tab:priors}. 
The prior labelled NS-2 is the one used in the paper. 
The prior labelled NS-1 is closest to the prior used in \citet{Ingebrigtsen2014}.
There, the prior means were set to $\mu_{\kappa,1} = -4$,  $\mu_{\kappa,h} = 0$,  $\mu_{\tau,1} = 4$, and
$\mu_{\tau,h} = 0$, i.e.\ approximately the same as in this paper. 
The prior variances were set to $\sigma_{\kappa,1}^2 = 0.1$, $\sigma_{\kappa,h}^2 = 1$, 
$\sigma_{\tau,1}^2 = 0.1$ and $\sigma_{\tau,h}^2 = 1$, i.e.\ lower variances than what is used for the default prior in this paper.

Table~\ref{tab:estimates} contains posterior quantiles for all model 
parameters for the replicate model and Table~\ref{tab:estimates_2008-2009} 
for the individual model fitted to the 2008-2009 data. 
Cross-validated predictive scores are in Table~\ref{tab:CV_prior_sensitivity}. 
The results are stable with respect to choice of $c_\rho$ and 
$c_\sigma$, most for the replicate model.

\begin{table}[!h]
 \footnotesize
 \centering
 \caption{Priors with different variances for the non-stationarity parameters 
          $\theta_{\tau, h}$ and $\theta_{\kappa, h}$.}
 \label{tab:priors}
 \begin{tabular}{ccccccc}
 \toprule
                         & NS-1 & NS-2 & NS-3 & NS-4 & NS-5 & NS-6 \\
 \midrule
  $c_\rho$               &  0.4 &  0.8 &  1.6 &  0.8 &  0.8 & 3.5 \\
  $c_\sigma$             &  0.6 &  1.3 &  3.4 &  1.0 &  2.0 & 13 \\
  $\sigma_{\tau, h}^2$   &  0.99 & 3.09 & 7.88 & 1.24 & 6.97 & 15.95 \\
  $\sigma_{\kappa, h}^2$ &  0.93 & 3.09 & 7.94 & 3.09 & 3.09 & 16.15\\
 \bottomrule
 \end{tabular}
\end{table}

\begin{table}
 \footnotesize
 \centering
 \caption{Posterior quantiles for the model parameters. 
          The replicate model was fitted to the southern 
          Norway 2008-2013 dataset, with stationary dependence 
          structure (S) and non-stationary dependence structure (NS-). 
          Six different priors were used for the non-stationary model.}
 \label{tab:estimates}
 \begin{tabular}{ccccccccccccccccccccccc}
  \toprule
    
                        && S & NS-1 & NS-2 & NS-3 & NS-4 & NS-5 & NS-6 \\
   parameter & quantiles \\
                    & \emph{\footnotesize{0.025}} & 0.25 & 0.13 & 0.14 & 0.15 & 0.14 & 0.15 & 0.15 \\ 
 $\beta_1$          & \emph{\footnotesize{0.500}} & 0.70 & 0.35 & 0.37 & 0.37 & 0.36 & 0.37 & 0.37 \\ 
                    & \emph{\footnotesize{0.975}} & 1.10 & 0.56 & 0.57 & 0.57 & 0.57 & 0.57 & 0.58 \\ 
 \hline
                    & \emph{\footnotesize{0.025}} & 0.11 & -0.01& 0.00 & 0.00 & -0.01& 0.00 & 0.00 \\ 
 $\beta_2$          & \emph{\footnotesize{0.500}} & 0.55 & 0.21 & 0.22 & 0.22 & 0.21 & 0.22 & 0.22 \\ 
                    & \emph{\footnotesize{0.975}} & 0.95 & 0.41 & 0.42 & 0.42 & 0.42 & 0.42 & 0.42 \\ 
\hline
                    & \emph{\footnotesize{0.025}} & 0.25 & 0.16 & 0.17 & 0.17 & 0.16 & 0.17 & 0.18 \\ 
 $\beta_3$          & \emph{\footnotesize{0.500}} & 0.70 & 0.38 & 0.39 & 0.40 & 0.39 & 0.40 & 0.40 \\ 
                    & \emph{\footnotesize{0.975}} & 1.11 & 0.59 & 0.59 & 0.60 & 0.59 & 0.60 & 0.60 \\ 
\hline
                    & \emph{\footnotesize{0.025}} & 0.55 & 0.23 & 0.25 & 0.25 & 0.24 & 0.25 & 0.25 \\ 
 $\beta_4$          & \emph{\footnotesize{0.500}} & 1.03 & 0.47 & 0.49 & 0.49 & 0.48 & 0.49 & 0.50 \\ 
                    & \emph{\footnotesize{0.975}} & 1.47 & 0.70 & 0.71 & 0.72 & 0.71 & 0.72 & 0.72 \\ 
\hline
                    & \emph{\footnotesize{0.025}} & 0.31 & 0.10 & 0.11 & 0.12 & 0.11 & 0.12 & 0.12 \\ 
 $\beta_5$          & \emph{\footnotesize{0.500}} & 0.75 & 0.32 & 0.33 & 0.34 & 0.33 & 0.34 & 0.34 \\ 
                    & \emph{\footnotesize{0.975}} & 1.16 & 0.52 & 0.53 & 0.53 & 0.52 & 0.53 & 0.53 \\ 
\hline
                    & \emph{\footnotesize{0.025}} & 0.29 & 0.19 & 0.18 & 0.18 & 0.19 & 0.18 & 0.18 \\ 
 $\beta_h$          & \emph{\footnotesize{0.500}} & 0.49 & 0.33 & 0.32 & 0.32 & 0.33 & 0.32 & 0.32 \\ 
                    & \emph{\footnotesize{0.975}} & 0.70 & 0.49 & 0.48 & 0.48 & 0.48 & 0.48 & 0.48 \\ 
\hline
                    & \emph{\footnotesize{0.025}} & 28.63 & 35.14 & 34.86 & 34.87 & 34.94 & 34.94 & 34.86 \\ 
 $\tau_\epsilon$    & \emph{\footnotesize{0.500}} & 35.45 & 43.62 & 43.51 & 43.29 & 43.42 & 43.39 & 43.33 \\ 
                    & \emph{\footnotesize{0.975}} & 43.90 & 54.18 & 53.83 & 53.75 & 53.90 & 53.92 & 53.78 \\ 
\hline
                    & \emph{\footnotesize{0.025}} & 3.37 & 3.80 & 3.81 & 3.82 & 3.81 & 3.82 & 3.82 \\ 
 $\theta_{\tau,1}$  & \emph{\footnotesize{0.500}} & 3.49 & 3.94 & 3.96 & 3.96 & 3.95 & 3.96 & 3.97 \\ 
                    & \emph{\footnotesize{0.975}} & 3.60 & 4.08 & 4.10 & 4.11 & 4.10 & 4.11 & 4.11 \\ 
\hline
                    & \emph{\footnotesize{0.025}} &      & -1.55 & -1.62 & -1.65 & -1.60 & -1.63 & -1.65 \\ 
 $\theta_{\tau,h}$  & \emph{\footnotesize{0.500}} &      & -1.23 & -1.29 & -1.31 & -1.27 & -1.30 & -1.31 \\ 
                    & \emph{\footnotesize{0.975}} &      & -0.93 & -0.97 & -0.98 & -0.96 & -0.97 & -0.98 \\ 
\hline
                    & \emph{\footnotesize{0.025}} & -4.78 & -6.16 & -6.22 & -6.24 & -6.23 & -6.22 & -6.24 \\ 
 $\theta_{\kappa,1}$& \emph{\footnotesize{0.500}} & -4.50 & -5.84 & -5.89 & -5.91 & -5.89 & -5.89 & -5.91 \\ 
                    & \emph{\footnotesize{0.975}} & -4.23 & -5.53 & -5.57 & -5.58 & -5.57 & -5.57 & -5.59 \\ 
\hline
                    & \emph{\footnotesize{0.025}} &      & 2.66 & 2.75 & 2.77 & 2.74 & 2.75 & 2.78 \\ 
 $\theta_{\kappa,h}$& \emph{\footnotesize{0.500}} &      & 3.02 & 3.12 & 3.15 & 3.11 & 3.12 & 3.15 \\ 
                    & \emph{\footnotesize{0.975}} &      & 3.39 & 3.51 & 3.54 & 3.50 & 3.51 & 3.55 \\ 

  \bottomrule
 \end{tabular}
\end{table}

\begin{table}
 \footnotesize
 \centering
 \caption{Quantiles of the posterior marginal distributions for all model parameters. 
          The individual model was fitted to annual precipitation data from 2008-2009, 
          with stationary dependence structure (S) and non-stationary dependence 
          structure (NS-). Six different priors were used for the non-stationary model.}
 \label{tab:estimates_2008-2009}
 \begin{tabular}{ccccccccccccccccccccccc}
  \toprule
                        && S & NS-1 & NS-2 & NS-3 & NS-4 & NS-5 & NS-6 \\
   parameter & quantiles \\
                    & \emph{\footnotesize{0.025}}  & 0.02 & -0.29 & -0.26 & -0.25 & -0.28 & -0.25 & -0.24 \\  
 $\beta_1$          & \emph{\footnotesize{0.500}}  & 0.54 &  0.13 &  0.15 &  0.16 &  0.13 &  0.16 &  0.16 \\  
                    & \emph{\footnotesize{0.975}}  & 0.96 &  0.48 &  0.49 &  0.50 &  0.48 &  0.50 &  0.50 \\ 
\hline
                    & \emph{\footnotesize{0.025}}  & 0.35 & 0.26 & 0.24 & 0.23 & 0.25 & 0.23 & 0.23 \\ 
 $\beta_h$          & \emph{\footnotesize{0.500}}  & 0.78 & 0.55 & 0.53 & 0.51 & 0.54 & 0.52 & 0.51 \\ 
                    & \emph{\footnotesize{0.975}}  & 1.22 & 0.88 & 0.85 & 0.84 & 0.87 & 0.85 & 0.84 \\ 
\hline
                    & \emph{\footnotesize{0.025}}  & 32.21 & 43.35 & 43.60 & 43.62 & 43.54 & 43.67 & 43.58 \\ 
 $\tau_\epsilon$    & \emph{\footnotesize{0.500}}  & 48.85 & 65.91 & 66.30 & 66.31 & 66.20 & 66.34 & 66.29 \\ 
                    & \emph{\footnotesize{0.975}}  & 73.61 & 99.69 & 99.96 & 99.99 & 99.83 & 100.08 & 99.95 \\ 
\hline
                    & \emph{\footnotesize{0.025}}  & 3.28 & 3.63 & 3.67 & 3.68 & 3.66 & 3.67 & 3.69 \\ 
 $\theta_{\tau,1}$  & \emph{\footnotesize{0.500}}  & 3.45 & 3.86 & 3.90 & 3.92 & 3.89 & 3.91 & 3.92 \\ 
                    & \emph{\footnotesize{0.975}}  & 3.64 & 4.09 & 4.14 & 4.16 & 4.12 & 4.15 & 4.16 \\ 
\hline
                    & \emph{\footnotesize{0.025}}  &      & -1.57 & -1.73 & -1.79 & -1.68 & -1.75 & -1.81 \\ 
 $\theta_{\tau,h}$  & \emph{\footnotesize{0.500}}  &      & -1.09 & -1.21 & -1.25 & -1.17 & -1.22 & -1.26 \\ 
                    & \emph{\footnotesize{0.975}}  &      & -0.60 & -0.70 & -0.72 & -0.67 & -0.71 & -0.73 \\ 
\hline
                    & \emph{\footnotesize{0.025}}  & -5.19 & -6.11 & -6.30 & -6.34 & -6.30 & -6.29 & -6.36 \\
 $\theta_{\kappa,1}$& \emph{\footnotesize{0.500}}  & -4.61 & -5.60 & -5.74 & -5.77 & -5.74 & -5.73 & -5.79 \\ 
                    & \emph{\footnotesize{0.975}}  & -4.11 & -5.09 & -5.20 & -5.23 & -5.20 & -5.19 & -5.24 \\ 
\hline
                    & \emph{\footnotesize{0.025}}  &       & 2.01 & 2.23 & 2.29 & 2.21 & 2.24 & 2.31 \\ 
 $\theta_{\kappa,h}$& \emph{\footnotesize{0.500}}  &       & 2.61 & 2.85 & 2.92 & 2.83 & 2.86 & 2.95 \\ 
                    & \emph{\footnotesize{0.975}}  &       & 3.21 & 3.51 & 3.60 & 3.49 & 3.52 & 3.64 \\ 
  \bottomrule
 \end{tabular}
\end{table}

\begin{table}
 \footnotesize
 \centering
 \caption{Cross-validated predictive scores.  
          The individual and replicate models were fitted to annual precipitation data from southern Norway, 
          with stationary dependence structure (S) and non-stationary dependence 
          structure (NS-). Six different priors were used for the non-stationary model.}
 \label{tab:CV_prior_sensitivity}
 \begin{tabular}{ccccccccccccccccccccccc}
  \toprule
                        && S & NS-1 & NS-2 & NS-3 & NS-4 & NS-5 & NS-6 \\
   CRPS &  \\
              & individual  & 0.1436 & 0.1383 & 0.1382 & 0.1383 & 0.1380 & 0.1383 & 0.1385 \\  
              & replicate   & 0.1438 & 0.1435 & 0.1436 & 0.1437 & 0.1435 & 0.1436 & 0.1437 \\  
 \\
   RMSE &  \\
              & individual  & 0.2655 & 0.2537 & 0.2531 & 0.2534 & 0.2530 & 0.2532 & 0.2536 \\ 
              & replicate   & 0.2646 & 0.2565 & 0.2567 & 0.2569 & 0.2566 & 0.2568 & 0.2569 \\
  \bottomrule
 \end{tabular}
\end{table}

\end{appendices}

\end{document}